\documentclass[11pt,a4paper]{article}

\usepackage[utf8]{inputenc}
\usepackage[T1]{fontenc}
\usepackage{lmodern}
\usepackage{amsmath,amssymb,amsthm}
\usepackage{enumerate}
\usepackage{xcolor}
\usepackage{graphicx}
\usepackage{booktabs}
\usepackage{array}

\usepackage{hyperref}
\hypersetup{pdftitle={Tableau Consequence under Positive Contexts: A Reconstruction of Walukiewicz's Completeness Proof},pdfauthor={Kuniaki Tamura}}
\usepackage[margin=2.5cm]{geometry}

\theoremstyle{plain}
\newtheorem{theorem}{Theorem}[section]
\newtheorem{lemma}[theorem]{Lemma}
\newtheorem{proposition}[theorem]{Proposition}
\newtheorem{corollary}[theorem]{Corollary}
\newtheorem{claim}[theorem]{Claim}

\theoremstyle{definition}
\newtheorem{definition}[theorem]{Definition}
\newtheorem{example}[theorem]{Example}
\newtheorem{construction}[theorem]{Construction}

\theoremstyle{remark}
\newtheorem{remark}[theorem]{Remark}

\newcommand{\Prop}{\mathsf{Prop}}
\newcommand{\Koz}{\mathsf{Koz}}
\newcommand{\Cl}{\mathsf{Cl}}
\newcommand{\Sub}{\mathsf{Sub}}
\newcommand{\Free}{\mathsf{Free}}
\newcommand{\Bound}{\mathsf{Bound}}
\newcommand{\Lit}{\mathsf{Lit}}
\newcommand{\Dis}{\mathsf{Dis}}
\newcommand{\Out}{\mathsf{Out}}
\newcommand{\Reg}{\mathsf{Reg}}
\newcommand{\ANF}{\mathsf{ANF}}
\newcommand{\anf}{\mathsf{anf}}
\newcommand{\ret}{\mathsf{ret}}
\newcommand{\pri}{\mathsf{pri}}
\newcommand{\Lab}{\mathsf{Lab}}
\newcommand{\own}{\mathsf{own}}
\newcommand{\Ch}{\mathsf{Ch}}
\newcommand{\Mod}{\mathsf{Mod}}
\newcommand{\RefPlayer}{\mathsf{Ref}}
\newcommand{\ModelPlayer}{\mathsf{Modl}}
\newcommand{\Graph}{\mathsf{Graph}}
\newcommand{\Dep}{\mathsf{Dep}}
\newcommand{\Fix}{\mathsf{Fix}}
\newcommand{\llangle}{\langle\!\langle}
\newcommand{\rrangle}{\rangle\!\rangle}
\let\relsim\sim
\renewcommand{\sim}{\mathord{\relsim}}

\title{Tableau Consequence under Positive Contexts:\\ A Reconstruction of Walukiewicz's Completeness Proof}

\author{Kuniaki Tamura\\
\small Independent Researcher, Tokyo, Japan\\
\small \texttt{kuniaki.tamura@gmail.com}}

\date{}

\begin{document}

\maketitle

\begin{abstract}
We establish a positive-context lifting theorem for tableau consequence in the modal $\mu$-calculus: under explicit syntactic hypotheses, $\mathcal T_B\Rightarrow\mathcal T_C$ induces $\mathcal T_{A(B)}\Rightarrow\mathcal T_{A(C)}$.
The proof orders instances of the given relation by birth time and assigns overlapping formula occurrences to the oldest supporting instance.
A corollary gives bidirectional replacement by automaton normal form inside positive contexts.
Together with a separate unfolding consequence lemma and ANF bisimulation, the lifting theorem yields a self-contained reconstruction of Walukiewicz's completeness proof for Kozen's axiomatisation and makes its least-fixpoint step more transparent.
We also give a finite parity-game proof of regular-refutation existence and a two-sided transfer from tableau consequence to thin refutations.
\end{abstract}

{\sloppy\noindent\textbf{Keywords:} modal $\mu$-calculus, Kozen's axiomatisation, completeness, canonical tableaux, tableau consequence, automaton normal form, parity automata.\par}


\section{Introduction}

\subsection{Background}

The modal $\mu$-calculus, developed by Kozen~\cite{Kozen83} into its modern form, extends basic modal logic with least and greatest fixpoint operators. It provides a powerful framework for specifying and verifying properties of labelled transition systems and subsumes many modal and temporal logics, including $\mathsf{LTL}$, $\mathsf{CTL}$, $\mathsf{CTL}^*$ and $\mathsf{PDL}$. Its close connections with automata on infinite trees and parity games make it a central subject in theoretical computer science~\cite{BradfieldStirling07,GradelThomasWilke02}.

Kozen~\cite{Kozen83} proposed a finitary Hilbert-style axiomatisation $\Koz$, obtained by extending the minimal normal modal logic $\mathsf{K}$ with the prefixpoint axiom and Park's fixpoint induction rule:
\[
\alpha(\mu x.\alpha(x)) \to \mu x.\alpha(x) \quad (\mathrm{Prefix}), \qquad
\frac{\alpha(\beta) \to \beta}{\mu x.\alpha(x) \to \beta} \quad (\mathrm{Ind}).
\]
Although these two principles directly mirror the Knaster--Tarski characterisation of the least fixpoint as the least prefixpoint, proving the completeness of $\Koz$ turned out to be difficult. Kozen himself established completeness only for the \emph{aconjunctive} fragment of the language, in which no active $\mu$-variable is simultaneously unfolded along both branches of a conjunction (see Definition~\ref{def:activity}(4) for the formal definition). Completeness for the full modal $\mu$-calculus remained open for more than a decade, until Walukiewicz~\cite{Walukiewicz00} finally established it.

His landmark proof combines tableaux and refutations~\cite{NiwinskiWalukiewicz96}, the disjunctive normal form of~\cite{JaninWalukiewicz95}, a game for \emph{tableau consequence}, and combinatorial devices controlling infinite traces.

\subsection{Walukiewicz's reduction}

We first outline the overall structure of Walukiewicz's completeness proof. Most of the technical notions used in this subsection will be defined precisely, and the relevant results established or cited, in Section~\ref{sec:preliminaries} and the subsequent sections; here we give only an informal account. Automaton normal form (ANF)\footnote{Automaton normal form is the class of \emph{disjunctive formulas} of Janin and Walukiewicz~\cite{JaninWalukiewicz95}, also used in~\cite{Walukiewicz00}. We use the name ANF because, in Section~\ref{sec:anf}, the formulas are read off a deterministic parity automaton.} is a disjunctive syntax in which each tableau branch carries a single trace. A refutation is a tableau all of whose infinite branches contain a least-fixpoint trace; it is \emph{thin} when an active least-fixpoint variable is never retained in both conjunctive continuations. Tableau consequence is a trace-preserving relation between two tableaux. With these notions, the completeness reduction decomposes into eight claims:
\begin{enumerate}[(a)]
\item If $\varphi$ is unsatisfiable, then $\vdash \neg\varphi$. \hfill (completeness, Theorem~\ref{thm:completeness})
\item Every $\varphi$ is equivalent to a formula $\anf(\varphi)$ in automaton normal form (ANF). \hfill\cite{JaninWalukiewicz95} (Theorem~\ref{thm:anf})
\item If an ANF formula $\widehat{\varphi}$ is unsatisfiable, then $\vdash \neg\widehat{\varphi}$.
\item Every $\varphi$ has an equivalent ANF formula $\widehat{\varphi}$ with $\vdash \varphi \to \widehat{\varphi}$. \hfill (main lemma, Lemma~\ref{lem:main-lemma})
\item Every unsatisfiable $\varphi$ has a refutation. \hfill\cite{NiwinskiWalukiewicz96} (Theorem~\ref{thm:unsat-refutation})
\item If $\varphi$ has a \emph{thin} refutation, then $\vdash \neg\varphi$. \hfill (thin-refutation theorem, Theorem~\ref{thm:thin-refutation})
\item For an ANF formula $\widehat{\alpha}(x)$ and $\widehat{\varphi} = \anf(\mu x.\widehat{\alpha}(x))$, the tableau of $\widehat{\varphi}$ is a tableau consequence of the tableau of $\widehat{\alpha}(\widehat{\varphi})$. \hfill (fixpoint step, Proposition~\ref{prop:fixpoint-step})
\item If the tableau of an ANF formula $\widehat\varphi$ is a tableau consequence of the tableau of an aconjunctive formula $\alpha$, then $\alpha\vdash\widehat\varphi$ has a two-sided thin refutation, and hence $\vdash\alpha\to\widehat\varphi$ by (f). \hfill (transfer, Theorem~\ref{thm:transfer})
\end{enumerate}

\begin{figure}[t]
\centering
\includegraphics{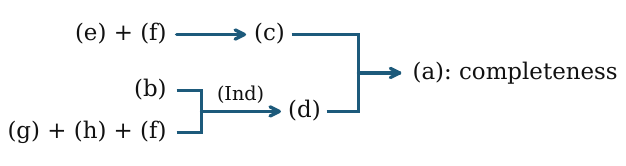}
\caption{Dependencies among Claims (a)--(h).}
\label{fig:claims}
\end{figure}

Claims~(c) and (d) give (a) at once, and (c) follows from (e) and (f) because every ANF formula is aconjunctive. Claim~(d) is proved by induction on $\varphi$; the only hard case is $\varphi=\mu x.\alpha(x)$, since $\mu x.\widehat{\alpha}(x)$ need not be aconjunctive even when $\widehat{\alpha}$ is in ANF. One sets $\widehat{\varphi}:=\anf(\mu x.\widehat{\alpha}(x))$, obtains $\vdash\widehat{\alpha}(\widehat{\varphi})\to\widehat{\varphi}$ from (g), (h) and (f), and concludes by $(\mathrm{Ind})$ (Figure~\ref{fig:claims}).

Three features make the proof in~\cite{Walukiewicz00} hard to verify:
(1)~tableaux depend on the order in which non-modal rules are applied;
(2)~tableau consequence is a game with ``all-sons'' moves, whose strategies must be composed and converted into refutations;
(3)~in set-labelled nodes, distinct threads producing the same formula merge, so one must decide whose history survives---token lists for Claim~(f), and priorities with node assignments for Claim~(g).

\subsection{Related work}

Enqvist, Seifan and Venema~\cite{EnqvistSeifanVenema18} reformulated Claims~(f)--(h) in terms of alternating modal parity automata, one-step logic, a satisfiability game and a consequence game; the framework extends to coalgebraic $\mu$-calculi~\cite{EnqvistSeifanVenema19}. A second line embeds determinisation into annotated tableau systems~\cite{Jungteerapanich09,Jungteerapanich10,Stirling14}. Afshari and Leigh~\cite{AfshariLeigh17} proposed a cut-free completeness proof based on the cyclic system $\mathsf{Clo}$, from which they also derived the completeness of $\Koz$. Subsequently, Kloibhofer~\cite{Kloibhofer23} showed that $\mathsf{Clo}$ is incomplete, revealing a gap in this approach (see also~\cite{Kloibhofer26}); Afshari, Leigh and Men\'endez Turata~\cite{AfshariLeighMenendezTurata25} analysed ill-founded and cyclic proofs through homomorphic admissibility and normal forms, explaining the counterexample and proving completeness with cut.
Their connections between cyclic proofs, well-founded proofs and Kozen's axiomatisation provide a different reconstruction of the delicate completeness argument.

Walukiewicz's original proof and its automata-theoretic reformulations, notably that of Enqvist, Seifan and Venema~\cite{EnqvistSeifanVenema18}, remain central to our understanding of the completeness of $\Koz$. In particular, it remains open whether completeness can be established without reducing formulas to disjunctive normal form. Related annotated and reset systems~\cite{DekkerKloibhoferMartiVenema23,LeighWehr24,AfshariEnqvistLeigh24,AfshariEnqvistLeighMartiVenema25,AfshariKloibhofer25} do not yet yield completeness of $\Koz$ for the full language. 

This paper revises and supersedes the author's earlier preprint~\cite{Tamura14}, whose rule-based traces were not well-defined, whose trace arguments contained errors, and whose explicit index appearance record did not simplify verification.
We replace these by label-based edge trace relations (Definition~\ref{def:traces}), black-box determinisation (Section~\ref{sec:buchi}), one-step saturation, a finite-game proof of Claim~(e), two-sided transfer, and a least-fixpoint argument based on positive-context lifting, aged copies and box isolation.

\subsection{Contributions}\label{sec:intro-contrib}

The eight claims of Walukiewicz's reduction are due to him. Our main contribution is an independent structural theorem for tableau consequence, together with a modular reconstruction of the completeness proof (Table~\ref{tab:comparison}).
\paragraph{Positive-context lifting of tableau consequence.}
The positive-context lifting theorem (Theorem~\ref{thm:positive-context-closure}) shows that, under explicit syntactic hypotheses, any tableau consequence relation $\mathcal T_B\Rightarrow\mathcal T_C$ lifts through a positive context $A(X)$ to $\mathcal T_{A(B)}\Rightarrow\mathcal T_{A(C)}$. Neither $B$ nor $C$ is required to be in ANF. The proof controls overlapping copies of the given relation by an older-wins rule, ensuring eventual stabilisation of trace ownership.
As a direct application, Corollary~\ref{cor:pcc-anf} gives bidirectional ANF replacement inside positive contexts, with the qualifications stated there. This is a tableau-theoretic analogue of the two-directional statement in Enqvist, Seifan and Venema~\cite[Theorem~4]{EnqvistSeifanVenema18}, not a formal generalisation of their automata-theoretic result.
\paragraph{Finite parity-game construction of regular refutations.}
The deterministic parity monitor used for ANF also yields a finite parity game whose positional strategies give regular refutations (Theorem~\ref{thm:finite-game-refutations}). This provides a self-contained proof of Claim~(e) by a standard method, without the Borel determinacy and separate Rabin-tree-automata arguments of~\cite{NiwinskiWalukiewicz96}.
\paragraph{One-step saturated canonical tableaux and two-sided transfer.}
Canonical tableaux resolve non-modal rules in one step, independently of their application order, and record traces explicitly (Section~\ref{sec:tableaux}). We prove their adequacy directly from evaluation games and transfer tableau consequence to thin two-sided refutations (Claim~(h), Theorem~\ref{thm:transfer}). Appendix~\ref{app:thin} supplies the corresponding proof of Claim~(f).
\paragraph{Completeness as the principal application.}
The reconstruction is not shorter overall; its advantage is modularity. The central difficulty of the least-fixpoint step is isolated in the lifting theorem, independently of the completeness reduction. Using it as a black box, together with the unfolding consequence lemma and ANF bisimulation, gives a short derivation of Claim~(g) (Proposition~\ref{prop:fixpoint-step}) and makes this step substantially more transparent. The remaining claims yield the completeness of $\Koz$ (Theorem~\ref{thm:completeness}).
\begin{table}[t]
\centering
\small
\renewcommand{\arraystretch}{1.15}
\begin{tabular}{@{}p{2.7cm}p{3.5cm}p{3.8cm}p{4.2cm}@{}}
\toprule
\textbf{Aspect} & \textbf{Walukiewicz (2000)} & \textbf{Enqvist et al.\ (2018)} & \textbf{Present paper}\\
\midrule
Primary objects & Rule-by-rule tableaux & Alternating parity automata & Canonical tableaux with one-step saturation\\
Traces & Path functions & Streams of state relations & Explicit edge trace relations\\
Parity & Variable dependency order & State priorities & Binder nesting order\\
Consequence & Two-player tableau game & Automata consequence game & Node relation $Z$; maps as functional cases\\
Normal form & Back-edge tree; dummy conjuncts & Disjunctive companion and translation & Formula read from the finite prefix tree $\mathcal G_\Gamma$\\
Claim~(f): context control & Token lists & $\mathsf{lrd}/\mathsf{mst}$ and reset & Recorded occurrences\\
Claim~(g): merging histories & Priority functions and node assignments & Birth dates $\mathsf{tb}$ in Proposition~7.4 & Aged copies and older-wins ownership\\
\bottomrule
\end{tabular}
\caption{Comparison of the proof architectures in~\cite{Walukiewicz00}, \cite{EnqvistSeifanVenema18}, and the present paper.}
\label{tab:comparison}
\end{table}

\paragraph{Reading guide and sources.}
All Claims~(a)--(h) are established here; Appendix~\ref{app:thin} reproduces the thin-refutation argument of Kozen and Walukiewicz~\cite{Kozen83,Walukiewicz00}.
We quote evaluation-game adequacy~\cite{Venema24,EmersonJutla91}, B\"uchi determinisation~\cite{Safra88,Piterman07}, and positional determinacy~\cite{EmersonJutla91,Venema24}.
After the preliminaries, Section~\ref{sec:tableaux} develops canonical tableaux and consequence, Section~\ref{sec:anf} constructs ANF and regular refutations, and Section~\ref{sec:transfer} proves two-sided transfer.
Section~\ref{sec:completeness} derives completeness from closure and unfolding; Section~\ref{sec:pcc} proves these structural results using only Lemma~\ref{lem:diamond-macro-step} from the transfer section.
Readers interested in the structural argument may pass directly from tableau consequence to those statements and proofs; the finite-game ANF construction is needed only for the completeness application.

\section{Preliminaries}\label{sec:preliminaries}
\subsection{Notation}

For a set $X$, $\mathcal{P}(X)$ denotes its power set. A \emph{tree} $(T,C,r)$ has a child relation $C$ such that every node is reached from the root $r$ by exactly one finite $C$-path; $C(t)$ is the set of children of $t$, ancestors, descendants, subtrees and branches (maximal $C$-paths from $r$) are as usual, and an ancestor $s$ of $t$ is \emph{proper} if $s\neq t$. The \emph{path from $s$ to $t$} includes both endpoints, and a labelled tree and its set of nodes are denoted by the same letter. Trees are drawn as derivations are, with the root at the bottom: the children of a node lie above it, and a construction proceeding from the root towards the leaves is called \emph{bottom-up}.

\subsection{Syntax}\label{sec:syntax}

Fix a countably infinite set $\Prop$ of propositional variables. We take the standard modal operators $\Diamond$ and $\Box$ as primitive, and introduce the \emph{cover modality} $\nabla$ as a structured abbreviation.

\begin{definition}[Formulas]\label{def:formulas}
Formulas of the modal $\mu$-calculus in positive normal form are generated by the grammar
\[
\varphi ::= \top \mid \bot \mid p \mid \neg p \mid (\varphi \lor \varphi) \mid (\varphi \land \varphi) \mid \Diamond\varphi \mid \Box\varphi \mid \mu x.\varphi \mid \nu x.\varphi,
\]
where $p, x \in \Prop$, and in $\eta x.\varphi$ ($\eta \in \{\mu, \nu\}$) the literal $\neg x$ does not occur in $\varphi$ (positivity condition). Parentheses are omitted when no ambiguity arises. \emph{Literals} are formulas of the form $\top, \bot, p, \neg p$.
\end{definition}

For a finite set $\Phi$ of formulas we write $\bigvee\Phi$ and $\bigwedge\Phi$ for the disjunction and conjunction of its elements, nested to the right according to a fixed global linear order on formulas, with the conventions $\bigvee\emptyset := \bot$, $\bigwedge\emptyset := \top$, and $\bigvee\{\psi\} = \bigwedge\{\psi\} := \psi$. In a conjunction with a further conjunct, an empty conjunction is omitted: $\varphi\land\bigwedge\emptyset$ stands for $\varphi$.

\paragraph{Cover modality.}
Following~\cite{JaninWalukiewicz95,Walukiewicz00}, for a finite set $\Phi$ we use the abbreviation
\[
\nabla\Phi := \Big(\bigwedge_{\psi \in \Phi} \Diamond\psi\Big) \land \Box\bigvee\Phi,
\]
so that $\Diamond\varphi \equiv \nabla\{\varphi, \top\}$, $\Box\varphi \equiv \nabla\emptyset \lor \nabla\{\varphi\}$, and $\nabla\emptyset$ says that there are no successors. This makes the modal tableau rule uniform (Definition~\ref{def:modal-successors}). Throughout, $\Diamond$ and $\Box$ are written in $\nabla$-form; formulas $\nabla\Phi$ are called \emph{$\nabla$-formulas}.

\begin{definition}[Negation]\label{def:negation}
The De Morgan dual (negation normal form) $\sim\varphi$ of a formula $\varphi$ is defined inductively by: $\sim\top := \bot$, $\sim\bot := \top$, $\sim p := \neg p$, $\sim\neg p := p$; $\sim(\alpha \lor \beta) := \sim\alpha \land \sim\beta$, $\sim(\alpha \land \beta) := \sim\alpha \lor \sim\beta$;
\[
\sim\Diamond\varphi := \Box\sim\varphi, \qquad \sim\Box\varphi := \Diamond\sim\varphi;
\]
$\sim\mu x.\alpha(x) := \nu x.\sim\alpha(\neg x)$ and $\sim\nu x.\alpha(x) := \mu x.\sim\alpha(\neg x)$, where in $\sim\alpha(\neg x)$ any resulting double negation $\neg\neg x$ is read as $x$. By induction on formulas of the primitive syntax, $\sim\sim\varphi=\varphi$ holds as a syntactic identity, and if $x$ is positive in $\alpha(x)$, then $x$ is positive in $\sim\alpha(\neg x)$.

For a $\nabla$-formula we use the following $\nabla$-form of its De Morgan dual:
\[
\sim\nabla\Phi := \bigvee_{\psi \in \Phi} (\nabla\emptyset \lor \nabla\{\sim\psi\}) \;\lor\; \nabla\big\{\bigwedge\{\sim\psi \mid \psi \in \Phi\}, \top\big\}.
\]
This is provably equivalent in $\Koz$ to $\bigvee_{\psi\in\Phi}\Box\sim\psi \lor \Diamond\bigwedge_{\psi\in\Phi}\sim\psi$ (Section~\ref{sec:koz}).
\end{definition}

We write $\Free(\varphi)$ and $\Bound(\varphi)$ for the sets of free and bound variables of $\varphi$, and $\Sub(\varphi)$ for the \emph{set} of subformulas of $\varphi$. When formulas are written in $\nabla$-form, we treat $\nabla\Phi$ as a single syntactic construct with $\Sub(\nabla\Phi) = \{\nabla\Phi\} \cup \bigcup_{\psi \in \Phi} \Sub(\psi)$. We write $\varphi(x)$ to indicate that $x$ may occur free in $\varphi$, and $\varphi(\psi)$ (or $\varphi[\psi/x]$) for capture-free substitution, i.e.\ substitution in which no free variable of $\psi$ becomes bound by a fixpoint binder $\mu y$ or $\nu y$ of $\varphi$.

\paragraph{Naming fixpoints.}
As usual~\cite{Kozen83,Walukiewicz00}, each bound variable $x$ has a unique defining subformula $\eta_x x.\varphi_x$; unfolding replaces $\eta_x x.\varphi_x$ by its body $\varphi_x$, and a later occurrence of $x$ is \emph{regenerated} to $\varphi_x$.

\begin{definition}[Clean and guarded formulas]\label{def:clean-guarded}
A formula $\varphi$ is \emph{clean} (or \emph{well-named}) if $\Free(\varphi) \cap \Bound(\varphi) = \emptyset$ and for every $x \in \Bound(\varphi)$ there is a unique subformula of $\varphi$ of the form $\eta x.\psi$; we denote this unique subformula by $\eta_x x.\varphi_x$ (so $\eta_x \in \{\mu, \nu\}$ is the fixpoint type of $x$ and $\varphi_x$ is its body). A clean formula $\varphi$ is \emph{guarded} if every occurrence of $x$ in $\varphi_x$ lies within the scope of some modal operator $\Diamond$ or $\Box$ (or, in $\nabla$-form, some $\nabla$) inside $\varphi_x$. A finite set $\Gamma$ of formulas is clean (guarded) if its conjunction $\bigwedge\Gamma$ is.
\end{definition}

\begin{remark}[Multiple occurrences of substituted formulas]\label{rem:sub-set}
Because $\Sub(\varphi)$ is defined as a \emph{set} of formulas rather than a set of syntax-tree positions, multiple identical occurrences of a fixpoint formula $\psi$ inside a larger formula count as a single element of $\Sub$. Consequently, the substituted formula $\alpha(\psi)$ is clean whenever $\alpha(x)$ and $\psi$ are clean, $\Bound(\alpha) \cap (\Bound(\psi) \cup \Free(\psi)) = \emptyset$, and $\Bound(\psi) \cap \Free(\alpha) = \emptyset$, \emph{regardless of how many times $x$ occurs free in $\alpha(x)$}. This observation spares us from having to rename bound variables apart across different copies of $\psi$ in the fixpoint step $\widehat{\alpha}(\widehat{\varphi})$ (Section~\ref{sec:completeness}).
\end{remark}

\begin{definition}[Nesting, dependency, activity, and aconjunctivity]\label{def:activity}
Let $\varphi$ be clean and $x, y \in \Bound(\varphi)$.
\begin{enumerate}
\item \emph{Nesting order:} $x \sqsubseteq y$ ($x$ is \emph{outer} to $y$) iff $\eta_y y.\varphi_y \in \Sub(\eta_x x.\varphi_x)$. This is a partial order on $\Bound(\varphi)$; we write $x \sqsubset y$ for its strict part.
\item \emph{Dependency order:} $x \preceq y$ ($y$ \emph{depends} on $x$) is the reflexive--transitive closure of the relation ``$x \in \Free(\eta_y y.\varphi_y)$''.
\item \emph{Activity:} $x$ is \emph{active} in a subformula $\psi$ if there is $y \in \Bound(\varphi)$ occurring free in $\psi$ with $x \preceq y$.
\item \emph{Aconjunctivity:} $\varphi$ is \emph{aconjunctive} if for every $\mu$-variable $x \in \Bound(\varphi)$ (i.e.\ $\eta_x = \mu$) and every conjunction $\alpha \land \beta \in \Sub(\varphi_x)$ (excluding the conjunctions hidden inside the $\nabla\Phi$ abbreviation), the variable $x$ is active in at most one of $\alpha$ and $\beta$.
\end{enumerate}
\end{definition}

\begin{example}[Nesting, indirect dependency, and activity]\label{ex:activity}
Consider the clean guarded formula
\[
\varphi := \mu x.\bigl(p \lor \Diamond\nu y.
((x\land q)\lor\Diamond\mu z.((y\land r)\lor\Diamond z))\bigr).
\]
Its binders are properly nested, so $x\sqsubset y\sqsubset z$. Moreover, $x$ occurs free in the defining formula for $y$, and $y$ occurs free in the defining formula for $z$. Hence
\[
x\preceq y\preceq z,
\qquad\text{and therefore}\qquad x\preceq z.
\]
The last dependency is indirect: $x$ does not occur free in $\eta_z z.\varphi_z$. For the subformula $\psi:=\Diamond z$, we have $z\in\Free(\psi)$ and $x\preceq z$, so $x$ is active in $\psi$ although $x\notin\Free(\psi)$. The displayed formula is aconjunctive: in each explicit conjunction, $x$ is active on at most one side. A conjunction such as $\Diamond y\land\Diamond z$ inside $\varphi_x$ would be forbidden, because $x$ would be active in both conjuncts.
\end{example}

If $x\in\Free(\eta_y y.\varphi_y)$ in a clean formula, then $x\sqsubset y$; hence ${\preceq}\subseteq{\sqsubseteq}$. Aconjunctivity ensures that threads of an active $\mu$-variable never split at an explicit conjunction (they may still split at the modal successors of a $\nabla$).

\subsection{Semantics}

A \emph{Kripke model} $\mathbb{S} = (S, R, \lambda)$ consists of a set $S$ of worlds, a relation $R \subseteq S \times S$, and a valuation $\lambda : \Prop \to \mathcal{P}(S)$; the denotation $[\![\varphi]\!]^{\mathbb{S}}_\rho \subseteq S$ under an environment $\rho$ is defined as usual~\cite{BradfieldStirling07}. For fixpoints, put $F_{\alpha,\rho}(U):=[\![\alpha]\!]^{\mathbb{S}}_{\rho[x\mapsto U]}$ for $U\subseteq S$. By positivity $F_{\alpha,\rho}$ is monotone, so the Knaster--Tarski theorem gives
\[
[\![\mu x.\alpha]\!]^{\mathbb{S}}_{\rho}
:=\operatorname{lfp}(F_{\alpha,\rho})
=\bigcap\{U\subseteq S\mid F_{\alpha,\rho}(U)\subseteq U\},
\]
\[
[\![\nu x.\alpha]\!]^{\mathbb{S}}_{\rho}
:=\operatorname{gfp}(F_{\alpha,\rho})
=\bigcup\{U\subseteq S\mid U\subseteq F_{\alpha,\rho}(U)\}.
\]
We have
\[
s \in [\![\nabla\Phi]\!]^{\mathbb{S}}_\rho \iff \forall \psi \in \Phi\,\exists t\,(sRt \land t \in [\![\psi]\!]^{\mathbb{S}}_\rho) \;\text{ and }\; \forall t\,(sRt \Rightarrow \exists \psi \in \Phi\, t \in [\![\psi]\!]^{\mathbb{S}}_\rho),
\]
and $[\![\sim\varphi]\!]^{\mathbb{S}}_\rho = S \setminus [\![\varphi]\!]^{\mathbb{S}}_\rho$. For closed formulas we omit the environment $\rho$. We write $\mathbb{S}, s \models \varphi$ for $s \in [\![\varphi]\!]^{\mathbb{S}}$, and $\varphi \models \psi$ and $\varphi \equiv \psi$ for global semantic consequence and semantic equivalence.

\subsection{Kozen's axiomatisation}\label{sec:koz}

Kozen~\cite{Kozen83} presented $\Koz$ as a Hilbert-style system: the minimal normal modal logic $\mathsf K$ extended by $(\mathrm{Prefix})$ and $(\mathrm{Ind})$ as displayed in the introduction. We use an equivalent set-based, Tait-style presentation, written in two-sided form. Its judgements are sequents
\[
\Gamma\vdash\Delta,
\]
where $\Gamma$ and $\Delta$ are finite sets of formulas, read as $\bigwedge\Gamma\to\bigvee\Delta$. We write $\alpha,\Gamma$ for $\{\alpha\}\cup\Gamma$ and similarly on the right. Since sequents are set-labelled, exchange and contraction are implicit.

Tableaux and refutations are written one-sidedly, using
\[
\Gamma\vdash\quad:=\quad\Gamma\vdash\emptyset.
\]
The two-sided and one-sided forms are interderivable under
\begin{equation}\label{eq:two-one-sided}
\Gamma\vdash\Delta
\quad\Longleftrightarrow\quad
\Gamma,\sim\Delta\vdash,
\qquad
\sim\Delta:=\{\sim\delta\mid\delta\in\Delta\}.
\end{equation}

\paragraph{Initial sequents and structural rules.}
\[
\varphi,\Gamma\vdash\Delta,\varphi\ (\mathrm{Id}),
\qquad
\bot,\Gamma\vdash\Delta\ (\bot L),
\qquad
\Gamma\vdash\Delta,\top\ (\top R),
\]
\[
\varphi,\sim\varphi,\Gamma\vdash\Delta\ (\mathrm{Tau}L),
\qquad
\Gamma\vdash\Delta,\varphi,\sim\varphi\ (\mathrm{Tau}R),
\]
\[
\frac{\Gamma\vdash\Delta}{\alpha,\Gamma\vdash\Delta}\ (\mathrm{Weak}L)
\qquad
\frac{\Gamma\vdash\Delta}{\Gamma\vdash\Delta,\alpha}\ (\mathrm{Weak}R)
\qquad
\frac{\Gamma\vdash\Delta,\alpha\qquad\alpha,\Gamma'\vdash\Delta'}
{\Gamma,\Gamma'\vdash\Delta,\Delta'}\ (\mathrm{Cut}).
\]

\paragraph{Propositional rules.}
\[
\frac{\alpha,\Gamma\vdash\Delta\qquad\beta,\Gamma\vdash\Delta}
{\alpha\lor\beta,\Gamma\vdash\Delta}\ (\lor L)
\qquad
\frac{\Gamma\vdash\Delta,\alpha,\beta}
{\Gamma\vdash\Delta,\alpha\lor\beta}\ (\lor R),
\]
\[
\frac{\alpha,\beta,\Gamma\vdash\Delta}
{\alpha\land\beta,\Gamma\vdash\Delta}\ (\land L)
\qquad
\frac{\Gamma\vdash\Delta,\alpha\qquad\Gamma\vdash\Delta,\beta}
{\Gamma\vdash\Delta,\alpha\land\beta}\ (\land R).
\]

\paragraph{Modal rules.}
For finite sets, put $\Box\Gamma:=\{\Box\gamma\mid\gamma\in\Gamma\}$ and $\Diamond\Delta:=\{\Diamond\delta\mid\delta\in\Delta\}$. The normal modal rule and its dual presentation are
\[
\frac{\psi,\Gamma\vdash\Delta}
{\Diamond\psi,\Box\Gamma\vdash\Diamond\Delta}\ (\Diamond L)
\qquad
\frac{\Gamma\vdash\Delta,\psi}
{\Box\Gamma\vdash\Diamond\Delta,\Box\psi}\ (\Box R).
\]
Additional side formulas are supplied by weakening. These are two-sided dual presentations of the same normal modal rule; under~\eqref{eq:two-one-sided}, both reduce to the following one-sided rule used below:
\[
\frac{\psi,\{\alpha\mid\Box\alpha\in\Gamma\}\vdash}
{\Diamond\psi,\Gamma\vdash}\ (\Diamond).
\]

\paragraph{Fixpoint principles.}
For $x$ positive in $\alpha(x)$, with all substitutions understood to be capture-free,
\[
\alpha(\mu x.\alpha(x))\vdash\mu x.\alpha(x)\ (\mathrm{Prefix}),
\qquad
\frac{\alpha(\beta)\vdash\beta}{\mu x.\alpha(x)\vdash\beta}\ (\mathrm{Ind}).
\]
Their contextual forms follow by $(\mathrm{Weak}L)$, $(\mathrm{Weak}R)$ and $(\mathrm{Cut})$; the corresponding $\nu$-principles follow by duality.

A \emph{derivation} is a finite tree built from these initial sequents and rules. For arbitrary formulas we use $\alpha\to\beta:=\sim\alpha\lor\beta$ and write $\vdash\varphi$ for $\emptyset\vdash\varphi$. By~\eqref{eq:two-one-sided}, the propositional rules and $(\mathrm{Cut})$, the three judgements $\vdash\alpha\to\beta$, $\alpha\vdash\beta$ and $\alpha,\sim\beta\vdash$ are interderivable. We write $\vdash\alpha\leftrightarrow\beta$, and say that $\alpha$ and $\beta$ are \emph{provably equivalent} in $\Koz$, if both $\alpha\vdash\beta$ and $\beta\vdash\alpha$. For an arbitrary formula $\varphi$, $\neg\varphi$ means its De Morgan dual $\sim\varphi$ (Definition~\ref{def:negation}); hence $\vdash\neg\varphi$ and $\varphi\vdash$ are interderivable.

When the succedent is empty, we write $(\mathrm{Bot})$ and $(\mathrm{Tau})$ for the $\Delta=\emptyset$ instances of $(\bot L)$ and $(\mathrm{Tau}L)$, and abbreviate the corresponding instances of $(\lor L)$, $(\land L)$ and $(\mathrm{Weak}L)$ to $(\lor)$, $(\land)$ and $(\mathrm{Weak})$. We also use the following one-sided forms, all derivable by~\eqref{eq:two-one-sided}:
\[
\frac{\Gamma,\sim\alpha\vdash\qquad\alpha,\Gamma'\vdash}{\Gamma,\Gamma'\vdash}\ (\mathrm{Cut}),
\qquad
\alpha(\mu x.\alpha(x)),\sim\mu x.\alpha(x)\vdash\ (\mathrm{Prefix}),
\]
\[
\frac{\alpha(\beta),\sim\beta\vdash}{\mu x.\alpha(x),\sim\beta\vdash}\ (\mathrm{Ind}).
\]
We call this derived one-sided presentation the \emph{one-sided calculus}; it is the form used for tableaux and refutations.

\begin{remark}[Equivalence of the three presentations]\label{rem:hilbert-tait}
For finite $\Gamma,\Delta$, $\Gamma\vdash\Delta$ iff $\Gamma,\sim\Delta\vdash$ iff $\bigwedge\Gamma\to\bigvee\Delta$ is derivable in Kozen's Hilbert system, by the structural rules and the Hilbert reading of the sequent rules.
In the $\nabla$-form of Definition~\ref{def:negation}, $\sim\sim\varphi$ is provably equivalent rather than syntactically identical to $\varphi$ (Lemma~\ref{lem:standard}(3)), so $(\mathrm{Tau}L)$, $(\mathrm{Tau}R)$ and~\eqref{eq:two-one-sided} remain derivable.
\end{remark}

\begin{lemma}[Standard facts]\label{lem:standard}
\begin{enumerate}
\item $\Koz$ is sound with respect to Kripke semantics: if $\Gamma\vdash$, then $\bigwedge\Gamma$ is unsatisfiable. Derivability is closed under capture-free substitution of formulas for free variables.
\item \emph{(Unfolding and coinduction)} $\vdash\eta x.\alpha(x)\leftrightarrow\alpha(\eta x.\alpha(x))$ for $\eta\in\{\mu,\nu\}$. Dually to $(\mathrm{Ind})$, if $\beta\vdash\alpha(\beta)$ then $\beta\vdash\nu x.\alpha(x)$. Consequently the \emph{unfolding rule}
\[
\frac{\alpha(\eta x.\alpha(x)),\Gamma\vdash}{\eta x.\alpha(x),\Gamma\vdash}\ (\eta\text{-}\mathrm{unf})
\]
is derivable.
\item \emph{(Monotonicity)} If $x$ is positive in $\chi(x)$ and $\alpha\vdash\beta$, then $\chi(\alpha)\vdash\chi(\beta)$. If $\alpha(x)\vdash\beta(x)$, then $\eta x.\alpha(x)\vdash\eta x.\beta(x)$.
\item \emph{(The $\nabla$-rule)} Let $\Gamma=\{\nabla\Psi_1,\dots,\nabla\Psi_i\}\cup\Delta$, let $1\le k\le i$ and $\psi\in\Psi_k$. Then the rule
\[
\frac{\psi,\ \{\bigvee\Psi_n\mid n\neq k\}\vdash}{\Gamma\vdash}\ (\nabla)
\]
is derivable, and so is its variant in which the premise also contains $\bigvee\Psi_k$. The following two-sided rule is also derivable. Let $A_1,\dots,A_n,B$ be finite sets of formulas and, for every $b\in B$, choose $1\le k_b\le n$ and $a_b\in A_{k_b}$:
\[
\frac{
\bigl\{\,a_b,\{\bigvee A_j\mid j\neq k_b\}\vdash b\;\bigm|\;b\in B\,\bigr\}
\qquad
\bigvee A_1,\dots,\bigvee A_n\vdash\bigvee B
}{
\nabla A_1,\dots,\nabla A_n\vdash\nabla B
}\ (\mathrm{Cover}).
\]
Likewise, if $\bot\in\Gamma$ or $\{p,\neg p\}\subseteq\Gamma$, then $\Gamma\vdash$.
\item Every formula $\varphi$ is provably equivalent in $\Koz$ to a clean guarded formula (written in $\nabla$-form) with the same free variables, in which every free variable that is positive (guarded) in $\varphi$ remains positive (guarded).
\item \emph{(Renaming)} If $\varphi'$ is obtained from $\varphi$ by a capture-free renaming of bound variables, then $\vdash\varphi\leftrightarrow\varphi'$.
\end{enumerate}
\end{lemma}

\begin{proof}
Items~(1)--(3), (5), and~(6) are standard facts about Kozen's axiomatisation~\cite{Kozen83,Walukiewicz00}.
(4) Since $\nabla\Psi_k$ contains the conjunct $\Diamond\psi$ and each $\nabla\Psi_n$ the conjunct $\Box\bigvee\Psi_n$, repeated $(\land)$ reduces $\Gamma\vdash$ to $\Diamond\psi,\ \{\Box\bigvee\Psi_n\mid n\neq k\},\ \Gamma''\vdash$ for some finite $\Gamma''$. (For the variant, keep also $\Box\bigvee\Psi_k$.) Rule $(\Diamond)$ with principal formula $\Diamond\psi$ reduces this to $\psi,\ \{\bigvee\Psi_n\mid n\neq k\},\ A\vdash$, where $A$ consists of the remaining box formulas, and $(\mathrm{Weak})$ reduces it to the premise. For $(\mathrm{Cover})$, expand the occurrences of $\nabla$ and use the propositional and modal rules: the family of left premises provides the diamond obligations of $\nabla B$, while the last premise provides its box obligation. The final claim is $(\mathrm{Bot})$ or $(\mathrm{Tau})$ followed by $(\mathrm{Weak})$.
\end{proof}

All rules of the canonical tableaux and refutations of Section~\ref{sec:tableaux} are thus primitive or derivable rules of $\Koz$, read bottom-up: $(\lor)$, $(\land)$ and $(\mathrm{Weak})$ are primitive, $(\nabla)$ is Lemma~\ref{lem:standard}(4), and the unfolding and regeneration steps are instances of $(\eta\text{-}\mathrm{unf})$ under the expansion $\llangle\cdot\rrangle$ of Definition~\ref{def:expansion}. The only difference between a refutation and a derivation is that a refutation may have infinite branches; Kozen's context-recording rule (Lemma~\ref{lem:kozen} in Appendix~\ref{app:thin}) is the tool for cutting them off.

\subsection{Conventions for $\omega$-automata and $\omega$-games}\label{sec:omega-conventions}

We assume familiarity with $\omega$-automata and infinite two-player games; see~\cite{GradelThomasWilke02}. Since conventions differ between texts---acceptance on states or on transitions, maximal or minimal priorities, the treatment of dead ends---we fix those used in this paper.

For an infinite sequence $\pi$, $\mathrm{Inf}(\pi)$ is the set of elements occurring infinitely often in $\pi$, and a map $g$ is applied to $\pi$ letterwise.

\paragraph{Automata.} An $\omega$-automaton is a tuple
\[
\mathcal A=(Q,\Sigma,Q_I,\delta,\mathsf{Acc}),
\]
where $Q$ is a finite set of states, $\Sigma$ a finite alphabet, $Q_I\subseteq Q$ the set of initial states, $\delta\subseteq Q\times\Sigma\times Q$ a transition relation, and $\mathsf{Acc}$ an acceptance condition for infinite runs. A run on $a_0a_1\cdots\in\Sigma^\omega$ is an infinite sequence $q_0q_1\cdots$ with $q_0\in Q_I$ and $(q_n,a_n,q_{n+1})\in\delta$ for every $n$. The automaton is \emph{deterministic} if $Q_I=\{q_0\}$ is a singleton (written $q_0$ in the tuple) and every pair $(q,a)$ has exactly one successor.

We use two kinds of $\omega$-automata~\cite{GradelThomasWilke02}. An $\mathsf{NBA}$ is a nondeterministic B\"uchi automaton: a set $F\subseteq\delta$ of accepting transitions is specified, and a run is accepting iff it uses transitions in $F$ infinitely often. A $\mathsf{DPA}$ is a deterministic parity automaton: each state $q$ has a priority $\pri(q)\in\mathbb N$, with finite range, and a run $\pi$ is accepting iff $\max\mathrm{Inf}(\pri(\pi))$ is even. Throughout, \emph{even is good for the refuter}: an even branch carries a $\mu$-trace (Definition~\ref{def:even-branches}), and an even priority yields a $\mu$-binder (Definition~\ref{def:anf-construction}).

\paragraph{Games.} We use two-player games on arenas $(V_{\ModelPlayer},V_{\RefPlayer},E)$, where $V:=V_{\ModelPlayer}\cup V_{\RefPlayer}$ and $E\subseteq V\times V$; put $E(v):=\{u\in V\mid(v,u)\in E\}$. A play is a maximal path in the arena. A strategy for a player $P$ assigns to every finite play ending at a position $v\in V_P$ with $E(v)\neq\emptyset$ a successor in $E(v)$. A play is consistent with the strategy if every move made at a $P$-position is prescribed by it, and the strategy is winning if every consistent play is won by $P$. A strategy is \emph{positional} if its choice depends only on the current position; we then identify it with a map $f:V_P\to V$ satisfying $f(v)\in E(v)$ whenever $E(v)\neq\emptyset$. A finite play ends at a \emph{dead end} ($E(v)=\emptyset$) and is lost by its owner. In a \emph{parity game}, given by $\Omega:V\to\mathbb N$ with finite range, $\RefPlayer$ wins an infinite play $\pi$ iff $\max\mathrm{Inf}(\Omega(\pi))$ is even~\cite{GradelThomasWilke02}.

\begin{theorem}[Positional determinacy {\cite{EmersonJutla91}}]\label{thm:positional-determinacy}
In a parity game, from every position exactly one player has a winning strategy, and each player has a positional strategy that is winning from every position of their winning region. For finite arenas see also~\cite{BjorklundSandbergVorobyov04}; for arbitrary arenas see~\cite[Theorem~5.24]{Venema24}.
\end{theorem}

Dead ends are covered by replacing a dead end owned by $P$ with a self-loop whose priority has the parity favourable to the opponent of $P$.

Three games occur in this paper. The evaluation game of Section~\ref{sec:evaluation-games} is a parity game on a possibly infinite arena; the tableau game of Definition~\ref{def:tableau-game} is in general infinite and its winning condition is not a parity condition; and the game of Section~\ref{sec:finite-refutation-game} is a finite parity game. Only the first and third use parity-game determinacy directly.

\subsection{Evaluation games}\label{sec:evaluation-games}

We use the standard evaluation game of the modal $\mu$-calculus~\cite[Definition~2.21 and Tables~3--4]{Venema24}; its adequacy theorem will be used as a black box.

\begin{definition}[Closure]\label{def:closure}
For a formula $\xi$ in $\nabla$-form, $\Cl(\xi)$ is the least set of formulas such that
\[
\Sub(\xi)\subseteq\Cl(\xi),
\qquad
\nabla\Psi\in\Cl(\xi)
\Longrightarrow
\Sub(\bigvee\Psi)\subseteq\Cl(\xi).
\]
For a finite set $\Gamma$, put $\Cl(\Gamma):=\bigcup_{\gamma\in\Gamma}\Cl(\gamma)$.
\end{definition}

Let $\xi$ be a clean formula in $\nabla$-form. The players are $\exists$ and $\forall$; they play the roles of $\ModelPlayer$ and $\RefPlayer$ in the conventions above, so a dead end is lost by its owner.

\begin{definition}[Evaluation game]\label{def:evaluation-game}
Let $\mathbb S=(S,R,\lambda)$ be a Kripke model. The \emph{evaluation game} $\mathrm{EG}(\xi,\mathbb S)$ has positions $(\varphi,w)$ with $\varphi\in\Cl(\xi)$ and $w\in S$, and intermediate positions $\langle\nabla\Psi,w,\psi\rangle_\Diamond$ with $\psi\in\Psi$; owners and moves are:
\begin{itemize}
\item $(\varphi_1\lor\varphi_2,w)$: owner $\exists$; moves to $(\varphi_1,w)$, $(\varphi_2,w)$.
\item $(\varphi_1\land\varphi_2,w)$: owner $\forall$; moves to $(\varphi_1,w)$, $(\varphi_2,w)$.
\item $(\eta_xx.\varphi_x,w)$ and $(x,w)$ with $x\in\Bound(\xi)$: owner $\exists$; one move, to $(\varphi_x,w)$; the move from $(x,w)$ is a \emph{regeneration} of $x$.
\item $(l,w)$, $l$ a literal: no moves; owner $\exists$ if $l$ fails at $w$, owner $\forall$ if $l$ holds at $w$.
\item $(\nabla\Psi,w)$: owner $\forall$; moves to $\langle\nabla\Psi,w,\psi\rangle_\Diamond$ for $\psi\in\Psi$, and to $(\bigvee\Psi,v)$ for $wRv$.
\item $\langle\nabla\Psi,w,\psi\rangle_\Diamond$: owner $\exists$; moves to $(\psi,v)$ for $wRv$.
\end{itemize}
$\forall$ wins an infinite play iff the $\sqsubseteq$-least variable regenerated infinitely often is a $\mu$-variable.
\end{definition}

\begin{theorem}[Adequacy of evaluation games {\cite[Theorem~3.30]{Venema24}, \cite{EmersonJutla91}}]\label{thm:eval-adequacy}
Let $\xi$ be a clean formula in $\nabla$-form, $\mathbb S$ a Kripke model and $s$ a world. Then $\mathbb S,s\models\xi$ if and only if $\exists$ has a winning strategy in $\mathrm{EG}(\xi,\mathbb S)$ from $(\xi,s)$. Moreover, each player has a positional strategy that is winning from every position of their winning region.
\end{theorem}

\begin{proof}
In~\cite{Venema24} the modalities $\Diamond$ and $\Box$ are primitive; write $\mathrm{EG}^\circ(\xi,\mathbb S)$ for that game, with $\nabla\Psi$ read as $\bigwedge_{\psi\in\Psi}\Diamond\psi\land\Box\bigvee\Psi$. Positions are formulas rather than occurrences, so the two occurrences of an element $\psi$ in this reading give the same positions (cf.\ Remark~\ref{rem:sub-set}), and the proof of~\cite{Venema24} applies to clean formulas in this reading.

\emph{Plays.} In $\mathrm{EG}^\circ$, a play through $(\nabla\Psi,w)$ continues with moves of $\forall$ through the conjunctions of this reading to $(\Diamond\psi,w)$ or $(\Box\bigvee\Psi,w)$; then $\exists$ moves to some $(\psi,v)$, or $\forall$ to some $(\bigvee\Psi,v)$, with $wRv$. Contracting each such segment to the moves of Definition~\ref{def:evaluation-game} (through $\langle\nabla\Psi,w,\psi\rangle_\Diamond$ in the first case) is a bijection between the plays of the two games from $(\xi,s)$. The contracted segments contain no fixpoint formula and no bound variable, so the bijection preserves the sequence of unfoldings and regenerations, finiteness of plays, and the owner of a final dead end. A strategy for either player in one game therefore induces a strategy in the other whose consistent plays are exactly the images of the consistent plays of the original, so a player wins from $(\xi,s)$ in one game iff the same player wins in the other, provided the winning conditions agree.

\emph{Winning conditions.} The winning condition of~\cite{Venema24} uses the variable of highest rank in the dependency order among those unfolded infinitely often; such a variable exists by~\cite[Proposition~2.24]{Venema24}. For the $\preceq$-least such variable, unfolding infinitely often is equivalent to regeneration infinitely often, since its binder is entered only finitely often. It is therefore $\preceq$-least among the variables regenerated infinitely often, hence also $\sqsubseteq$-least because ${\preceq}\subseteq{\sqsubseteq}$, so the two winning conditions agree. The first claim follows.

\emph{Positional strategies.} The condition of Definition~\ref{def:evaluation-game} is a parity condition: let $d(x)$ be the number of strict $\sqsubseteq$-predecessors of $x\in\Bound(\xi)$, $N:=|\Bound(\xi)|$, and put $\Omega(x,w):=2(N-d(x))$ if $\eta_x=\mu$, $\Omega(x,w):=2(N-d(x))+1$ if $\eta_x=\nu$, and $\Omega:=0$ at all other positions. By the argument of Lemma~\ref{lem:outermost}, which applies verbatim to the formulas along an infinite play, infinitely many regenerations occur and the variables regenerated infinitely often have a $\sqsubseteq$-least element $x$, all others lying strictly $\sqsubseteq$-above it. Hence $\max\mathrm{Inf}(\Omega(\pi))=\Omega(x,\cdot)$, which is even iff $\eta_x=\mu$. The second claim is now positional determinacy of parity games (Theorem~\ref{thm:positional-determinacy}).
\end{proof}

\section{Canonical tableaux, refutations and consequence}\label{sec:tableaux}

Throughout this section, $\Gamma_0$ is a fixed finite, clean, guarded set of formulas written in $\nabla$-form, and all formulas considered are taken from its syntactic closure $\Cl(\Gamma_0)$. Two-sided sequents are encoded as one-sided ones by tagging formulas with their side (Section~\ref{sec:thin-refutation-proof}).

\begin{definition}[Expansion]\label{def:expansion}
Because we decompose fixpoints by stepping from $\eta_x x.\varphi_x$ to $\varphi_x$, a formula $\psi \in \Cl(\Gamma_0)$ may contain variables from $\Bound(\Gamma_0)$ free. Its \emph{expansion} $\llangle\psi\rrangle$ is obtained by repeatedly replacing every free occurrence of a variable $x\in\Bound(\Gamma_0)$ by its defining fixpoint formula $\eta_x x.\varphi_x$, processing variables in decreasing $\sqsubseteq$-order (innermost first), until no variable from $\Bound(\Gamma_0)$ remains free. (For example, let $\Gamma_0=\{\mu x.(p\lor\Diamond\nu y.(x\land\Box y))\}$, so that $\varphi_x=p\lor\Diamond\nu y.(x\land\Box y)$ and $\varphi_y=x\land\Box y$. Then $\llangle y\rrangle=\nu y.\bigl(\mu x.(p\lor\Diamond\nu y.(x\land\Box y))\land\Box y\bigr)$: first $y$ is replaced by its defining fixpoint formula $\nu y.(x\land\Box y)$, and then the free outer variable $x$ occurring in it is expanded.) For a set $\Gamma \subseteq \Cl(\Gamma_0)$, we put $\llangle\Gamma\rrangle := \{\llangle\psi\rrangle \mid \psi \in \Gamma\}$. We say that $\Gamma \subseteq \Cl(\Gamma_0)$ is \emph{satisfiable} if $\bigwedge\llangle\Gamma\rrangle$ is satisfiable, and we write $\Gamma \vdash$ to mean $\llangle\Gamma\rrangle \vdash$.
\end{definition}

A formula of $\Cl(\Gamma_0)$ is called \emph{modal} if it is either a literal over $\Free(\Gamma_0)$ or a $\nabla$-formula $\nabla\Psi$; otherwise it is a non-modal formula (a disjunction, a conjunction, a fixpoint formula $\eta_x x.\varphi_x$, or a bound variable $x \in \Bound(\Gamma_0)$). A set $M$ of modal formulas is \emph{consistent} if $\bot \notin M$ and $\{p, \neg p\} \nsubseteq M$ for all $p \in \Prop$, and \emph{$\nabla$-free} if it contains no $\nabla$-formula $\nabla\Psi$; otherwise it is \emph{$\nabla$-non-free}. For such $M$ we write $\Lit(M)$ for the set of literals in $M$, so that $M=\{\nabla\Psi\mid\nabla\Psi\in M\}\cup\Lit(M)$.

\subsection{Saturation}

Walukiewicz's tableaux apply non-modal rules one formula at a time, and the order affects the branching structure~\cite{Walukiewicz00}. We instead resolve all non-modal connectives between a choice node and its modal children in one step, driven by a choice function.

Let $\Dis$ denote the set of all disjunctions in $\Cl(\Gamma_0)$. A \emph{choice function} is a map $c$ assigning to each disjunction $\alpha \lor \beta \in \Dis$ one of its disjuncts $\alpha$ or $\beta$.

\begin{definition}[Saturation and trace relations]\label{def:saturation}
For a set $\Gamma \subseteq \Cl(\Gamma_0)$ and a choice function $c$, the \emph{decomposition graph} $E_c$ on $\Cl(\Gamma_0)$ is the directed graph with edges
\[
\alpha \land \beta \to \alpha, \quad \alpha \land \beta \to \beta, \quad \alpha \lor \beta \to c(\alpha \lor \beta), \quad \eta_x x.\varphi_x \to \varphi_x, \quad x \to \varphi_x \;\;(x \in \Bound(\Gamma_0)),
\]
and with no outgoing edges from modal formulas. We write $S_c(\Gamma)$ for the set of formulas reachable from $\Gamma$ by paths in $E_c$ (including $\Gamma$ itself), and $M_c(\Gamma)$ for the set of modal formulas in $S_c(\Gamma)$. For an $E_c$-path $\pi$, let $\Reg(\pi) \subseteq \Bound(\Gamma_0)$ be the set of bound variables $x$ such that $\pi$ traverses the regeneration edge $x \to \varphi_x$. The \emph{trace relation} of $(\Gamma, c)$ is the ternary relation $R_c(\Gamma) \subseteq \Gamma \times M_c(\Gamma) \times \mathcal{P}(\Bound(\Gamma_0))$ defined by
\[
R_c(\Gamma) := \{(\gamma, m, \Reg(\pi)) \mid \gamma \in \Gamma,\; m \in M_c(\Gamma),\; \pi \text{ is an $E_c$-path from $\gamma$ to $m$}\}.
\]
The \emph{outcomes} of $\Gamma$ are the distinct pairs $\Out(\Gamma) := \{(M_c(\Gamma), R_c(\Gamma)) \mid c \text{ is a choice function}\}$.
\end{definition}

Thus the construction below may be viewed as an ordinary tableau with each finite non-modal segment contracted; $R_c(\Gamma)$ keeps the transitions and regenerations of the contracted segment (see the realisation remark after Definition~\ref{def:traces}).

\begin{lemma}[Acyclicity of saturation]\label{lem:acyclic}
For every choice function $c$, the graph $E_c$ is acyclic. In particular, no variable $x \in \Bound(\Gamma_0)$ is regenerated more than once along an $E_c$-path. Every maximal $E_c$-path is finite and terminates in a modal formula, and $\Out(\Gamma)$ is finite.
\end{lemma}

\begin{proof}
Every edge of $E_c$ other than a regeneration $x \to \varphi_x$ steps to a strict subformula, so any cycle in $E_c$ must contain at least one regeneration edge. Suppose for a contradiction that $E_c$ contains a cycle, and let $z$ be a $\sqsubseteq$-minimal variable among those regenerated on the cycle. Immediately after traversing $z \to \varphi_z$, the cycle is at $\varphi_z$. It cannot leave $\Sub(\varphi_z)$: doing so would require regenerating a variable $w$ whose binder $\eta_w w.\varphi_w$ strictly encloses $\eta_z z.\varphi_z$ (i.e.\ $w \sqsubset z$), contradicting the minimality of $z$. However, to complete the cycle and return to $z$, the path must reach a free occurrence of $z$ inside $\varphi_z$. By guardedness, every occurrence of $z$ in $\varphi_z$ lies strictly inside the scope of some $\nabla$; since modal formulas $\nabla\Psi$ have no outgoing edges in $E_c$, no $E_c$-path can ever enter a $\nabla$, a contradiction. Since $\Cl(\Gamma_0)$ is finite, acyclicity makes every maximal $E_c$-path finite and hence terminating at a modal formula; it also prevents a regeneration edge from occurring twice on one path. Finiteness of $\Out(\Gamma)$ follows from finiteness of $\Cl(\Gamma_0)$.
\end{proof}

\begin{example}[One-step saturation and path merging]\label{ex:saturation}
Let $\Gamma=\{x,\nu y.(\nabla\{\psi\}\lor p)\}$, where $x$ is a $\mu$-variable with binding definition $\mu x.((\nabla\{\psi\}\land q)\lor r)$, and let $c$ select left disjuncts. Then $x$ regenerates and reaches $\nabla\{\psi\}$ and $q$, while $\nu y.(\nabla\{\psi\}\lor p)$ unfolds to $\nabla\{\psi\}$ without regeneration. The two paths merge at $\nabla\{\psi\}$ in $M_c(\Gamma)=\{\nabla\{\psi\},q\}$, while
\[
R_c(\Gamma)=\big\{(x,\nabla\{\psi\},\{x\}),\ (x,q,\{x\}),\ (\nu y.(\nabla\{\psi\}\lor p),\nabla\{\psi\},\emptyset)\big\}
\]
retains their distinct origins and regenerated-variable sets.
\end{example}

\subsection{The canonical tableau}

By the semantics of $\nabla$, a world satisfying a consistent label $\{\nabla\Psi_1, \dots, \nabla\Psi_i\} \cup \Lit(M)$ has, for each $k$ and $\psi\in\Psi_k$, a successor satisfying $\psi$ and every box part $\bigvee\Psi_n$, $n\neq k$.

\begin{definition}[Modal successors]\label{def:modal-successors}
Let $M$ be a consistent, $\nabla$-non-free set of modal formulas containing the $\nabla$-formulas $\nabla\Psi_1, \dots, \nabla\Psi_i$ ($i \ge 1$). For each $1 \le k \le i$ and each $\psi \in \Psi_k$, the \emph{$(k, \psi)$-successor} of $M$ is the pair $(\Gamma_{k,\psi}, R_{k,\psi})$ defined by
\[
\Gamma_{k,\psi} := \{\psi\} \cup \big\{\bigvee\Psi_n \;\big|\; n \neq k\big\}, \qquad R_{k,\psi} := \{(\nabla\Psi_k, \psi, \emptyset)\} \cup \big\{(\nabla\Psi_n, \bigvee\Psi_n, \emptyset) \;\big|\; n \neq k\big\}.
\]
The \emph{successors} of $M$ are the distinct pairs $(\Gamma_{k,\psi}, R_{k,\psi})$ obtained in this way. (If $\Psi_k = \emptyset$ for all $k$, then $M$ has no successors.)
\end{definition}

\begin{definition}[Canonical tableau]\label{def:tableau}
For a finite set $\Gamma \subseteq \Cl(\Gamma_0)$, the \emph{canonical tableau} $\mathcal{T}_\Gamma$ is the bipartite labelled tree with two alternating kinds of nodes, \emph{choice nodes} and \emph{modal nodes}, constructed bottom-up as follows:
\begin{itemize}
\item The root is a choice node labelled $\Gamma$.
\item A choice node $t$ labelled $\Delta$ has one modal child $m$ for each outcome $(M, R) \in \Out(\Delta)$, with label $L(m) := M$; the edge $(t, m)$ carries the trace relation $R$.
\item A modal node $m$ labelled $M$ is a leaf if $M$ is inconsistent, $M$ is $\nabla$-free, or every $\nabla$-formula in $M$ is $\nabla\emptyset$; otherwise it has one choice child $u$ for each modal successor $(\Gamma_{k,\psi}, R_{k,\psi})$ of $M$, with label $L(u) := \Gamma_{k,\psi}$; the edge $(m, u)$ carries the trace relation $R_{k,\psi}$.
\end{itemize}
We write $\mathcal{T}_\varphi$ for $\mathcal{T}_{\{\varphi\}}$, $L(t)$ for the label of a node $t$, and $R(x,y)$ (or $R_e$) for the trace relation on an edge $e=(x,y)$.
\end{definition}

\begin{remark}[Uniqueness and label determinacy]\label{rem:label-determinacy}
Because $\Out(\Delta)$ and the set of modal successors of $M$ are determined solely by the set label of the parent node, $\mathcal{T}_\Gamma$ is finitely branching and unique up to label- and relation-preserving tree isomorphism. Moreover, for any choice node $t$ of $\mathcal{T}_\Gamma$, the subtree rooted at $t$ is canonically isomorphic to $\mathcal{T}_{L(t)}$. We refer to this property as \emph{label determinacy}.
\end{remark}

\subsection{Traces}

\begin{definition}[Traces]\label{def:traces}
Let $\xi = t_0 t_1 t_2 \cdots$ be a (finite or infinite) branch of $\mathcal{T}_\Gamma$, and let $R_n$ be the trace relation on the edge $(t_n, t_{n+1})$. A \emph{trace} $\tau$ on $\xi$ consists of formulas $\gamma_n\in L(t_n)$ and sets $V_n\subseteq\Bound(\Gamma_0)$ such that $(\gamma_n,\gamma_{n+1},V_n)\in R_n$ whenever $t_{n+1}$ is defined. For an infinite trace $\tau$, we define $\Reg^\infty(\tau) := \{x \in \Bound(\Gamma_0) \mid x \in V_n \text{ for infinitely many } n\}$ as the set of variables regenerated infinitely often along $\tau$.
\end{definition}

By Definition~\ref{def:saturation} and Definition~\ref{def:modal-successors}, every triple $(\gamma_n, \gamma_{n+1}, V_n) \in R_n$ is witnessed by a finite sequence of single steps (an $E_c$-path for choice-to-modal edges, or a single modal step $\nabla\Psi_k \to \psi$ or $\nabla\Psi_n \to \bigvee\Psi_n$ for modal-to-choice edges). Concatenating these witnesses yields a \emph{realisation} of $\tau$ as a sequence of formulas linked by single decomposition steps, and $\Reg^\infty(\tau)$ is precisely the set of bound variables regenerated infinitely often in any such realisation.

Although $\sqsubseteq$ is only partial, cleanness and guardedness ensure that the variables regenerated infinitely often on a trace have a least element:

\begin{lemma}[Outermost variable]\label{lem:outermost}
Let $\tau$ be an infinite trace. Then $\Reg^\infty(\tau) \neq \emptyset$, and $\Reg^\infty(\tau)$ has a unique $\sqsubseteq$-least element $x$. Moreover, from some position onwards, every formula in any realisation of $\tau$ belongs to $\Cl(\varphi_x)$, and every variable regenerated from that point on is $\sqsubseteq$-above $x$ (i.e.\ satisfies $x \sqsubseteq y$).
\end{lemma}

\begin{proof}
Measure formulas in $\Cl(\Gamma_0)$ by syntactic size, arranged so that the size of $\nabla\Psi$ strictly exceeds that of $\bigvee\Psi$. Every single step in a realisation other than a regeneration $z \to \varphi_z$ strictly decreases formula size (in particular, modal steps go from $\nabla\Psi$ either to $\psi \in \Psi$ or to $\bigvee\Psi$). Consequently, an infinite realisation must contain infinitely many regeneration steps, and since $\Bound(\Gamma_0)$ is finite, $\Reg^\infty(\tau) \neq \emptyset$.

Pick a $\sqsubseteq$-minimal element $x$ of the non-empty finite set $\Reg^\infty(\tau)$ and choose a position $N$ after which no variable outside $\Reg^\infty(\tau)$ is ever regenerated on $\tau$. Consider the first regeneration $x \to \varphi_x$ occurring after position $N$; at that moment, the realisation is at $\varphi_x \in \Cl(\varphi_x)$. We claim that the realisation never leaves $\Cl(\varphi_x)$ thereafter. Any non-regeneration step (subformula decomposition or modal step) from a formula in $\Cl(\varphi_x)$ remains in $\Cl(\varphi_x)$. Suppose a variable $z$ is regenerated ($z \to \varphi_z$) from a formula $z \in \Cl(\varphi_x)$. Because $\varphi_x$ is clean, the unique binder $\eta_z z.\varphi_z$ in $\Gamma_0$ must stand in one of three relations to $\eta_x x.\varphi_x$:
\begin{enumerate}[(i)]
\item $\eta_z z.\varphi_z$ lies strictly inside $\varphi_x$: then $x \sqsubset z$ and $\varphi_z \in \Sub(\varphi_x) \subseteq \Cl(\varphi_x)$;
\item $\eta_z z.\varphi_z$ is $\eta_x x.\varphi_x$ itself: then $z = x$ and $\varphi_z = \varphi_x \in \Cl(\varphi_x)$;
\item $\eta_z z.\varphi_z$ strictly encloses $\eta_x x.\varphi_x$: then $z \sqsubset x$.
\end{enumerate}
Case~(iii) is impossible: since this regeneration of $z$ occurs after position $N$, we would have $z \in \Reg^\infty(\tau)$ with $z \sqsubset x$, contradicting the $\sqsubseteq$-minimality of $x$ in $\Reg^\infty(\tau)$. So the realisation stays inside $\Cl(\varphi_x)$, and every variable $y$ regenerated after this point satisfies $x \sqsubseteq y$; in particular $x\sqsubseteq y$ for all $y\in\Reg^\infty(\tau)$, so $x$ is its unique $\sqsubseteq$-least element.
\end{proof}

\begin{definition}[$\mu$-traces and even/odd branches]\label{def:even-branches}
An infinite trace $\tau$ is a \emph{$\mu$-trace} if the $\sqsubseteq$-least element of $\Reg^\infty(\tau)$ is a $\mu$-variable, and a \emph{$\nu$-trace} otherwise. An infinite branch $\xi$ of a tableau is \emph{even} if it carries at least one $\mu$-trace, and \emph{odd} otherwise.
\end{definition}
With this convention an even branch is favourable to the refuter; some texts on automata games use the opposite convention.

\subsection{Tableau games and refutations}\label{sec:games-refutations}

The semantics of a canonical tableau is captured by a game between a \emph{model player} $\ModelPlayer$ and a \emph{refuter} $\RefPlayer$, played on the tableau and on a Kripke model at the same time: $\ModelPlayer$ resolves choice nodes, and $\RefPlayer$ challenges modal nodes. A refutation (Definition~\ref{def:refutations}) has the dual shape: it keeps all children of a choice node and one child of a modal node, and requires every infinite branch to be even. We use the game terminology of Section~\ref{sec:omega-conventions}. A literal $l$ \emph{holds} at a world $w$ if $\mathbb S,w\models l$.

\begin{definition}[Tableau game]\label{def:tableau-game}
Let $\mathcal T$ be a canonical tableau, $\mathbb S=(S,R,\lambda)$ a Kripke model, and $s\in S$. The \emph{tableau game} $\mathcal E(\mathcal T,\mathbb S,s)$ has the following positions, owners and moves; here $t,m,u$ range over nodes of $\mathcal T$, and $w,v$ over $S$.
\begin{itemize}
\item $(t,w)$, $t$ a choice node: owner $\ModelPlayer$; moves to $(m,w)$ for $m\in C(t)$.
\item $(m,w)$, $m$ a modal node:
\begin{itemize}
\item if some literal of $L(m)$ does not hold at $w$: owner $\ModelPlayer$, no moves;
\item otherwise, if $L(m)$ is $\nabla$-free: owner $\RefPlayer$, no moves;
\item otherwise: owner $\RefPlayer$; moves to $\langle m,w,u\rangle_{\Diamond}$ for $u\in C(m)$, and to $\langle m,w,v\rangle_{\Box}$ for $wRv$.
\end{itemize}
\item $\langle m,w,u\rangle_{\Diamond}$, $u\in C(m)$: owner $\ModelPlayer$; moves to $(u,v)$ for $wRv$.
\item $\langle m,w,v\rangle_{\Box}$, $wRv$: owner $\ModelPlayer$; moves to $(u,v)$ for $u\in C(m)$.
\end{itemize}
The initial position is $(\mathrm{root}_{\mathcal T},s)$. The tableau nodes occurring in an infinite play $\pi$ form, in order, an infinite branch $\xi(\pi)$ of $\mathcal T$; $\RefPlayer$ wins $\pi$ iff $\xi(\pi)$ is even.
\end{definition}

An inconsistent label always contains a literal that fails, so it is a loss for $\ModelPlayer$. If every $\nabla$-formula of $L(m)$ is $\nabla\emptyset$, then $C(m)=\emptyset$, and $\ModelPlayer$ loses at $(m,w)$ iff $w$ has an $R$-successor.

\begin{theorem}[Adequacy of canonical tableaux]\label{thm:adequacy}
Let $\Gamma$ be a finite clean guarded set and $\Gamma_0:=\Gamma$. Then $\mathbb{S}, s \models \bigwedge\Gamma$ if and only if $\ModelPlayer$ has a winning strategy in $\mathcal E(\mathcal T_\Gamma,\mathbb S,s)$.
\end{theorem}

The idea is that a positional strategy of $\exists$ in the evaluation game is used as a guide: at a world $w$ it resolves every disjunction, i.e.\ it is a choice function, and hence selects a modal child of a choice node.

\begin{proof}
Write $\mathcal E:=\mathcal E(\mathcal T_\Gamma,\mathbb S,s)$ and $\mathcal V:=\mathrm{EG}(\bigwedge\Gamma,\mathbb S)$, and let $W_\exists$ be the winning region of $\exists$ in $\mathcal V$. Since $\forall$ chooses the conjunct at a conjunction, Theorem~\ref{thm:eval-adequacy} gives: $\mathbb S,s\models\bigwedge\Gamma$ iff $(\gamma,s)\in W_\exists$ for every $\gamma\in\Gamma$. A set $\Theta\subseteq\Cl(\Gamma_0)$ is \emph{winning at} $w$ if $(\theta,w)\in W_\exists$ for every $\theta\in\Theta$. A literal $l$ with $(l,w)\in W_\exists$ holds at $w$, since otherwise $(l,w)$ is a dead end owned by $\exists$; in particular $\bot$ is in no winning set. A position of $\forall$ in $W_\exists$ has all its successors in $W_\exists$. No move of $\mathcal V$ other than one from $(x,w)$ with $x\in\Bound(\Gamma_0)$ is a regeneration.

($\Rightarrow$) Fix a positional strategy $f$ of $\exists$ in $\mathcal V$ winning from every position of $W_\exists$. $\ModelPlayer$ keeps the invariant that the label of the current node is winning at the current world; initially $L(\mathrm{root})=\Gamma$ is winning at $s$.
\begin{itemize}
\item At $(t,w)$: let $c$ be a choice function with $(c(D),w)=f(D,w)$ whenever $(D,w)\in W_\exists$. Every $E_c$-edge from a formula $\varphi$ with $(\varphi,w)\in W_\exists$ is a move of $\mathcal V$ that stays in $W_\exists$: at a disjunction it is the move of $f$, at a conjunction a move of $\forall$, and otherwise the only move. Hence $S_c(L(t))$ is winning at $w$. $\ModelPlayer$ moves to the child $m$ of $t$ with outcome $(M_c(L(t)),R_c(L(t)))$; $L(m)$ is winning at $w$, so every literal of $L(m)$ holds at $w$, and $\RefPlayer$ cannot move if $L(m)$ is $\nabla$-free.
\item At $(m,w)$ with $L(m)=\{\nabla\Psi_1,\dots,\nabla\Psi_i\}\cup\Lit(L(m))$, $i\ge1$: each $(\nabla\Psi_n,w)$ is a position of $\forall$ in $W_\exists$, so $(\bigvee\Psi_n,v)\in W_\exists$ whenever $wRv$, and $\langle\nabla\Psi_n,w,\psi\rangle_\Diamond\in W_\exists$ for $\psi\in\Psi_n$.
\begin{itemize}
\item At $\langle m,w,u\rangle_\Diamond$, with $u$ the $(k,\theta)$-child of $m$: $f$ moves from $\langle\nabla\Psi_k,w,\theta\rangle_\Diamond$ to some $(\theta,v)\in W_\exists$ with $wRv$, and $\ModelPlayer$ moves to $(u,v)$.
\item At $\langle m,w,v\rangle_\Box$: since $(\bigvee\Psi_1,v)\in W_\exists$ and $\bigvee\emptyset=\bot$, $\Psi_1\neq\emptyset$; following $f$ from $(\bigvee\Psi_1,v)$ along the box chain of $\nabla\Psi_1$, through positions of $W_\exists$, reaches some $(\theta,v)$ with $\theta\in\Psi_1$. $\ModelPlayer$ moves to $(u,v)$ with $u$ the $(1,\theta)$-child of $m$.
\end{itemize}
In both cases $L(u)=\{\theta\}\cup\{\bigvee\Psi_n\mid n\neq k\}$ (with $k=1$ in the second case) is winning at $v$.
\end{itemize}
Thus no finite play consistent with this strategy is lost by $\ModelPlayer$. Let $\pi$ be an infinite play with tableau branch $\xi$, and suppose that $\xi$ carries a $\mu$-trace $\tau$, starting at some $\gamma\in\Gamma$. Realise $\tau$ by single steps (the remark after Definition~\ref{def:traces}): along a choice-to-modal edge at world $w$ it is an $E_c$-path for the $c$ chosen there, which consists of moves of $\mathcal V$ consistent with $f$; along a modal-to-choice edge from $(m,w)$ to $(u,v)$ it is a step $\nabla\Psi_n\to\bigvee\Psi_n$ with $n\neq k$, which is a move of $\forall$, or the principal step $\nabla\Psi_k\to\theta$, which we replace by the moves through $\langle\nabla\Psi_k,w,\theta\rangle_\Diamond$ in the first case and by the moves through $(\bigvee\Psi_1,v)$ and its box chain in the second, all consistent with $f$. The result is an infinite play of $\mathcal V$ from $(\gamma,s)\in W_\exists$, consistent with $f$, whose regenerations are those of $\tau$; its $\sqsubseteq$-least infinitely regenerated variable is a $\mu$-variable, so $\forall$ wins it, contradicting the choice of $f$. Hence every infinite branch of a play is odd, and the strategy is winning.

($\Leftarrow$) Suppose $\mathbb S,s\not\models\bigwedge\Gamma$. Then $(\gamma,s)\notin W_\exists$ for some $\gamma\in\Gamma$, and by Theorem~\ref{thm:eval-adequacy} $\forall$ has a winning strategy $g$ from $(\gamma,s)$. Let $\sigma$ be any strategy of $\ModelPlayer$ in $\mathcal E$. $\RefPlayer$ maintains a formula $\varphi$ of the current label and a finite play $\rho$ of $\mathcal V$ from $(\gamma,s)$, consistent with $g$, ending at $(\varphi,w)$ for the current world $w$; initially $\varphi=\gamma$.
\begin{itemize}
\item At $(t,w)$, $\sigma$ moves to some $(m,w)$; let $c$ be a choice function realising the edge $(t,m)$. Extend $\rho$ from $(\varphi,w)$ by moving at a disjunction $D$ to $c(D)$ (a move of $\exists$) and at a conjunction as $g$ prescribes. This follows an $E_c$-path from $\varphi$, which is finite by Lemma~\ref{lem:acyclic} and ends at a modal formula $\varphi'\in M_c(L(t))=L(m)$; it witnesses a triple $(\varphi,\varphi',V)\in R_c(L(t))$, where $V$ is the set of variables regenerated by the extension. Put $\varphi:=\varphi'$.
\item At $(m,w)$: if some literal of $L(m)$ fails at $w$, $\ModelPlayer$ has lost. If $\varphi$ were a literal, $\rho$ would end at the dead end $(\varphi,w)$, which is not lost by $\forall$, so $\varphi$ would fail at $w$. Hence $\varphi=\nabla\Psi_k$, and $g$ moves from $(\nabla\Psi_k,w)$, since a dead end of $\forall$ is lost by him.
\begin{itemize}
\item If $g$ moves to $\langle\nabla\Psi_k,w,\theta\rangle_\Diamond$, $\RefPlayer$ moves to $\langle m,w,u\rangle_\Diamond$ for the $(k,\theta)$-child $u$ of $m$. When $\sigma$ answers $(u,v)$ with $wRv$, extend $\rho$ by the move of $\exists$ to $(\theta,v)$, and put $\varphi:=\theta\in L(u)$.
\item If $g$ moves to $(\bigvee\Psi_k,v)$, $\RefPlayer$ moves to $\langle m,w,v\rangle_\Box$. Let $\sigma$ answer $(u,v)$, with $u$ the $(k',\theta')$-child of $m$. If $k'\neq k$, put $\varphi:=\bigvee\Psi_k\in L(u)$. If $k'=k$, extend $\rho$ by moves of $\exists$ along the box chain of $\nabla\Psi_k$ to $(\theta',v)$, and put $\varphi:=\theta'\in L(u)$.
\end{itemize}
In each case $(\nabla\Psi_k,\varphi,\emptyset)\in R(m,u)$, and the extension of $\rho$ regenerates nothing.
\end{itemize}
Thus $\RefPlayer$ can maintain the construction unless $\ModelPlayer$ has already lost. In an infinite play, the formulas $\varphi$ form an infinite trace $\tau$ on its tableau branch, and the plays $\rho$ form an infinite play of $\mathcal V$ from $(\gamma,s)$, consistent with $g$, whose regenerations are those of $\tau$. Since $g$ is winning, the $\sqsubseteq$-least variable regenerated infinitely often is a $\mu$-variable, so $\tau$ is a $\mu$-trace, the branch is even, and $\RefPlayer$ wins. Hence $\sigma$ is not winning.
\end{proof}

Niwi\'nski and Walukiewicz~\cite{NiwinskiWalukiewicz96} prove the corresponding statement for their tableau game directly by signatures; here the ordinal argument is confined to Theorem~\ref{thm:eval-adequacy}.

\begin{definition}[Refutations and thinness]\label{def:refutations}
A \emph{refutation} of a finite set $\Gamma \subseteq \Cl(\Gamma_0)$ is a labelled tree $\mathcal{R}$ of alternating choice and modal nodes whose edges carry trace relations, with a choice root labelled $\Gamma$, such that:
\begin{enumerate}
\item a choice node labelled $\Delta$ has, for every choice function $c$, a modal child labelled $K_c \cap M_c(\Delta)$, where $K_c \subseteq S_c(\Delta)$ is a \emph{thinning} of the saturation $S_c(\Delta)$: every element of $K_c$ is reachable from $K_c \cap \Delta$ by an $E_c$-path staying entirely inside $K_c$, and for every non-modal formula $\psi \in K_c$:
\begin{itemize}
\item if $\psi = \alpha \land \beta$, then $K_c \cap \{\alpha, \beta\} \neq \emptyset$ (at least one conjunct is kept);
\item if $\psi = \alpha \lor \beta$, then $c(\psi) \in K_c$;
\item if $\psi = \eta_x x.\varphi_x$, then $\varphi_x \in K_c$;
\item if $\psi = x \in \Bound(\Gamma_0)$, then $\varphi_x \in K_c$.
\end{itemize}
The edge carries the trace relation defined exactly as $R_c(\Delta)$ in Definition~\ref{def:saturation}, but with $E_c$-paths restricted to $K_c$;
\item a modal node labelled $M$ is a leaf if $M$ is inconsistent; otherwise it is consistent and $\nabla$-non-free, and has \emph{exactly one} child, which is one $(k, \psi)$-successor $(\Gamma_{k,\psi}, R_{k,\psi})$ of $M$;
\item every leaf of $\mathcal{R}$ is an inconsistent modal node, and every infinite branch of $\mathcal{R}$ is \emph{even}.
\end{enumerate}
A refutation is \emph{thin} if, whenever a thinning $K_c$ contains a conjunction $\alpha \land \beta$ together with \emph{both} conjuncts $\alpha$ and $\beta$, there is no $\mu$-variable $x$ such that $\alpha \land \beta \in \Sub(\varphi_x)$ and $x$ is active in both $\alpha$ and $\beta$.
\end{definition}

A thinning is \emph{full} if $K_c=S_c(\Delta)$.

\begin{lemma}[Acyclicity of the global decomposition graph]\label{lem:global-acyclic}
Let $E:=\bigcup_cE_c$ on $\Cl(\Gamma_0)$; thus both $\alpha$ and $\beta$ are successors of $\alpha\lor\beta$. Then $E$ is acyclic.
\end{lemma}
\begin{proof}
The proof of Lemma~\ref{lem:acyclic} does not use the choice of disjuncts and applies verbatim to $E$.
\end{proof}

\begin{definition}[Sequential refutations]\label{def:sequential}
Let $E:=\bigcup_cE_c$ and let $\prec$ be a \emph{topological order} of $E$.\footnote{A topological order of a finite acyclic directed graph is a linear order of its vertices in which every edge points forward. It exists here by Lemma~\ref{lem:global-acyclic}.}
At a choice node labelled $\Delta$, let $K_c$ be the thinning associated with a choice function $c$. For a formula $\psi$ put
\[
H_c(\psi):=\{(D,c(D))\mid D\in K_c\cap\Dis,\ D\prec\psi\},
\]
and define the \emph{thinning decision} at $\psi$ by
\[
d_c(\psi):=
\begin{cases}
\mathbf 1_{\psi\in K_c}, & \psi\in\Dis,\\[1mm]
\bigl(\mathbf 1_{\psi\in K_c},\ \{\chi\mid (\psi,\chi)\in E_c,\ \chi\in K_c\}\bigr), & \psi\notin\Dis.
\end{cases}
\]
The refutation is \emph{sequential with respect to $\prec$} if, at every choice node and for all choice functions $c,c'$ and formulas $\psi\in S_c(\Delta)\cap S_{c'}(\Delta)$,
\[
H_c(\psi)=H_{c'}(\psi)\quad\Longrightarrow\quad d_c(\psi)=d_{c'}(\psi).
\]
It is \emph{sequential} if it is sequential with respect to some topological order of $E$.
\end{definition}

\begin{remark}[Sequentiality as non-anticipation]\label{rem:sequential-intuition}
Sequentiality means that thinning decisions follow $\prec$, depending only on earlier retained disjunctions, never on future choices.
Full thinnings are sequential for every topological order.
Construction~\ref{con:expansion} uses this non-anticipation; Example~\ref{ex:non-sequential} shows its failure.
\end{remark}

\begin{theorem}[Existence of refutations; Claim (e)]\label{thm:unsat-refutation}
Let $\Gamma$ be a finite clean guarded set and take $\Gamma_0:=\Gamma$. If $\Gamma$ is unsatisfiable, then $\Gamma$ has a refutation in which every thinning is the full saturation $K_c=S_c(\Delta)$ (and which is therefore sequential). Moreover, the refutation may be chosen regular: up to isomorphism it has only finitely many rooted labelled subtrees.
\end{theorem}

The proof is given in Theorem~\ref{thm:finite-game-refutations}, after the canonical tableau has been turned into a finite parity game (Section~\ref{sec:finite-refutation-game}). In~\cite{NiwinskiWalukiewicz96} existence follows from Martin's Borel determinacy, and regularity, with an exponential bound, is obtained separately via Rabin tree automata~\cite{NiwinskiWalukiewicz96}; here both follow from positional determinacy of a finite parity game.

\begin{lemma}\label{lem:aconj-thin}
Suppose that $\bigwedge\Gamma_0$ is aconjunctive.
Then, for every finite set $\Gamma\subseteq\Cl(\Gamma_0)$,
every refutation of $\Gamma$ in which all thinnings are full
is thin.
\end{lemma}

\begin{proof}
A violation of thinness at $\alpha\land\beta\in K_c$ is a $\mu$-variable $x$ with $\alpha\land\beta\in\Sub(\varphi_x)$ active in both conjuncts, which aconjunctivity excludes.
\end{proof}

\subsection{Two-sided refutations and Claim (f)}\label{sec:thin-refutation-proof}

Section~\ref{sec:transfer} constructs refutations of two-sided sequents $\Gamma\vdash\Delta$. A formula may occur on both sides, and the two occurrences must be decomposed independently. We therefore tag every formula with its side.

\begin{definition}[Tagged closures and two-sided refutations]\label{def:two-sided}
Let $\Gamma_0$ and $\Delta_0$ be finite clean guarded sets in $\nabla$-form such that no bound variable of either occurs in the other, and put $\sim\Delta_0:=\{\sim\delta\mid\delta\in\Delta_0\}$, again clean and guarded, with the same bound variables and $\mu$, $\nu$ interchanged. The \emph{tagged closure} is the disjoint union
\[
\Cl(\Gamma_0;\Delta_0):=\bigl(\{L\}\times\Cl(\Gamma_0)\bigr)\cup\bigl(\{R\}\times\Cl(\sim\Delta_0)\bigr).
\]
A tagged formula $(s,\chi)$ is of the same kind as $\chi$ (literal, $\nabla$-formula, disjunction, \dots); its $E_c$-edges, modal successors and trace steps are those of $\chi$, with the tag $s$ kept; a choice function is a map on tagged disjunctions, so that $(L,D)$ and $(R,D)$ are resolved independently; a set of tagged modal formulas is \emph{consistent} if the set of underlying formulas is; and the fixpoint type, $\varphi_x$, $\sqsubseteq$, $\preceq$ and activity of a bound variable are those of the side containing it. With these readings, Definitions~\ref{def:saturation}--\ref{def:sequential} apply verbatim to finite subsets $\Theta\subseteq\Cl(\Gamma_0;\Delta_0)$. We write $\llangle\Theta\rrangle:=\{\llangle\chi\rrangle\mid(s,\chi)\in\Theta\}$ and $\Theta\vdash$ for $\llangle\Theta\rrangle\vdash$; by~\eqref{eq:two-one-sided}, $\Theta\vdash$ iff
\[
\{\llangle\chi\rrangle\mid(L,\chi)\in\Theta\}\vdash\{\sim\llangle\chi\rrangle\mid(R,\chi)\in\Theta\}.
\]
A \emph{two-sided refutation} of $\Gamma\vdash\Delta$, for $\Gamma\subseteq\Cl(\Gamma_0)$ and $\Delta\subseteq\Delta_0$, is a refutation of $(\{L\}\times\Gamma)\cup(\{R\}\times\sim\Delta)$.
\end{definition}

For $\Delta_0=\emptyset$ we identify $(L,\chi)$ with $\chi$. A right formula $(R,\chi)$ stands for the succedent formula $\sim\chi$: a right disjunction is a conjunction of the succedent, decomposed by $(\land R)$, a right conjunction is a disjunction of the succedent, decomposed by $(\lor R)$ and thinned by $(\mathrm{Weak}R)$, and a right $\mu$-variable is a $\nu$-variable of $\Delta_0$.

\begin{lemma}[Tagged closures]\label{lem:tagged}
For finite subsets of $\Cl(\Gamma_0;\Delta_0)$, Lemmas~\ref{lem:acyclic}, \ref{lem:outermost} and~\ref{lem:global-acyclic} hold.
\end{lemma}

\begin{proof}
Every $E_c$-edge, modal step and trace step joins two formulas with the same tag, and on each side it is an edge or step of $\Cl(\Gamma_0)$ or $\Cl(\sim\Delta_0)$. The three lemmas concern single paths and follow from the one-sided statements; in Lemma~\ref{lem:outermost}, a trace stays on one side, whose bound variables are disjoint from those of the other.
\end{proof}

\begin{theorem}[Thin refutations; Claim (f), Walukiewicz {\cite[Theorem~31]{Walukiewicz00}}]\label{thm:thin-refutation}
Let $\Theta\subseteq\Cl(\Gamma_0;\Delta_0)$ be finite, and suppose that no formula of $\Theta$ contains a variable of $\Bound(\Gamma_0)\cup\Bound(\Delta_0)$ free. If $\Theta$ has a thin sequential refutation, then $\Theta\vdash$. In particular, for $\Gamma\subseteq\Gamma_0$ and $\Delta\subseteq\Delta_0$, if $\Gamma\vdash\Delta$ has a thin sequential two-sided refutation, then $\Gamma\vdash\Delta$ is derivable in $\Koz$.
\end{theorem}

Walukiewicz proved Claim~(f) for refutations of a single well-named formula built from single rule applications; we expand our saturated refutations of tagged sets, whose inner labels may contain bound variables free, into single-rule format (Construction~\ref{con:expansion}) and give the proof in Appendix~\ref{app:thin}, which uses nothing from Section~\ref{sec:maps} or Sections~\ref{sec:anf}--\ref{sec:pcc}.

\begin{construction}[Single-rule expansion]\label{con:expansion}
Let $\mathcal R$ be a thin refutation of a finite set $\Gamma\subseteq\Cl(\Gamma_0;\Delta_0)$, sequential with respect to a topological order $\prec$ of $E=\bigcup_cE_c$ (Definition~\ref{def:sequential}); the construction uses this $\prec$. The \emph{single-rule expansion} $\mathcal R^+$ replaces each saturated step of $\mathcal R$, from a choice node labelled $\Delta$ to its modal children $\{K_c\cap M_c(\Delta)\}_c$, by the following finite tree of single rule applications, built upwards from $\Delta$. At a sequent $\Theta$, let $\theta$ be the $\prec$-least non-modal formula of $\Theta$; if there is none, the modal formulas of $\Theta$ outside $K_c$ are removed by $(\mathrm{Weak})$, and the block ends.
\begin{itemize}
\item If $\theta\notin K_c$, it is removed by $(\mathrm{Weak})$.
\item $\theta=\alpha\lor\beta$: two premises, with $\alpha$ and with $\beta$ in place of $\theta$ $(\lor)$.
\item $\theta=\alpha\land\beta$: one premise containing those of $\alpha,\beta$ that belong to $K_c$ $(\land)$.
\item $\theta=\eta_xx.\varphi_x$ or $\theta=x$: one premise with $\varphi_x$ in place of $\theta$ $(\eta)$, $(\mathrm{reg})$.
\end{itemize}
A branch of the block fixes a partial choice function on the disjunctions decomposed so far; the decisions at $\theta$ (whether $\theta\in K_c$ and, for a conjunction, which conjuncts belong to $K_c$) are read off from $K_c$ for any total extension $c$. By Lemma~\ref{lem:expansion-well-defined} below they do not depend on the extension, so the construction is well-defined; since the formulas decomposed along a branch form a $\prec$-increasing sequence, no decision is revised later in the block. By Lemma~\ref{lem:global-acyclic} every block is finite, and a branch of the block extending $c$ ends in the modal sequent $K_c\cap M_c(\Delta)$. Choice functions that lead to the same branch of a block have the same thinning and the same trace relation (Lemma~\ref{lem:expansion-well-defined}(3)); for each branch we fix one of them, $c$, and continue above the branch with the subtree of $\mathcal R$ at the child for $c$, beginning with the modal rule $(\nabla)$ of $\mathcal R$ unless that child is a leaf of $\mathcal R$.
\end{construction}

\begin{lemma}[Well-definedness of the expansion]\label{lem:expansion-well-defined}
Consider a block of Construction~\ref{con:expansion} at a choice node of $\mathcal R$ labelled $\Delta$, and a sequent $\Theta$ on a branch of the block. Let $p$ be the partial choice function fixed by the disjunctions decomposed on the branch before $\Theta$, and let $\theta$ be the $\prec$-least non-modal formula of $\Theta$, if there is one. Then, for all choice functions $c,c'$ extending $p$:
\begin{enumerate}
\item every formula decomposed on the branch before $\Theta$ belongs to $K_c$, $\Theta\subseteq S_c(\Delta)$, and every non-modal $\psi\in K_c$ with $\psi\prec\theta$ (every non-modal $\psi\in K_c$, if $\Theta$ contains no non-modal formula) has been decomposed on the branch before $\Theta$;
\item if $\theta$ exists, then $H_c(\theta)=H_{c'}(\theta)$, and hence $d_c(\theta)=d_{c'}(\theta)$;
\item if $\Theta$ contains no non-modal formula, then $K_c=K_{c'}$, the modal formulas of $\Theta$ that belong to $K_c$ form the set $K_c\cap M_c(\Delta)=K_{c'}\cap M_{c'}(\Delta)$, and the trace relations of $\mathcal R$ at the modal children for $c$ and for $c'$ coincide.
\end{enumerate}
\end{lemma}

\begin{proof}
Every formula introduced by decomposing a formula $\chi$ is an $E$-successor of $\chi$, hence $\succ\chi$. When $\chi$ is decomposed it is the $\prec$-least non-modal formula of the current sequent, so every non-modal formula present later is $\succ\chi$; thus the formulas decomposed along a branch form a $\prec$-increasing sequence, and all of them precede $\theta$.

We prove (1) and (2) simultaneously by induction on the position of $\Theta$ on the branch. Let $\Theta'$ be a sequent strictly before $\Theta$, with $\prec$-least non-modal formula $\theta'$ and partial choice function $p'\subseteq p$. By~(2) for $\Theta'$, the decision taken at $\theta'$ is $d_c(\theta')$ for every $c$ extending $p$, so it agrees with $K_c$: $\theta'$ is weakened iff $\theta'\notin K_c$; if it is decomposed, a disjunction is followed along the premise of $p(\theta')=c(\theta')$, a conjunction keeps exactly its conjuncts in $K_c$, and a fixpoint formula or variable passes to $\varphi_x$, which lies in $K_c$ by Definition~\ref{def:refutations}(1). This proves the first assertion of~(1), and $\Theta\subseteq S_c(\Delta)$ because every introduced formula is reached from $\Delta$ by $E_c$-edges consistent with $p$.
For the last assertion of~(1), let $\psi\in K_c$ be non-modal with $\psi\prec\theta$, and choose an $E_c$-path $\chi_0\chi_1\cdots\chi_l=\psi$ inside $K_c$ with $\chi_0\in\Delta$; all $\chi_i$ are non-modal and $\preceq\psi\prec\theta$. By induction on $i$, $\chi_i$ has been introduced on the branch ($\chi_0\in\Delta$, and $\chi_{i+1}$ is introduced when $\chi_i$ is decomposed, by the agreement just shown). A formula leaves the sequent only by being decomposed or weakened, and $\chi_i$ is no longer present in $\Theta$, since $\chi_i\prec\theta$; as $\chi_i\in K_c$, it was not weakened, so it was decomposed.
For~(2), by~(1) and the first paragraph, the disjunctions $D\in K_c$ with $D\prec\theta$ are exactly the disjunctions decomposed on the branch, and $c(D)=p(D)$ for them. Hence $H_c(\theta)=\{(D,p(D))\mid D\text{ a disjunction decomposed on the branch}\}=H_{c'}(\theta)$. Since $\theta\in S_c(\Delta)\cap S_{c'}(\Delta)$ by~(1), sequentiality with respect to $\prec$ gives $d_c(\theta)=d_{c'}(\theta)$.
For~(3), by~(1) every non-modal formula of $K_c$ has been decomposed, so all disjunctions of $K_c$ have been decomposed and $c=p$ on them. Every modal $m\in K_c$ lies in $\Theta$: either $m\in\Delta$, or $m$ is an $E_c$-successor kept in $K_c$ of a decomposed formula of $K_c$, and was introduced at that step; modal formulas are not removed before the end of the block. Conversely, the modal formulas of $\Theta$ lie in $M_c(\Delta)$ by~(1). For a modal $m\in\Theta$, $H_c(m)=\{(D,p(D))\mid D\text{ decomposed},\ D\prec m\}=H_{c'}(m)$, so $d_c(m)=d_{c'}(m)$ by sequentiality, i.e.\ $m\in K_c$ iff $m\in K_{c'}$. Thus $K_c$ and $K_{c'}$ consist of the same decomposed non-modal formulas and the same modal formulas of $\Theta$, so $K_c=K_{c'}$; since $c$ and $c'$ agree on the disjunctions of $K_c$, the $E_c$- and $E_{c'}$-paths inside $K_c$ coincide, and so do the trace relations.
\end{proof}

\begin{lemma}[Properties of the expansion]\label{lem:thin-reduction}
$\mathcal R^+$ is a finitely branching tree of finite subsets of $\Cl(\Gamma_0;\Delta_0)$ with root $\Gamma$, and:
\begin{enumerate}
\item each block ends at a node with the label of a node of $\mathcal R$, contracting each block maps $\mathcal R^+$ onto a subtree of $\mathcal R$ preserving labels, and every infinite branch of $\mathcal R^+$ projects to an infinite branch of $\mathcal R$;
\item every trace of $\mathcal R^+$ projects to a trace of $\mathcal R$ with the same set of variables regenerated infinitely often, and every trace of the corresponding branch of $\mathcal R$ is such a projection; in particular every infinite branch of $\mathcal R^+$ carries a $\mu$-trace;
\item $\mathcal R^+$ is thin (cf.\ \cite{Walukiewicz00}): no conjunction $\alpha\land\beta$ with a $\mu$-variable active in both conjuncts is decomposed into both $\alpha$ and $\beta$; and every leaf of $\mathcal R^+$ is an inconsistent sequent.
\end{enumerate}
\end{lemma}

\begin{proof}
(1) holds by construction; the leaves of $\mathcal R^+$ are the inconsistent leaves of $\mathcal R$.
(2) Within a block, consecutive formulas on a branch are linked by single $E_c$-edges, and the occurrences surviving to the end of the block are exactly those reachable inside $K_c$; so a trace through the block is an $E_c$-path inside $K_c$ from a formula of $\Delta$ to a modal formula, with the regenerated variables $V$ of the corresponding triple of $\mathcal R$, and conversely every triple $(\gamma,m,V)$ is realised by such a path. Modal edges are unchanged and contain no regeneration. Hence traces correspond block by block with identical regeneration sets, and the last claim follows from Definition~\ref{def:refutations}(3).
(3) Suppose a block decomposes $\alpha\land\beta$ into both conjuncts and a $\mu$-variable $x$ is active in both. Then $\alpha\land\beta,\alpha,\beta\in K_c$, and $\alpha\land\beta\in\Sub(\varphi_x)$: if $x$ is active in $\alpha$ through a variable $y$ free in $\alpha$, then $x\sqsubseteq y$ (since ${\preceq}\subseteq{\sqsubseteq}$) and, by cleanness of the side of $\alpha\land\beta$, the formula $\alpha\land\beta$, in which $y$ occurs free, lies in $\Sub(\varphi_y)\subseteq\Sub(\varphi_x)$. This contradicts thinness of $\mathcal R$.
\end{proof}

\begin{example}[Why sequentiality]\label{ex:non-sequential}
Let $P:=p\lor\neg p$, $Q:=q\lor\neg q$, $X:=P\land r$, $Y:=Q\land s$ and $\Delta=\{X,Y\}$. Let $K_c$ keep $X,Y,r,s$, keep $P$ (and $c(P)$) iff $c(Q)=q$, and keep $Q$ (and $c(Q)$) iff $c(P)=p$. These are thinnings, but $\mathcal R$ is not sequential for any topological order $\prec$. Take choice functions $c,c'$ with $c(P)=c'(P)=\neg p$, $c(Q)=q$ and $c'(Q)=\neg q$. Then $Q\notin K_c$ and $Q\notin K_{c'}$, so no retained disjunction precedes $P$ and $H_c(P)=H_{c'}(P)=\emptyset$; yet $P\in K_c$ and $P\notin K_{c'}$. In a rule-by-rule expansion, whichever disjunction is reached first would have to be kept or weakened before the other has been decomposed.
\end{example}

\subsection{Tableau consequence relations}\label{sec:maps}

We present tableau consequence as a relation between the nodes of two canonical tableaux; maps are the functional special case.

\begin{figure}[t]
\centering
\includegraphics[width=0.85\textwidth]{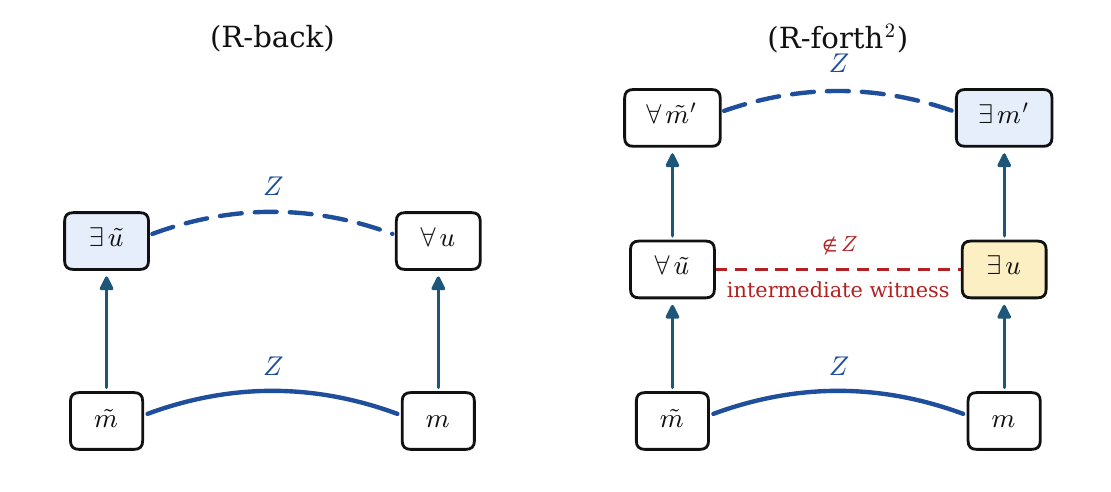}
\caption{Clauses (R-back) and (R-forth$^2$) of Definition~\ref{def:consequence-relation}.}
\label{fig:rconds}
\end{figure}

\begin{definition}[Consequence relation]\label{def:consequence-relation}
Let $\widetilde{\mathcal T}$ and $\mathcal T$ be canonical tableaux, with choice-node sets $\Ch(\widetilde{\mathcal T}),\Ch(\mathcal T)$ and modal-node sets $\Mod(\widetilde{\mathcal T}),\Mod(\mathcal T)$. A \emph{consequence relation from $\widetilde{\mathcal T}$ to $\mathcal T$} is a relation
\[
Z\subseteq(\Ch(\widetilde{\mathcal T})\times\Ch(\mathcal T))\cup
(\Mod(\widetilde{\mathcal T})\times\Mod(\mathcal T))
\]
which relates the two roots and satisfies the following conditions (see Figure~\ref{fig:rconds}).
\begin{description}
\item[(R-ch)] If $(\tilde t,t)\in Z$, then for every $\tilde m\in C(\tilde t)$ there is $m\in C(t)$ with $(\tilde m,m)\in Z$.
\item[(R-loc)] If $(\tilde m,m)\in Z$ and $L(\tilde m)$ is consistent, then $L(m)$ is consistent, $\Lit(L(m))\subseteq\Lit(L(\tilde m))$, and, if $L(m)$ is $\nabla$-non-free, then $L(\tilde m)$ is $\nabla$-non-free.
\item[(R-back)] If $(\tilde m,m)\in Z$, $L(\tilde m)$ is consistent, and $L(m)$ is $\nabla$-non-free, then for every $u\in C(m)$ there is $\tilde u\in C(\tilde m)$ with $(\tilde u,u)\in Z$.
\item[(R-forth$^2$)] Under the same hypotheses, for every $\tilde u\in C(\tilde m)$ and every $\tilde m'\in C(\tilde u)$ such that $L(\tilde m')$ is consistent, there are $u\in C(m)$ and $m'\in C(u)$ with $(\tilde m',m')\in Z$.
\item[(R-par)] Call a finite or infinite sequence $(\tilde m_0,m_0),(\tilde m_1,m_1),\dots$ of modal pairs of $Z$ a \emph{$Z$-path} if $\tilde m_0$ and $m_0$ are children of the two roots and $\tilde m_{j+1}$, $m_{j+1}$ are grandchildren of $\tilde m_j$, $m_j$; its \emph{left} and \emph{right projections} are the paths $\tilde t_0\tilde m_0\tilde u_1\tilde m_1\cdots$ and $t_0m_0u_1m_1\cdots$ (branches when the $Z$-path is infinite), where $\tilde t_0,t_0$ are the roots and $\tilde u_{j+1},u_{j+1}$ the parents of $\tilde m_{j+1},m_{j+1}$. If the right projection of an infinite $Z$-path is even, then so is its left projection.
\end{description}
The right choice node $u$ in (R-forth$^2$) may depend on $\tilde m'$, and $(\tilde u,u)$ is not required to belong to $Z$. We write $\widetilde{\mathcal T}\Rightarrow_Z\mathcal T$, or $Z:\widetilde{\mathcal T}\Rightarrow\mathcal T$, if $Z$ is a consequence relation from $\widetilde{\mathcal T}$ to $\mathcal T$, and $\widetilde{\mathcal T}\Rightarrow\mathcal T$ if one exists. The pairs of a $Z$-path have equal depths, and the pairs of ancestors of a pair on a $Z$-path again lie on it.
\end{definition}

\begin{remark}[Design choices]\label{rem:back-direction}
(1)~(R-back) must challenge every right child: reversing its quantifiers would incorrectly relate $\nabla\{p\}$ to $\nabla\{p,q\}$.
(2)~Requiring $(\tilde u,u)\in Z$ in (R-forth$^2$) would impose (R-ch) for all children of $\tilde u$, although $u$ may depend on $\tilde m'$ (Section~\ref{sec:pcc}; Example~\ref{ex:consequence-relation}).
(3)~On canonical trees, consequence relations represent positional strategies in Walukiewicz's game~\cite{Walukiewicz00}: (R-back) answers single-successor moves and (R-forth$^2$) all-sons moves.
\end{remark}

\begin{figure}[t]
\centering
\includegraphics[width=\textwidth]{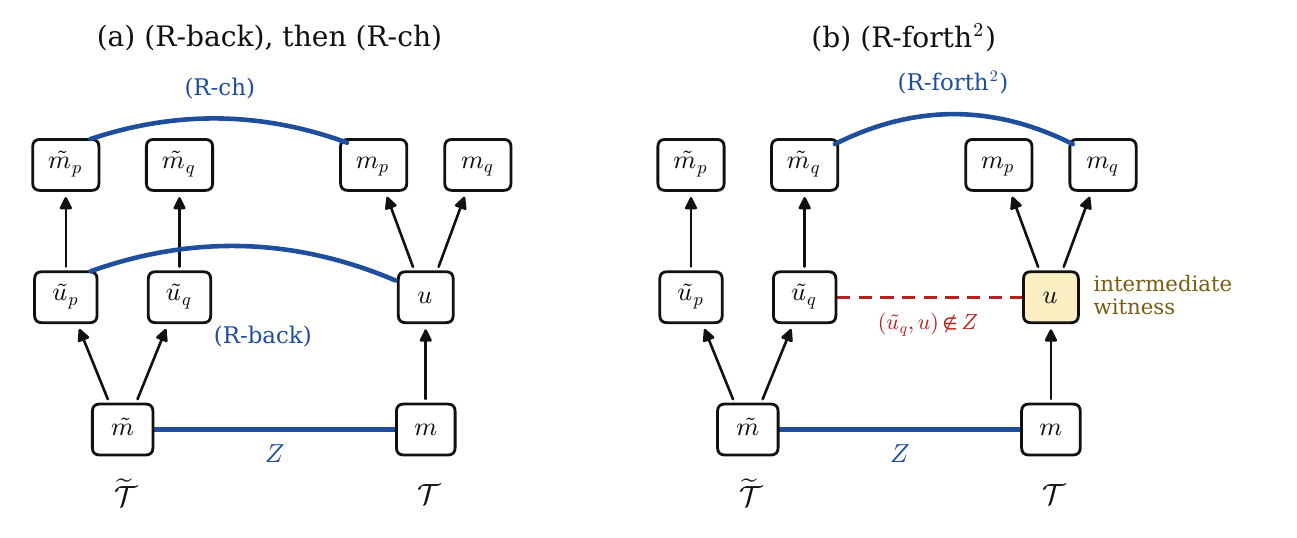}
\caption{Example~\ref{ex:consequence-relation}. Thin arrows are tableau edges; thick lines are pairs of $Z$.}
\label{fig:back-forth}
\end{figure}

\begin{example}[A consequence relation]\label{ex:consequence-relation}
Let $\alpha:=\nabla\{p,q\}$ and $\widehat\varphi:=\nabla\emptyset\lor\nabla\{p\lor q\}$; then $\alpha\models\widehat\varphi$. On the left, the modal node $\tilde m$ is labelled $\{\nabla\{p,q\}\}$, with choice children $\tilde u_p,\tilde u_q$ labelled $\{p\},\{q\}$ and $\nabla$-free modal children $\tilde m_p,\tilde m_q$ with the same labels. On the right, the root has modal children labelled $\{\nabla\emptyset\}$ and $\{\nabla\{p\lor q\}\}$; the latter, $m$, has the choice child $u$ labelled $\{p\lor q\}$, with modal children $m_p,m_q$ labelled $\{p\},\{q\}$. Relate $\tilde m$ to $m$. (R-back) is witnessed by $(\tilde u_p,u)$ and (R-ch) there by $(\tilde m_p,m_p)$; (R-forth$^2$) answers $\tilde m_p$ and $\tilde m_q$ by $(u,m_p)$ and $(u,m_q)$. Here $u$ serves only as an intermediate witness for $\tilde m_q$: the pair $(\tilde u_q,u)$ is not in $Z$. Altogether, with $\tilde t_0,t_0$ the roots,
\[
Z=\{(\tilde t_0,t_0),\ (\tilde m,m),\ (\tilde u_p,u),\ (\tilde m_p,m_p),\ (\tilde m_q,m_q)\}.
\]
There are no infinite branches, so (R-par) holds vacuously (Figure~\ref{fig:back-forth}).
\end{example}

\begin{definition}[Bisimulation maps]\label{def:tableau-map}
A \emph{bisimulation map} $\zeta:\widetilde{\mathcal T}\to\mathcal T$ is a map from the nodes of $\widetilde{\mathcal T}$ to those of $\mathcal T$ that sends the root to the root, maps $C(\tilde x)$ into $C(\zeta(\tilde x))$ for every node $\tilde x$, surjectively at choice nodes and bijectively at consistent $\nabla$-non-free modal nodes, preserves consistency in both directions and, at modal nodes with a consistent label, literals and $\nabla$-freeness in both directions, and satisfies the parity condition in both directions: an infinite branch $\tilde\xi$ of $\widetilde{\mathcal T}$ is even iff $\zeta(\tilde\xi)$ is even. Its graph $\Graph(\zeta):=\{(\tilde x,\zeta(\tilde x))\}$ is a consequence relation each of whose modal pairs lies on a $\Graph(\zeta)$-path.
\end{definition}

\begin{lemma}[Composition]\label{lem:relation-map-composition}
Let $Z:\widetilde{\mathcal T}\Rightarrow\mathcal T$ and $Z':\mathcal T\Rightarrow\mathcal U$ be consequence relations each of whose modal pairs lies on a path of the respective relation. Then
\[
Z'\circ Z:=\{(\tilde x,y)\mid (\tilde x,x)\in Z\text{ and }(x,y)\in Z'\text{ for some }x\}
\]
is a consequence relation from $\widetilde{\mathcal T}$ to $\mathcal U$, and every modal pair of the composite lies on a finite path of the composite from its roots. By Definition~\ref{def:tableau-map}, this applies to $Z'=\Graph(f)$ for a bisimulation map $f$.
\end{lemma}

\begin{proof}
The roots are related; (R-ch), (R-loc), (R-back), and (R-forth$^2$) follow by composing witnesses, with (R-loc) ensuring consistency and $\nabla$-non-freeness at the intermediate modal node whenever required.
(R-par): let $(\tilde x_i,y_i)_i$ be an infinite $(Z'\circ Z)$-path whose right projection is even, and fix $n$. Choose $x_n$ with $(\tilde x_n,x_n)\in Z$ and $(x_n,y_n)\in Z'$. By hypothesis these pairs lie on a $Z$-path and on a $Z'$-path, which consist of the pairs of ancestors at equal depths (Definition~\ref{def:consequence-relation}); so the ancestors $x_i$ of $x_n$ satisfy $(\tilde x_i,x_i)\in Z$ and $(x_i,y_i)\in Z'$ for all $i\le n$. These finite sequences $(x_i)_{i\le n}$ form a non-empty set for each $n$; it is finite because $\mathcal T$ is finitely branching, and closed under prefixes. By K\"onig's lemma there is an infinite sequence $(x_i)_i$ such that $(\tilde x_i,x_i)_i$ is a $Z$-path and $(x_i,y_i)_i$ a $Z'$-path. Since $(y_i)$ is even, so is $(x_i)$ by (R-par) for $Z'$, and so is $(\tilde x_i)$ by (R-par) for $Z$.
For the path property, take a composite modal pair $(\tilde m,n)$ with intermediate witness $m$. Its unique modal ancestors in the three trees are related level by level: the path property for $Z$ and $Z'$ supplies the ancestor pairs $(\tilde m_i,m_i)\in Z$ and $(m_i,n_i)\in Z'$ on paths ending at the chosen pairs. Thus $(\tilde m_i,n_i)\in Z'\circ Z$ at each depth, giving the required finite composite path.
\end{proof}

\begin{proposition}[Transport of model strategies]\label{prop:run-transport}
Let $Z:\widetilde{\mathcal T}\Rightarrow\mathcal T$ be a consequence relation and $(\mathbb S,s)$ a pointed model. If $\ModelPlayer$ has a winning strategy in $\mathcal E(\widetilde{\mathcal T},\mathbb S,s)$, then she has one in $\mathcal E(\mathcal T,\mathbb S,s)$. For a bisimulation map $f:\mathcal T\to\mathcal U$, $\ModelPlayer$ has a winning strategy in $\mathcal E(\mathcal T,\mathbb S,s)$ if and only if she has one in $\mathcal E(\mathcal U,\mathbb S,s)$.
\end{proposition}
\begin{proof}
Let $\tilde\sigma$ be a winning strategy in the left game. $\ModelPlayer$ plays the right game while simulating, at the same world, a play of the left game consistent with $\tilde\sigma$, and she maintains a pair $(\tilde x,x)\in Z$ of current choice or modal nodes, initially the root pair at $s$.
\begin{itemize}
\item At $(t,w)$ with $(\tilde t,t)\in Z$: $\tilde\sigma$ moves from $(\tilde t,w)$ to some $(\tilde m,w)$; (R-ch) gives $m\in C(t)$ with $(\tilde m,m)\in Z$, and she moves to $(m,w)$.
\item At $(m,w)$ with $(\tilde m,m)\in Z$: since $\tilde\sigma$ is winning, $(\tilde m,w)$ is not a loss for $\ModelPlayer$, so every literal of $L(\tilde m)$ holds at $w$; in particular $L(\tilde m)$ is consistent. By (R-loc), every literal of $L(m)$ holds at $w$, and $L(\tilde m)$ is $\nabla$-non-free if $L(m)$ is. If $L(m)$ is $\nabla$-free, $\RefPlayer$ cannot move. Otherwise:
\begin{itemize}
\item at $\langle m,w,u\rangle_{\Diamond}$: (R-back) gives $\tilde u\in C(\tilde m)$ with $(\tilde u,u)\in Z$; in the simulated play $\RefPlayer$ moves to $\langle\tilde m,w,\tilde u\rangle_{\Diamond}$, $\tilde\sigma$ answers some $(\tilde u,v)$ with $wRv$, and she moves to $(u,v)$; the play continues with $(\tilde u,u)\in Z$;
\item at $\langle m,w,v\rangle_{\Box}$: in the simulated play $\RefPlayer$ moves to $\langle\tilde m,w,v\rangle_{\Box}$, $\tilde\sigma$ answers some $(\tilde u,v)$ with $\tilde u\in C(\tilde m)$ and then moves to some $(\tilde m',v)$, where $L(\tilde m')$ is consistent because $\tilde\sigma$ is winning; (R-forth$^2$) gives $u\in C(m)$ and $m'\in C(u)$ with $(\tilde m',m')\in Z$; she moves to $(u,v)$ and then to $(m',v)$; the play continues with $(\tilde m',m')\in Z$.
\end{itemize}
\end{itemize}
$\ModelPlayer$ is never stuck, so a finite play ends at a position where $\RefPlayer$ cannot move. In an infinite play, the pairs at modal positions form an infinite $Z$-path, whose steps use (R-back) and (R-ch) after a challenge of the first kind and (R-forth$^2$) after one of the second kind; its left projection is the branch of the simulated play and its right projection the branch of the actual play. The former is odd because $\tilde\sigma$ is winning, so the latter is odd by (R-par). Hence the strategy is winning. For a bisimulation map $f$, the forward direction is the first claim applied to $\Graph(f)$. Conversely, lift a winning strategy along $f$, using surjectivity at choice nodes and bijectivity at consistent $\nabla$-non-free modal nodes to answer the two kinds of challenge.
Two-way preservation of consistency, literals, $\nabla$-freeness and parity ensures that every lifted play is winning.
\end{proof}

\begin{lemma}[Inverse of a bisimulation map]
\label{lem:pcc-inv-bisim}
Let $\zeta : \mathcal{U}_1 \to \mathcal{U}_2$ be a bisimulation map between
canonical tableaux (Definition~\ref{def:tableau-map}). Then its inverse graph
\[
\Graph(\zeta)^{-1} := \{(\zeta(s), s) \mid s \text{ is a node of } \mathcal{U}_1\}
\]
is a tableau consequence relation $\mathcal{U}_2 \Rightarrow \mathcal{U}_1$, every
modal pair of which lies on a $\Graph(\zeta)^{-1}$-path.
\end{lemma}

\begin{proof}
By Definition~\ref{def:tableau-map}, roots correspond.
Surjectivity at choice nodes gives (R-ch).
Bijectivity at consistent $\nabla$-non-free modal nodes, together with choice-node surjectivity, gives (R-back) and (R-forth$^2$).
Two-way preservation gives (R-loc) and (R-par), since the left projection is the image of the right projection under $\zeta$.
The image pairs of any modal node's ancestor chain form the required finite inverse-graph path.
\end{proof}

\section{Automaton normal form}\label{sec:anf}

Section~\ref{sec:anf-def} introduces automaton normal form. Sections~\ref{sec:buchi}--\ref{sec:finite-refutation-game} equip $\mathcal T_\Gamma$ with a deterministic parity monitor, cut it at loop nodes to a finite prefix tree $\mathcal G_\Gamma$, and read $\mathcal G_\Gamma$, with its loop nodes folded back, as a finite parity game; this proves Claim~(e). Section~\ref{sec:anf-construction} reads $\anf(\Gamma)$ off $\mathcal G_\Gamma$ (Claim~(b)), and Section~\ref{sec:box-isolation} adds a syntactic normalisation needed in Sections~\ref{sec:completeness} and~\ref{sec:pcc}.

\subsection{Automaton normal form}\label{sec:anf-def}

Automaton normal form (the \emph{disjunctive formulas} of~\cite{JaninWalukiewicz95}; cf.\ \cite{Walukiewicz00}) restricts explicit conjunctions to literal sets $\bigwedge L$ and to $\nabla\Phi\land\bigwedge L$, with $L$ a set of literals, so that each branch of an ANF tableau carries a single trace (Lemma~\ref{lem:anf-properties}).

\begin{example}\label{ex:anf}
The formula $\mu x.(\nabla\{p\lor x,\ q\}\land r)$ is in automaton normal form: the bound variable $x$ occurs only as a disjunct of $p\lor x$, which is an element of $\Phi$ under $\nabla$, and $p$, $q$, $r$, $x$ are singleton literal sets. The formula $\nu y.\mu x.\nabla\{x\land y\}$ is not: $x$ and $y$ occur together in an explicit conjunction, which clause~4 of Definition~\ref{def:anf-set} forbids. Saturating $\{x\land y\}$, the paths through $x$ and through $y$ both reach $\nabla\{x\land y\}$ but have different regeneration sets, so a branch can carry distinct infinite traces.
\end{example}

\begin{definition}[Automaton normal form]\label{def:anf-set}
The set $\ANF$ of formulas in \emph{automaton normal form} (or \emph{ANF formulas}) is the least set of clean guarded formulas such that:
\begin{enumerate}
\item $\bigwedge L \in \ANF$ for every finite set $L$ of literals (in particular $\top, \bot, p, \neg p$);
\item if $\alpha, \beta \in \ANF$ and $\alpha \lor \beta$ is clean, then $\alpha \lor \beta \in \ANF$;
\item if $\Phi \subseteq \ANF$ is finite, $L$ is a finite set of literals, and $\nabla\Phi \land \bigwedge L$ is clean, then $\nabla\Phi \land \bigwedge L \in \ANF$ (for $L=\emptyset$ this is $\nabla\Phi$, by the convention of Section~\ref{sec:syntax});
\item if $\alpha \in \ANF$, $x$ is positive and guarded in $\alpha$, and $x$ occurs in $\alpha$ only as the singleton ANF formula $\bigwedge\{x\}$, rather than as a member of a literal set used in an explicit conjunction (such a singleton may be a disjunct, an element of some $\Phi$, or the body of a binder), then $\eta x.\alpha \in \ANF$.
\end{enumerate}
\end{definition}

Free variables may occur in the literal sets $L$ (so a single variable $x$ is the ANF formula $\bigwedge\{x\}$), and clause~4 admits vacuous binders $\eta y.\alpha$ with $y \notin \Free(\alpha)$.

\begin{lemma}[Structure of ANF tableaux]\label{lem:anf-properties}
Let $\widehat{\varphi} \in \ANF$. Then every choice node of $\mathcal{T}_{\widehat{\varphi}}$ is labelled by a singleton $\{\hat{\gamma}\}$, every consistent modal node is labelled by a set $\{\nabla\Phi\} \cup L$ or by a set $L$ of literals, and each infinite branch of $\mathcal{T}_{\widehat{\varphi}}$ carries exactly one infinite trace. Moreover, $\widehat{\varphi}$ is aconjunctive.
\end{lemma}

\begin{proof}
Every subformula of an ANF formula that is not inside a literal set is again in ANF, with variables bound outside that subformula read as free variables, as clause~1 allows. By induction on Definition~\ref{def:anf-set}, saturating $\{\hat\gamma\}$ with $\hat\gamma\in\ANF$ follows one chain of disjunctions, binders and regenerations to $\bigwedge L$ or $\nabla\Phi\land\bigwedge L$; so modal labels are $L$ or $\{\nabla\Phi\}\cup L$, their successors are the singletons $\{\psi\}$ with $\psi\in\Phi$, and only the non-literal formula of a label has a successor in the next trace relation. Aconjunctivity holds because, by clause~4, no bound variable occurs in a literal set of a conjunction: in $\nabla\Phi\land\bigwedge L$ only the $\nabla\Phi$ conjunct can have an active bound variable, and in $\bigwedge L$ with $|L|\ge2$ neither conjunct can.
\end{proof}

While $\ANF$ is not closed under conjunction, $\nu x$, or substitution directly, applying these operations to ANF formulas preserves \emph{aconjunctivity} (corresponding to Walukiewicz's closure of weakly aconjunctive formulas~\cite{Walukiewicz00} and Enqvist et al.'s closure of semi-disjunctive automata~\cite{EnqvistSeifanVenema18}):

\begin{lemma}[Aconjunctive closure properties of ANF]\label{lem:anf-closure}
Let $\widehat{\alpha}(x), \widehat{\beta}, \widehat{\varphi} \in \ANF$ have pairwise disjoint sets of bound variables, none of which occurs free in the others (and assume $x$ is positive and guarded in $\widehat{\alpha}(x)$). Then $\widehat{\alpha} \land \widehat{\beta}$, $\nu x.\widehat{\alpha}(x)$, and $\widehat{\alpha}(\widehat{\varphi})$ are clean and aconjunctive.
\end{lemma}

\begin{proof}
Cleanness follows from Remark~\ref{rem:sub-set}.
Consider $\widehat\alpha(\widehat\varphi)$ and a $\mu$-variable $w$.
If $w\in\Bound(\widehat\alpha)$, no variable of $\Bound(\widehat\alpha)$ occurs in $\widehat\varphi$, so no dependency chain starting at $w$ enters a variable bound in $\widehat\varphi$.
Conjunctions originating in $\widehat\alpha$ therefore retain their original activity pattern, while $w$ is inactive inside substituted copies.
If $w\in\Bound(\widehat\varphi)$, its defining body lies inside $\widehat\varphi$; an outer conjunction cannot lie in that body, and inner conjunctions satisfy Lemma~\ref{lem:anf-properties}.
Repeated substitutions introduce the same subformula set, not additional defining binders (Remark~\ref{rem:sub-set}).
The cases $\widehat\alpha\land\widehat\beta$ and $\nu x.\widehat\alpha(x)$ are similar and simpler: the new conjunction lies outside all fixpoint bodies, and the new $\nu$-binder imposes no aconjunctivity constraint and cannot depend on an inner $\mu$-variable.
\end{proof}

\subsection{Recognising even branches via B\"uchi determinisation}\label{sec:buchi}

Throughout Sections~\ref{sec:buchi}--\ref{sec:anf-construction}, $\Gamma=\Gamma_0$ is a fixed finite clean guarded set. A branch of $\mathcal{T}_\Gamma$ may carry many traces and is even iff one of them is a $\mu$-trace, whereas a branch of $\mathcal{T}_{\anf(\Gamma)}$ carries one trace. We therefore recognise the even branches of $\mathcal{T}_\Gamma$ by a nondeterministic B\"uchi automaton $\mathsf{NBA}_\Gamma$, which guesses a $\mu$-trace and its $\sqsubseteq$-least infinitely often regenerated variable, and determinise it.

Specifically, every edge $e = (t, u)$ of $\mathcal{T}_\Gamma$ determines a letter $a_e := (L(t), R_e, L(u))$ from a finite alphabet $\Sigma_\Gamma$ (finiteness holds because $L(t), L(u) \subseteq \Cl(\Gamma_0)$ and $R_e \subseteq \Cl(\Gamma_0) \times \Cl(\Gamma_0) \times \mathcal{P}(\Bound(\Gamma_0))$). An infinite branch $\xi$ of $\mathcal{T}_\Gamma$ thus induces an infinite word $w(\xi) \in \Sigma_\Gamma^\omega$.

\begin{definition}[Branch B\"uchi automaton $\mathsf{NBA}_\Gamma$]\label{def:buchi}
The nondeterministic B\"uchi automaton $\mathsf{NBA}_\Gamma$ over $\Sigma_\Gamma$ has states of the form $\gamma$ (tracking a trace before committing to $x$) and $(\gamma, x)$ (tracking a trace committed to the $\mu$-variable $x$), for $\gamma \in \Cl(\Gamma_0)$ and $\mu$-variables $x \in \Bound(\Gamma_0)$, initial states $\gamma \in \Gamma$, and the following transitions on a letter $(\Delta, R, \Delta') \in \Sigma_\Gamma$:
\begin{itemize}
\item from $\gamma$ to $\gamma'$, and from $\gamma$ to $(\gamma', x)$, whenever $(\gamma, \gamma', V) \in R$ for some $V$;
\item from $(\gamma, x)$ to $(\gamma', x)$ whenever $(\gamma, \gamma', V) \in R$ for some $V$ such that $x \sqsubseteq y$ for all $y \in V$; this transition is \emph{accepting} if $x \in V$.
\end{itemize}
\end{definition}

\begin{lemma}\label{lem:buchi-even}
For every infinite branch $\xi$ of $\mathcal{T}_\Gamma$, $\mathsf{NBA}_\Gamma$ accepts $w(\xi)$ if and only if $\xi$ is even.
\end{lemma}

\begin{proof}
If $\xi$ is even, it carries a $\mu$-trace $\tau$ with $\sqsubseteq$-least infinitely often regenerated variable $x$ ($\eta_x = \mu$). Lemma~\ref{lem:outermost} provides a position after which every variable regenerated along $\tau$ is $\sqsubseteq$-above $x$; transitioning from $\gamma_n$ to $(\gamma_{n+1}, x)$ at the first edge after that position yields an accepting run of $\mathsf{NBA}_\Gamma$, since $x \in \Reg^\infty(\tau)$. Conversely, any accepting run of $\mathsf{NBA}_\Gamma$ on $w(\xi)$ eventually stays in states $(\gamma_n, x)$ for a fixed $\mu$-variable $x$; the states of the run, together with witnessing regeneration sets for its transitions, define a trace $\tau$ on $\xi$, along which all variables regenerated after that point are $\sqsubseteq$-above $x$, and $x$ itself is regenerated infinitely often; hence $x$ is the $\sqsubseteq$-least element of $\Reg^\infty(\tau)$ and $\tau$ is a $\mu$-trace.
\end{proof}

\begin{theorem}[B\"uchi determinisation {\cite{Safra88,Piterman07}}]\label{thm:determinisation}
For every nondeterministic B\"uchi word automaton there is an equivalent $\mathsf{DPA}$ (Section~\ref{sec:omega-conventions}).
\end{theorem}

Fix once and for all such a deterministic parity automaton $\mathsf{DPA}_\Gamma = (Q, \Sigma_\Gamma, q_0, \delta, \pri)$ equivalent to $\mathsf{NBA}_\Gamma$. For each node $t$ of $\mathcal{T}_\Gamma$, let $q_t \in Q$ denote the state reached by $\mathsf{DPA}_\Gamma$ after reading the letters on the unique path from the root of $\mathcal{T}_\Gamma$ to $t$. By Lemma~\ref{lem:buchi-even}, an infinite branch $\xi=t_0t_1\cdots$ of $\mathcal{T}_\Gamma$ is even iff $\max\mathrm{Inf}(\pri(q_{t_0})\pri(q_{t_1})\cdots)$ is even. Nothing else about the internal structure of $\mathsf{DPA}_\Gamma$ is used in the sequel.

\subsection{Cutting $\mathcal{T}_\Gamma$ to a finite prefix tree $\mathcal{G}_\Gamma$}\label{sec:folding}

We run $\mathsf{DPA}_\Gamma$ along $\mathcal{T}_\Gamma$ from the root and cut it at \emph{loop nodes}, which repeat kind, label and automaton state of an ancestor with no priority larger than theirs in between. Folding loop nodes onto their return nodes gives the arena of a finite parity game (Section~\ref{sec:finite-refutation-game}), while the resulting finite tree $\mathcal G_\Gamma$ is the skeleton of $\anf(\Gamma)$ (Section~\ref{sec:anf-construction}).

\begin{definition}[Loop nodes, return nodes, and the finite tree $\mathcal{G}_\Gamma$]\label{def:loop-return}
A node $t$ of $\mathcal{T}_\Gamma$ is a \emph{loop node} if it has a proper ancestor $t'$ of the same kind (both choice nodes or both modal nodes) such that
\[
L(t') = L(t), \quad q_{t'} = q_t, \quad\text{and}\quad \pri(q_u) \le \pri(q_t) \text{ for every node } u \text{ on the path from } t' \text{ to } t.
\]
The nearest (deepest) such proper ancestor $t'$ is called the \emph{return node} of $t$ and is denoted by $\ret(t)$. We define $\mathcal{G}_\Gamma$ to be the set of all nodes of $\mathcal{T}_\Gamma$ none of whose \emph{proper} ancestors is a loop node, and we write $\mathcal{G}_\Gamma^\circ$ for the set of nodes of $\mathcal G_\Gamma$ that are not loop nodes, i.e.\ the internal nodes of $\mathcal G_\Gamma$ and the leaves of $\mathcal T_\Gamma$ in $\mathcal G_\Gamma$.
\end{definition}

In~\cite[Lemma~22.1]{Walukiewicz00} back edges occur only at choice nodes, which requires choice states to dominate modal states in priority; loop nodes of both kinds avoid this preprocessing.

\begin{lemma}[Properties of $\mathcal{G}_\Gamma$]\label{lem:return-nodes}
\begin{enumerate}
\item The prefix tree $\mathcal{G}_\Gamma$ is finite.
\item For every node $t$ of $\mathcal{T}_\Gamma$, define $\rho(t) \in \mathcal{G}_\Gamma$ inductively by $\rho(\mathrm{root}) := \mathrm{root}$ and, if $u$ is a child of $t$, $\rho(u)$ is the child of $\rho'(t)$ in $\mathcal{G}_\Gamma$ with the same label and incoming edge trace relation as $u$ (the correspondence of Remark~\ref{rem:label-determinacy}), where $\rho'(t) := \ret(\rho(t))$ if $\rho(t)$ is a loop node and $\rho'(t) := \rho(t)$ otherwise. Then $\rho(t)$ is well-defined, $\rho'(t) \in \mathcal{G}_\Gamma^\circ$, $L(\rho'(t)) = L(\rho(t)) = L(t)$, and $q_{\rho'(t)} = q_{\rho(t)} = q_t$.
\item Let $\xi$ be an infinite branch of $\mathcal{T}_\Gamma$, and let $R_\xi$ be the set of return nodes $r$ with $r = \ret(\rho(t))$ for infinitely many $t \in \xi$. Then $R_\xi$ has a unique element $r$ of least depth, every element of $R_\xi$ lies in the subtree of $r$, and $\xi$ is even if and only if $\pri(q_r)$ is even.
\end{enumerate}
\end{lemma}

\begin{figure}[htbp]
\centering
\includegraphics[width=0.72\textwidth]{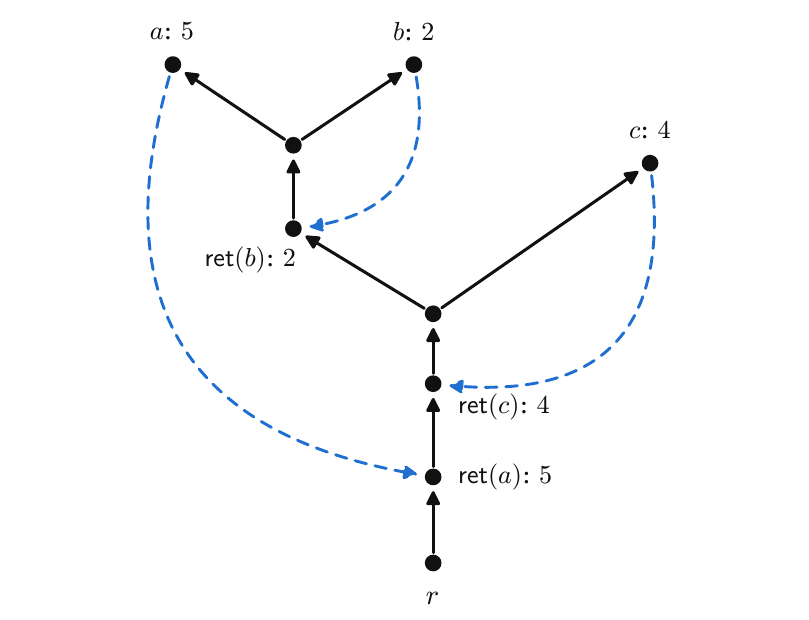}
\caption{Nested returns and their priorities.}
\label{fig:returns}
\end{figure}

Figure~\ref{fig:returns} illustrates part~(3); tree segments stand for paths, dashed arrows for returns, and numbers for $\pri(q_t)$. If the return from $a$ is used infinitely often, $\ret(a)$ is the shallowest recurrent return node and priority $5$ determines parity; if the walk eventually stays in the $b$-loop, priority $2$ does. In the ANF construction of Section~\ref{sec:anf-construction}, a return node closer to the root gives an outer binder.

\begin{proof}
(1) On any infinite branch, beyond the last position whose priority exceeds the maximal recurrent priority $p$, the pigeonhole principle gives two nodes of priority $p$ agreeing in kind, label and state; all priorities between them are at most $p$, so the later one is a loop node. Hence $\mathcal G_\Gamma$ has no infinite branch and is finite by K\"onig's lemma.
(2) Induct on depth: a loop node's return is a non-loop ancestor with the same kind, label and state; label determinacy matches children and edge relations, and determinism matches their states, proving all assertions.
(3) Write $\rho(\xi)$ for the infinite walk $\rho(t_0)\rho(t_1)\cdots$ in $\mathcal{G}_\Gamma$; it moves from a node to a child along tree edges of $\mathcal{G}_\Gamma$, except that after visiting a loop node $\ell$ it jumps to $\ret(\ell)$ and continues to a child of $\ret(\ell)$. By part~(2), $\xi$ and $\rho(\xi)$ visit the same sequence of automaton states and priorities. Since $\mathcal{G}_\Gamma$ is finite, $R_\xi$ is non-empty, and there is $N$ such that after position $N$ every jump of $\rho(\xi)$ goes to an element of $R_\xi$. Let $d$ be the least depth of an element of $R_\xi$ and $r\in R_\xi$ of depth $d$. After the first jump to $r$ beyond $N$, the walk never leaves the subtree of $r$: steps to children stay inside it, and a jump from a loop node $\ell$ in the subtree goes to $\ret(\ell)$, an ancestor of $\ell$; if $\ret(\ell)$ were outside the subtree, it would be a proper ancestor of $r$, an element of $R_\xi$ of depth less than $d$. Hence every element of $R_\xi$ lies in the subtree of $r$, and $r$ is the only one of depth $d$. From then on, $\rho(\xi)$ decomposes into a sequence of finite \emph{cycles} from $r$ back to $r$.

We claim that every node visited on such a cycle satisfies $\pri(q_n)\le\pri(q_r)$. A cycle from $r$ to $r$ is a sequence of tree paths $s_i\to\ell_i$ ($0\le i\le k$) from $s_i$ up to a loop node $\ell_i$, with $s_0=r$, $s_{i+1}=\ret(\ell_i)$ for $i<k$, and $\ret(\ell_k)=r$; all $s_i$ lie in the subtree of $r$. Since every node of the $i$-th path lies on the tree path from $r$ to $\ell_i$, it suffices to show, by induction on $i$ from $k$ down to $0$, that every node on the path from $r$ to $\ell_i$ has priority at most $\pri(q_r)$. For $i=k$ this is the loop condition of Definition~\ref{def:loop-return} for $\ell_k$ with $\ret(\ell_k)=r$. For $i<k$, the path from $r$ to $\ell_i$ is the path from $r$ to $\ret(\ell_i)=s_{i+1}$ followed by the path from $\ret(\ell_i)$ to $\ell_i$. The first part is a prefix of the path from $r$ to $\ell_{i+1}$ (as $s_{i+1}$ is an ancestor of $\ell_{i+1}$), so its priorities are at most $\pri(q_r)$ by the induction hypothesis; the second part has priorities at most $\pri(q_{\ret(\ell_i)})$ by Definition~\ref{def:loop-return}, and $\pri(q_{\ret(\ell_i)})\le\pri(q_r)$ since $\ret(\ell_i)=s_{i+1}$ lies on the first part.

Consequently, from the first jump to $r$ beyond $N$ on, all priorities occurring along $\xi$ are at most $\pri(q_r)$, while $\pri(q_r)$ occurs infinitely often, at positions $t$ for which $\rho(t)$ is a loop node $\ell$ with $\ret(\ell)=r$, since $q_t=q_\ell=q_r$. By Lemma~\ref{lem:buchi-even} and Theorem~\ref{thm:determinisation}, $\xi$ is even iff $\pri(q_r)$ is even.
\end{proof}

\subsection{The folded tableau as a finite parity game}\label{sec:finite-refutation-game}

We now read $\mathcal G_\Gamma$, with its loop nodes folded back, as a finite parity game and prove Theorem~\ref{thm:unsat-refutation}. As in Kirsten's parity games over tiles~\cite{Kirsten02}, the saturated step at a choice node plays the role of a tile; by label determinacy (Remark~\ref{rem:label-determinacy}) no separate notion of tile is needed.

For $g\in\mathcal G_\Gamma$ put $\hat g:=\ret(g)$ if $g$ is a loop node and $\hat g:=g$ otherwise; thus $\rho'(t)=\widehat{\rho(t)}$ in Lemma~\ref{lem:return-nodes}(2). For $t\in\mathcal G_\Gamma^\circ$, all children of $t$ in $\mathcal T_\Gamma$ belong to $\mathcal G_\Gamma$.

\begin{definition}[Folded tableau game]\label{def:finite-tableau-arena}
The parity game $\mathcal P_\Gamma$ is given by
\begin{align*}
V&:=\mathcal G_\Gamma^\circ,\\
V_{\ModelPlayer}&:=\{t\in V\mid t\text{ is a choice node, or a modal node with $L(t)$ inconsistent}\},\\
V_{\RefPlayer}&:=\{t\in V\mid t\text{ is a modal node with $L(t)$ consistent}\},\\
E&:=\{(t,\hat g)\mid t\in V,\ g\in C(t)\},\\
\Omega(t)&:=\pri(q_t),
\end{align*}
with initial position $\mathrm{root}_{\mathcal G_\Gamma}$.
\end{definition}

The dead ends of $\mathcal P_\Gamma$ are exactly the nodes of $\mathcal G_\Gamma^\circ$ that are leaves of $\mathcal T_\Gamma$. An inconsistent modal position is a loss for $\ModelPlayer$; a consistent modal position without children (its label is $\nabla$-free, or all its $\nabla$-formulas are $\nabla\emptyset$) is a loss for $\RefPlayer$. A choice position is never a dead end, since $\Out(\Delta)\neq\emptyset$.

\begin{lemma}[Plays and branches]\label{lem:arena-quotient}
For a branch $\xi=t_0t_1\cdots$ of $\mathcal T_\Gamma$ put $\rho'(\xi):=\rho'(t_0)\rho'(t_1)\cdots$.
\begin{enumerate}
\item $\rho'(\xi)$ is a play of $\mathcal P_\Gamma$ from the initial position, and $\Omega(\rho'(t_i))=\pri(q_{t_i})$ for all $i$.
\item If $v=\rho'(t)$ and $(v,v')\in E$, then $t$ has a child $u$ with $\rho'(u)=v'$. Hence every play from the initial position is $\rho'(\xi)$ for some branch $\xi$.
\item An infinite branch $\xi$ is even iff $\RefPlayer$ wins $\rho'(\xi)$.
\end{enumerate}
\end{lemma}
\begin{proof}
By Lemma~\ref{lem:return-nodes}(2), $u\mapsto\rho(u)$ is a bijection from $C(t)$ onto the children of $\rho'(t)$, and $\rho'(u)=\widehat{\rho(u)}$; hence the moves from $\rho'(t)$ are exactly the pairs $(\rho'(t),\rho'(u))$ with $u\in C(t)$. Moreover $L(\rho'(t))=L(t)$ and $q_{\rho'(t)}=q_t$, so $t$ is a leaf iff $\rho'(t)$ is a dead end. This proves (1) and (2). For (3), $\xi$ is even iff $\mathsf{DPA}_\Gamma$ accepts $w(\xi)$ (Lemma~\ref{lem:buchi-even} and Theorem~\ref{thm:determinisation}), i.e.\ iff $\max\mathrm{Inf}(\pri(q_{t_0})\pri(q_{t_1})\cdots)$ is even, i.e.\ iff $\RefPlayer$ wins $\rho'(\xi)$ by~(1).
\end{proof}

\begin{lemma}[Positional model strategies are finite models]\label{lem:model-run-correspondence}
Let $\sigma:V_{\ModelPlayer}\to V$ be a positional winning strategy of $\ModelPlayer$ from the initial position of $\mathcal P_\Gamma$, and let $W$ be the set of positions occurring in plays from the initial position consistent with $\sigma$. Define $\mathbb S_\sigma=(S_\sigma,R_\sigma,\lambda_\sigma)$ and $s_\sigma$ by
\begin{align*}
S_\sigma&:=\{n\in W\mid n\text{ is a modal node}\},\\
n\,R_\sigma\,n'&:\iff \sigma(u)=n'\text{ for some }u\text{ with }(n,u)\in E,\\
\lambda_\sigma(p)&:=\{n\in S_\sigma\mid p\in L(n)\},\\
s_\sigma&:=\sigma(\mathrm{root}_{\mathcal G_\Gamma}).
\end{align*}
Then $\mathbb S_\sigma$ is a finite Kripke model and $\mathbb S_\sigma,s_\sigma\models\bigwedge\llangle\Gamma\rrangle$.
\end{lemma}
\begin{proof}
No play consistent with $\sigma$ reaches a dead end of $\ModelPlayer$, so every $n\in S_\sigma$ has a consistent label, and every literal of $L(n)$ holds at $n$. If $n\in S_\sigma$ and $(n,u)\in E$, then $u\in W$ and $\sigma(u)\in S_\sigma$, so $R_\sigma\subseteq S_\sigma\times S_\sigma$. By Theorem~\ref{thm:adequacy}, it suffices to give $\ModelPlayer$ a winning strategy in $\mathcal E(\mathcal T_\Gamma,\mathbb S_\sigma,s_\sigma)$. She keeps the invariant that the tableau nodes visited so far project under $\rho'$ to a prefix of a play consistent with $\sigma$, that the world at a choice position $(t,w)$ is $w=\sigma(\rho'(t))$, and that the world at a modal position $(m,w)$ is $w=\rho'(m)$.
\begin{itemize}
\item At $(t,w)$: by Lemma~\ref{lem:arena-quotient}(2) there is $m\in C(t)$ with $\rho'(m)=\sigma(\rho'(t))=w$, and she moves to $(m,w)$.
\item At $(m,w)$ with $n:=w=\rho'(m)$: $L(m)=L(n)$, so all its literals hold at $w$; if $L(m)$ is $\nabla$-free, $\RefPlayer$ cannot move. Otherwise:
\begin{itemize}
\item at $\langle m,w,u\rangle_{\Diamond}$ she moves to $(u,v)$ with $v:=\sigma(\rho'(u))$, which satisfies $nR_\sigma v$ because $(n,\rho'(u))\in E$;
\item at $\langle m,w,v\rangle_{\Box}$, where $nR_\sigma v$, she takes $u'$ with $(n,u')\in E$ and $\sigma(u')=v$, and moves to $(u,v)$ for a child $u\in C(m)$ with $\rho'(u)=u'$, whose existence follows from Lemma~\ref{lem:arena-quotient}(2).
\end{itemize}
In both cases the play continues at $(u,v)$ with $v=\sigma(\rho'(u))$.
\end{itemize}
She is never stuck. For an infinite play with tableau branch $\xi$, the invariant makes $\rho'(\xi)$ a play consistent with $\sigma$, which $\RefPlayer$ loses; so $\xi$ is odd by Lemma~\ref{lem:arena-quotient}(3). Hence the strategy is winning.
\end{proof}

The construction is that of~\cite{Kirsten02}.

\begin{lemma}[Positional refuters and regular refutations]\label{lem:refuter-refutation-correspondence}
If $\RefPlayer$ has a positional winning strategy $\tau$ from the initial position of $\mathcal P_\Gamma$, then $\Gamma$ has a regular refutation in which every thinning is full.
\end{lemma}
\begin{proof}
For each position $n\in V_{\RefPlayer}$ that is not a dead end, fix a child $g_n$ of $n$ with $\hat g_n=\tau(n)$. Let $U$ be the subtree of $\mathcal T_\Gamma$ obtained by keeping all children of choice nodes and, at a consistent modal node $m$ with $C(m)\neq\emptyset$, only the child $u$ with $\rho(u)=g_{\rho'(m)}$. For every branch $\xi$ of $U$, $\rho'(\xi)$ is a play consistent with $\tau$, hence won by $\RefPlayer$. Consequently every leaf of $U$ is an inconsistent modal node (a consistent leaf would be a dead end of $\RefPlayer$), and every infinite branch of $U$ is even by Lemma~\ref{lem:arena-quotient}(3).

Turn $U$ into a refutation $\mathcal R$ by giving a choice node labelled $\Delta$, for every choice function $c$, a copy of the subtree of $U$ at its child with outcome $(M_c(\Delta),R_c(\Delta))$. Then every thinning is $K_c=S_c(\Delta)$, each consistent modal node is $\nabla$-non-free and has exactly one child, a modal successor; every leaf of $\mathcal R$ is inconsistent, and every infinite branch projects to an even branch of $U$; so $\mathcal R$ satisfies Definition~\ref{def:refutations}.

For regularity, give each copied node of $\mathcal R$ the value of $\rho'$ at its original node in $U$.
Nodes with the same value have the same label.
At choice nodes, match children indexed by the same choice function; at consistent modal nodes, the retained children correspond because $g_n$ was fixed for each game position $n$.
By Lemma~\ref{lem:return-nodes}(2), corresponding children have the same labels, edge trace relations and values of $\rho'$.
Recursively matching these children gives a label- and relation-preserving isomorphism of the rooted subtrees.
Hence $\mathcal R$ has at most $|\mathcal G_\Gamma^\circ|$ rooted labelled subtrees up to isomorphism.
\end{proof}

\begin{theorem}[Finite-game existence of regular refutations]\label{thm:finite-game-refutations}
If $\Gamma$ is unsatisfiable, then $\Gamma$ has a regular refutation with full thinnings. Hence Theorem~\ref{thm:unsat-refutation} holds.
\end{theorem}
\begin{remark}[Size of the extracted witnesses]\label{rem:finite-game-size}
The positional strategies in the finite game yield witnesses bounded by the number of positions of the explicitly constructed prefix game, in particular by a quantity proportional to $|\mathcal G_\Gamma^\circ|$ with the indicated local choices. This is not asserted to be a single-exponential bound in the input formula: an exponentially deep, exponentially branching prefix tree may already be doubly exponential. Obtaining a smaller quotient game requires a separate proof that identifying positions by their labels, node types and monitor states preserves the relevant strategy and trace information. No such quotient or improved bound is claimed here.
\end{remark}

\begin{proof}
By Theorem~\ref{thm:positional-determinacy}, one player has a positional winning strategy from the initial position of $\mathcal P_\Gamma$. If it were $\ModelPlayer$, Lemma~\ref{lem:model-run-correspondence} would make $\bigwedge\llangle\Gamma\rrangle$ satisfiable. Hence it is $\RefPlayer$, and Lemma~\ref{lem:refuter-refutation-correspondence} yields the refutation.
\end{proof}

\subsection{Reading off $\anf(\Gamma)$}\label{sec:anf-construction}

\begin{definition}[Construction of $\anf(\Gamma)$]\label{def:anf-construction}
Let $\Gamma\neq\emptyset$; then every label of $\mathcal T_\Gamma$ is non-empty, since saturating a non-empty set reaches a modal formula and modal successors are non-empty. Assign a fresh bound variable $x_t$ to every non-loop node $t \in \mathcal{G}_\Gamma^\circ$, and a fresh \emph{dummy} bound variable $y_\ell$ to every loop node $\ell \in \mathcal{G}_\Gamma \setminus \mathcal{G}_\Gamma^\circ$. For each $t \in \mathcal{G}_\Gamma^\circ$, let $\eta_t := \mu$ if $\pri(q_t)$ is even, and $\eta_t := \nu$ if $\pri(q_t)$ is odd. We define a formula $\anf(t)$ for every node $t \in \mathcal{G}_\Gamma$ by induction from the leaves of the finite tree $\mathcal{G}_\Gamma$ towards its root:
\begin{itemize}
\item if $t$ is a loop node: $\anf(t) := \nu y_t.\, x_{\ret(t)}$;
\item if $t \in \mathcal{G}_\Gamma^\circ$ is a modal node: $\anf^-(t) := \bot$ if $L(t)$ is inconsistent; $\anf^-(t) := \bigwedge L(t)$ if $L(t)$ is consistent and $\nabla$-free; and otherwise
\[
\anf^-(t) := \nabla\{\anf(u) \mid u \text{ is a child of } t \text{ in } \mathcal{G}_\Gamma\} \land \bigwedge\Lit(L(t));
\]
\item if $t \in \mathcal{G}_\Gamma^\circ$ is a choice node: $\anf^-(t) := \bigvee\{\anf(m) \mid m \text{ is a child of } t \text{ in } \mathcal{G}_\Gamma\}$;
\item for every non-loop node $t \in \mathcal{G}_\Gamma^\circ$: $\anf(t) := \eta_t x_t.\,\anf^-(t)$.
\end{itemize}
Finally, we set $\anf(\Gamma) := \anf(\mathrm{root}_{\mathcal{G}_\Gamma})$ and $\anf(\varphi) := \anf(\{\varphi\})$.
\end{definition}

\begin{figure}[t]
\centering
\includegraphics[width=\textwidth]{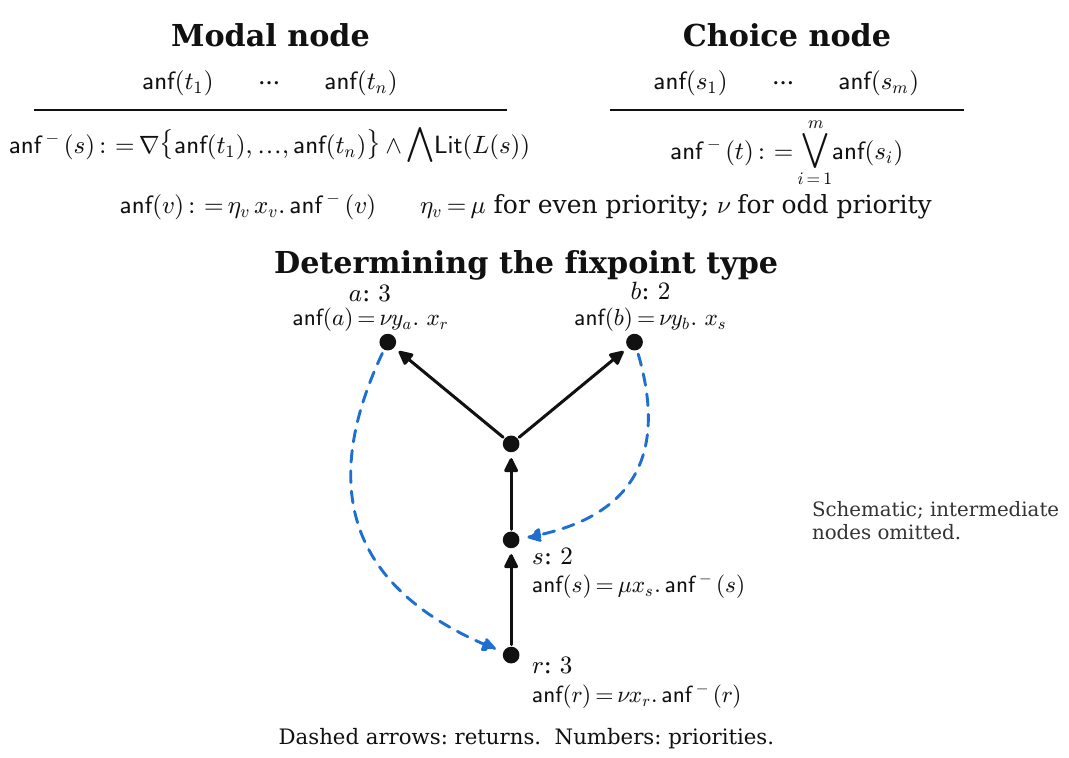}
\caption{Reading $\anf(\Gamma)$ from $\mathcal G_\Gamma$: modal and choice clauses, and determination of the fixpoint type.}
\label{fig:anf-reading}
\end{figure}

Figure~\ref{fig:anf-reading} illustrates the modal and choice clauses and the assignment of fixpoint types; intermediate nodes are omitted in the schematic return tree.

\begin{definition}[Modal-component label]\label{def:modal-component-label}
For a consistent modal node $n \in \mathcal{G}_\Gamma^\circ$, let $\Lab(n)$ denote the set of modal components obtained by saturating $\anf^-(n)$: namely, $\Lab(n):=L(n)$ if $L(n)$ is $\nabla$-free, and
\[
\Lab(n):=\{\nabla\{\anf(u)\mid u\in C_{\mathcal G_\Gamma}(n)\}\}\cup\Lit(L(n))
\]
if $L(n)$ is $\nabla$-non-free.
\end{definition}

\begin{remark}[Syntactic injectivity and the role of vacuous binders]\label{rem:node-injectivity}
In Definition~\ref{def:anf-construction} the binders $\nu y_\ell$, and the binders $\eta_tx_t$ of non-return nodes $t$, are vacuous; they are only unfolded, never regenerated. Since every node $v\in\mathcal G_\Gamma$ carries its own outermost binder, $v\mapsto\anf(v)$ is injective on $\mathcal G_\Gamma$. Since distinct nodes of a tree have disjoint sets of children, injectivity of $u\mapsto\anf(u)$ implies that distinct modal nodes with non-empty child sets have distinct sets $\{\anf(u)\mid u\in C_{\mathcal G_\Gamma}(n)\}$. Hence, for a consistent modal node $n\in\mathcal G_\Gamma^\circ$ with children, the $\nabla$-formula of $\Lab(n)$ (Definition~\ref{def:modal-component-label}) determines $n$; this injectivity plays the role of~\cite[Observation~22.2]{Walukiewicz00}.
\end{remark}

\begin{theorem}[Automaton normal form; Claim (b)]\label{thm:anf}
For every non-empty finite clean guarded set $\Gamma$, $\anf(\Gamma) \in \ANF$, $\Free(\anf(\Gamma)) \subseteq \Free(\Gamma)$, every free variable that is positive (guarded) in $\Gamma$ is positive (guarded) in $\anf(\Gamma)$, and there is a bisimulation map $\zeta_{\anf} : \mathcal{T}_\Gamma \to \mathcal{T}_{\anf(\Gamma)}$. Consequently, $\bigwedge\llangle\Gamma\rrangle \equiv \anf(\Gamma)$.
\end{theorem}

\begin{proof}
\emph{Syntactic shape.} Fresh, pairwise distinct binders ensure cleanness: each $x_r$ occurs only at loop nodes returning to $r$, inside its subtree.
A return-to-loop path joins distinct nodes of the same kind and crosses a modal node, so each $x_r$ is guarded.
Its occurrences are singleton literal sets under vacuous binders $\nu y_\ell$, themselves disjuncts or $\nabla$-elements; thus Definition~\ref{def:anf-set}(4) holds and $\anf(\Gamma)\in\ANF$.
Free variables come only from modal literal sets, giving $\Free(\anf(\Gamma))\subseteq\Free(\Gamma)$; decomposition preserves polarity, and any path to a guarded free literal crosses a modal edge, placing its ANF occurrence under $\nabla$.
\paragraph{The bisimulation map $\zeta_{\anf}$.} For a modal node $n \in \mathcal{G}_\Gamma^\circ$, let $\Lab(n)$ be as in Definition~\ref{def:modal-component-label} (with $\Lab(n) := \{\bot\}$ if $L(n)$ is inconsistent). By Lemma~\ref{lem:anf-properties}, every choice node of $\mathcal{T}_{\anf(\Gamma)}$ is labelled by a single formula. Using the projection $\rho$ of Lemma~\ref{lem:return-nodes}(2), we construct $\zeta_{\anf}$ inductively from the root, maintaining
\[
L(\zeta_{\anf}(t))=\{\anf(\rho(t))\}
\ \ \text{for choice nodes }t,
\qquad
L(\zeta_{\anf}(m))=\Lab(\rho'(m))
\ \ \text{for modal nodes }m.
\]
The root maps to the root.

At a choice node $t$ of $\mathcal{T}_\Gamma$, saturating the singleton $\{\anf(\rho(t))\}$ in $\mathcal{T}_{\anf(\Gamma)}$ first unfolds the dummy binder $\nu y_{\rho(t)}$ and regenerates $x_{\ret(\rho(t))} \to \anf^-(\rho'(t))$ (if $\rho(t)$ is a loop node), or unfolds $\eta_{\rho(t)} x_{\rho(t)} \to \anf^-(\rho'(t))$ (if $\rho(t) \in \mathcal{G}_\Gamma^\circ$); it then selects one disjunct $\anf(\tilde{m})$ for a modal child $\tilde{m} \in C_{\mathcal{G}_\Gamma}(\rho'(t))$, unfolds or regenerates $\anf(\tilde{m})$ to $\anf^-(\rho'(\tilde{m}))$, and splits it into the modal set $\Lab(\rho'(\tilde{m}))$. Define $\zeta_{\anf}(m)$ to be the child of $\zeta_{\anf}(t)$ determined by this saturation outcome, including its trace relation. By Lemma~\ref{lem:return-nodes}(2), the children $m \in C(t)$ in $\mathcal{T}_\Gamma$ correspond bijectively to the children $\rho(m) = \tilde{m} \in C_{\mathcal{G}_\Gamma}(\rho'(t))$. Since all choice functions are exhausted, this gives a surjection from $C(t)$ onto $C(\zeta_{\anf}(t))$; it need not be injective, because distinct children of $\rho'(t)$ may generate the same $(M_c,R_c)$ and are then merged into a single modal child of $\zeta_{\anf}(t)$.

At a consistent $\nabla$-non-free modal node $m$ of $\mathcal{T}_\Gamma$ with $n := \rho'(m)$, the choice children of $\zeta_{\anf}(m)$ (labelled $\Lab(n)$) are the singletons $\{\anf(\tilde{u})\}$ for $\tilde{u} \in C_{\mathcal{G}_\Gamma}(n)$ (which are pairwise distinct by Remark~\ref{rem:node-injectivity}), and the choice children $u \in C(m)$ in $\mathcal{T}_\Gamma$ correspond bijectively to $\rho(u) \in C_{\mathcal{G}_\Gamma}(n)$. Mapping $u$ to the child of $\zeta_{\anf}(m)$ labelled $\{\anf(\rho(u))\}$ gives a bijection from $C(m)$ onto $C(\zeta_{\anf}(m))$. Consistency, and at consistent modal nodes literals and $\nabla$-freeness, are preserved in both directions: $L(m)$ is inconsistent iff $L(\zeta_{\anf}(m))=\{\bot\}$, and for consistent $L(m)$ the literals and $\nabla$-freeness of $\Lab(\rho'(m))$ are those of $L(\rho'(m))=L(m)$ (Definition~\ref{def:modal-component-label}).

\emph{Two-way parity condition (R-par).} Let $\xi$ be an infinite branch of $\mathcal{T}_\Gamma$. By Lemma~\ref{lem:anf-properties}, the image branch $\zeta_{\anf}(\xi)$ in $\mathcal{T}_{\anf(\Gamma)}$ carries a unique infinite trace $\hat{\tau}$. The variables regenerated infinitely often along $\hat\tau$ are exactly
\[
\Reg^\infty(\hat\tau)=\{x_r\mid r\in R_\xi\},
\]
where $R_\xi$ is as in Lemma~\ref{lem:return-nodes}(3). Indeed, each visit to a loop node $\ell$ regenerates $x_{\ret(\ell)}$ in the corresponding saturation, and these are the only regenerations: non-return binders and dummy binders are only unfolded.
Let $r \in \mathcal{G}_\Gamma^\circ$ be the unique shallowest return node visited infinitely often along $\rho(\xi)$, as given by Lemma~\ref{lem:return-nodes}(3). Then $x_r \in \Reg^\infty(\hat{\tau})$, and every other variable $x_{r'} \in \Reg^\infty(\hat{\tau})$ belongs to a return node $r'$ in the subtree of $r$, so $\anf(r')$ is a subformula of $\anf(r)$ and hence $x_r \sqsubseteq x_{r'}$. Thus $x_r$ is the $\sqsubseteq$-least element of $\Reg^\infty(\hat{\tau})$. Therefore, $\zeta_{\anf}(\xi)$ is even iff $\eta_r = \mu$ iff $\pri(q_r)$ is even (by Definition~\ref{def:anf-construction}) iff $\xi$ is even (by Lemma~\ref{lem:return-nodes}(3)). Applying Proposition~\ref{prop:run-transport} and Theorem~\ref{thm:adequacy} with $\Gamma_0=\Gamma$ and with $\Gamma_0=\{\anf(\Gamma)\}$ completes the proof.
\end{proof}

\subsection{Box isolation}\label{sec:box-isolation}

Positive-context lifting (Theorem~\ref{thm:positive-context-closure}) and the unfolding lemma (Lemma~\ref{lem:pcc-unfold}) need one more syntactic normalisation. It makes the box parts $\bigvee\Psi$ produced by
modal steps independent of every other formula.

\begin{definition}[Box chains and box-isolated formulas]\label{def:box-isolated}
For a finite set $\Psi=\{\theta_1,\dots,\theta_s\}$ with $s\ge2$, listed in the fixed global order, the \emph{box chain} of $\nabla\Psi$ is
\[
\left\{\bigvee\{\theta_i,\dots,\theta_s\}\;\middle|\;1\le i<s\right\};
\]
its \emph{head} is $\bigvee\Psi$. For $|\Psi|\le1$ the box chain is empty.
A clean formula $\chi$ in $\nabla$-form is \emph{box-isolated} if, for every $\nabla\Psi\in\Sub(\chi)$:
\begin{enumerate}
\item every $\theta\in\Psi$ is a fixpoint formula $\eta_z z.\varphi_z$;
\item every occurrence in $\chi$ of any $\theta\in\Psi$ is an occurrence as an element of $\nabla\Psi$.
\end{enumerate}
\end{definition}

In particular, distinct $\nabla$-subformulas of a box-isolated formula
have disjoint sets of elements.

\begin{lemma}[Box isolation]\label{lem:box-isolation}
\begin{enumerate}
\item For every $\widehat\alpha(x)\in\ANF$ with $x$ positive and guarded
there is a box-isolated $\widehat\alpha^\dagger(x)\in\ANF$ with the
same free variables, in which $x$ is positive and guarded, such
that $\vdash\widehat\alpha(x)\leftrightarrow\widehat\alpha^\dagger(x)$.
\item For every non-empty finite clean guarded set $\Gamma$, the formula
$\anf(\Gamma)$ is box-isolated.
\item If $\widehat\alpha(x)$ and $\widehat\beta$ are box-isolated, $x$ is positive in $\widehat\alpha(x)$, they have
disjoint sets of bound variables, and no bound variable of either
occurs free in the other, then $\mu x.\widehat\alpha(x)$ and
$\widehat\alpha(\widehat\beta)$ are box-isolated.
\item Let $\chi$ be box-isolated, take $\Gamma_0:=\{\chi\}$, and let
$\nabla\Psi\in\Cl(\chi)$. Then:
(a) no disjunction of the box chain of $\nabla\Psi$ belongs to
$\Sub(\chi)$;
(b) in $E:=\bigcup_cE_c$, every predecessor of a disjunction of the
box chain lies in the same box chain, and the head $\bigvee\Psi$ has
no predecessor;
(c) no edge inside a box chain is a regeneration.
Consequently, if $\Delta\subseteq\Cl(\chi)$ contains no disjunction
of the box chain of $\nabla\Psi$, then for every choice function
$c$ the sets $S_c(\Delta)$ contain no such disjunction, and
$S_c(\Delta)$, $M_c(\Delta)$, and $R_c(\Delta)$ do not depend on
the values of $c$ on that box chain.
\end{enumerate}
\end{lemma}

\begin{proof}
(1) Rename bound variables apart and, from inner to outer $\nabla$-occurrences, replace each element $\theta$ by $\nu z.\theta$ with a fresh $z$; Lemma~\ref{lem:standard}(2),(3),(6) preserves provable equivalence, and vacuous binders preserve $\ANF$, free variables, positivity and guardedness while making each element unique to its $\nabla$.
(2) Each element $\anf(u)$ has its own binder and is used only in its parent's $\nabla$ (Definition~\ref{def:anf-construction}).
(3) Adding $\mu x$ leaves the $\nabla$-subformulas unchanged, and separated substitution preserves their uniquely identifying binders, with elements of $\widehat\beta$ confined to copies of $\widehat\beta$.
(4) The closure $\Cl(\chi)$ consists of $\Sub(\chi)$ and the box chains of
its $\nabla$-subformulas, because the subformulas of $\bigvee\Psi$ are the
disjunctions of its box chain and subformulas of elements.
(a) A disjunction $D$ of the box chain has an element $\theta$ as an
immediate disjunct. If $D\in\Sub(\chi)$, then $\theta$ would occur in
$\chi$ other than as an element of $\nabla\Psi$.
(b) An $E$-edge into $D$ starts at a formula with $D$ as an immediate
subformula, or at a variable $y$ with $\varphi_y=D$. By~(a), neither lies
in $\Sub(\chi)$, so it is a disjunction $D'$ of some box chain of
$\nabla\Psi'$ having $D$ as its tail. Then the elements of $D$ are
elements of $\Psi'$, so $\nabla\Psi'=\nabla\Psi$. The head is the
largest disjunction of its chain, so it has no predecessor.
(c) All edges inside a box chain are disjunction edges.
By~(b), no $E$-edge enters the box chain from outside. Since $\Delta$ is disjoint from the chain, no $E_c$-path from $\Delta$ reaches it. Thus saturation never consults $c$ on the chain, proving the final assertion.
\end{proof}

\section{From tableau consequence to thin refutations}\label{sec:transfer}

The following macro-step is the saturation calculation used at diamond nodes in the transfer construction and later in Section~\ref{sec:pcc}.

\begin{lemma}[Diamond macro-step]\label{lem:diamond-macro-step}
Let $\tilde m$ be a consistent $\nabla$-non-free modal node of $\mathcal T_\alpha$ with $L(\tilde m)=\{\nabla\Psi_1,\dots,\nabla\Psi_i\}\cup\Lit(L(\tilde m))$, put $B:=\{\bigvee\Psi_1,\dots,\bigvee\Psi_i\}$, and let $c$ be a choice function. For every $n$ with $\Psi_n\neq\emptyset$, $c$ leads along the box chain of $\bigvee\Psi_n$ to an element $\theta_n\in\Psi_n$, defined as the first member of $\Psi_n$ encountered on the $c$-selected path (including its starting formula, if applicable). This is a path-dependent choice, not an assertion that the syntactic expression uniquely identifies $n$. Fix $k$ with $\Psi_k\neq\emptyset$, let $\tilde u$ be the $(k,\theta_k)$-child of $\tilde m$ and $\tilde m'$ the modal child of $\tilde u$ with outcome $c$. Then
\[
M_c(B)=L(\tilde m')
\qquad\text{and}\qquad
\{(\nabla\Psi_n,\varpi,V)\mid(\textstyle\bigvee\Psi_n,\varpi,V)\in R_c(B)\}
= R(\tilde m,\tilde u)\circ R(\tilde u,\tilde m'),
\]
where $\circ$ is the composition of trace relations (with union of regeneration sets). If $\Psi_n=\emptyset$ for some $n$, then $\bot\in M_c(B)$.
\end{lemma}

\begin{proof}
For each $n$ with $\Psi_n=\emptyset$, put $\theta_n:=\bot$. Then, for every $n$, the $c$-selected path from $\bigvee\Psi_n$ to $\theta_n$ consists only of disjunction edges, possibly none. Thus $S_c(\bigvee\Psi_n)$ consists of a segment of the chain and of $S_c(\theta_n)$, and a triple $(\bigvee\Psi_n,\varpi,V)$ lies in $R_c(B)$ iff $(\theta_n,\varpi,V)\in R_c(\{\theta_n\})$. The same choice function $c$ is used for every $\bigvee\Psi_n$ on the single graph $E_c$. Thus any shared subformula is resolved identically, and reachability distributes over unions: $S_c(\bigcup_n\Gamma_n)=\bigcup_nS_c(\Gamma_n)$; the analogous decomposition holds for modal outcomes and for trace triples grouped by their starting formula. No disjointness assumption on the $\Psi_n$ is needed. Hence $M_c(B)=\bigcup_nM_c(\theta_n)=M_c(\{\theta_k\}\cup\{\bigvee\Psi_n\mid n\neq k\})=M_c(L(\tilde u))=L(\tilde m')$. The canonical composite consists of the triples $(\nabla\Psi_k,\varpi,V)$ with $(\theta_k,\varpi,V)\in R_c(L(\tilde u))$ and $(\nabla\Psi_n,\varpi,V)$ with $(\bigvee\Psi_n,\varpi,V)\in R_c(L(\tilde u))$ for $n\neq k$, which is the displayed set. Finally $\bigvee\emptyset=\bot$ is modal.
\end{proof}
Thus saturating all box parts at once, as the refutation will do at a diamond node, produces the canonical two-step successor $\tilde m'$ of $\tilde m$ together with its traces; the $(k,\theta_k)$-child $\tilde u$ is recovered from $c$ and need not be a node of the refutation.

Let $Z:\mathcal T_\alpha\Rightarrow\mathcal T_{\widehat\varphi}$.
We follow both tableaux upwards from their roots along $Z$ and record the walk as a two-sided refutation of $\alpha\vdash\widehat\varphi$ (Definition~\ref{def:two-sided}). At each pair of choice nodes, we retain the full left saturation and use $Z$ to select a matching right saturation outcome:
\[
L(\tilde t)\vdash L(t)
\ \xrightarrow{\text{decompose left}}\
L(\tilde m)\vdash L(t)
\ \xrightarrow{Z\text{, decompose right}}\
L(\tilde m)\vdash L(m).
\]
The right side is represented by tagged dual formulas $(R,\sim\hat\gamma)$; selecting a right disjunct therefore amounts to thinning its dual conjunction. In two-sided notation this is right weakening $(\mathrm{Weak}R)$ read upwards. Throughout, $\Gamma_0:=\{\alpha\}$ and $\Delta_0:=\{\widehat\varphi\}$.

\begin{figure}[htbp]
\centering
\includegraphics[width=\textwidth]{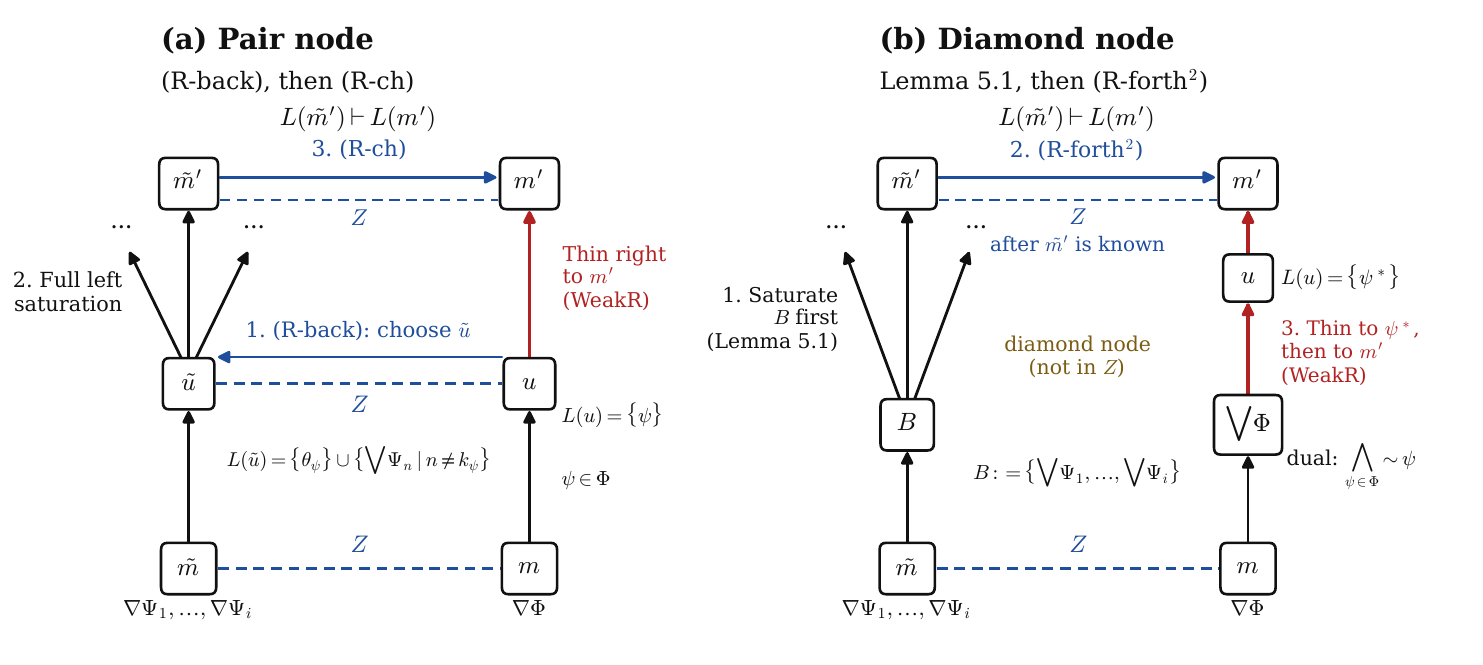}
\caption{The two continuations of $(\mathrm{Cover})$, read upwards. Black arrows retain the full left development; blue arrows select witnesses in $Z$; red arrows depict right thinning (upward $(\mathrm{Weak}R)$). In (b), the right witnesses are chosen only after the left saturation outcome is known. The right-hand formulas are displayed before dualisation; in the tagged refutation their duals are thinned.}
\label{fig:transfer}
\end{figure}

At a consistent modal pair $(\tilde m,m)\in Z$, write the cover formulas on the left as $\nabla\Psi_1,\dots,\nabla\Psi_i$ and that on the right as $\nabla\Phi$. The right literal alternatives close immediately by (R-loc). For the remaining cover part, recall the two-sided rule of Lemma~\ref{lem:standard}(4), here written in the notation of this section:
\[
\frac{
\overbrace{\bigl\{\,\theta_\psi,\{\bigvee\Psi_n\mid n\neq k_\psi\}\vdash\psi
\;\bigm|\;\psi\in\Phi\,\bigr\}}^{\text{\footnotesize (a) (R-back), then (R-ch)}}
\qquad
\overbrace{\bigvee\Psi_1,\dots,\bigvee\Psi_i\vdash\bigvee\Phi}^{\text{\footnotesize (b) (R-forth$^2$)}}
}{
\nabla\Psi_1,\dots,\nabla\Psi_i\vdash\nabla\Phi
}\ (\mathrm{Cover}),
\]
where $1\le k_\psi\le i$ and $\theta_\psi\in\Psi_{k_\psi}$.
Read upwards, its two kinds of premises explain the two continuations in Figure~\ref{fig:transfer}. In the one-sided tagged presentation, the disjunction defining $\sim\nabla\Phi$ (Definition~\ref{def:negation}) distributes these obligations among the modal children; inconsistent alternatives close immediately.

\emph{Pair nodes, Figure~\ref{fig:transfer}(a).}
For each $\psi\in\Phi$, the right child $u$ has label $\{\psi\}$. Clause (R-back) selects a left child $\tilde u$ with $(\tilde u,u)\in Z$, giving a premise in the left family of $(\mathrm{Cover})$. We then saturate the left side in full. For each resulting modal child $\tilde m'$, clause (R-ch) selects a matching $m'\in C(u)$, and the right dual decomposition is thinned to follow that outcome.

\emph{Diamond nodes, Figure~\ref{fig:transfer}(b).}
For the box premise, put $B:=\{\bigvee\Psi_1,\dots,\bigvee\Psi_i\}$.
First saturate $B$ in full. Lemma~\ref{lem:diamond-macro-step} identifies each consistent outcome with a modal node $\tilde m'$ two tableau steps above $\tilde m$.
Only then does (R-forth$^2$) select a right child $u$, with $L(u)=\{\psi^*\}$ for some $\psi^*\in\Phi$, and a modal child $m'\in C(u)$ with $(\tilde m',m')\in Z$.
The right side is thinned first to $\psi^*$ and then along the outcome $m'$. The intermediate diamond node is not a pair in $Z$: the correspondence is restored only at the modal level. The construction below also handles inconsistent outcomes and the empty-cover cases.

\begin{theorem}[Transfer from consequence relations to thin refutations; Claim (h)]\label{thm:transfer}
Let $\alpha$ be a clean guarded aconjunctive formula and $\widehat\varphi\in\ANF$, such that no bound variable of either formula occurs in the other. If $Z:\mathcal T_\alpha\Rightarrow\mathcal T_{\widehat\varphi}$ is a consequence relation, then $\alpha\vdash\widehat\varphi$ has a thin sequential two-sided refutation.
\end{theorem}

\begin{proof}
\emph{Conventions.} Fix Skolem witnesses for the three existential clauses of Definition~\ref{def:consequence-relation}: $\mathsf{ch}(\tilde t,t,\tilde m)\in C(t)$ for (R-ch), $\mathsf{bk}(\tilde m,m,u)\in C(\tilde m)$ for (R-back), and $\mathsf{fo}(\tilde m,m,\tilde u,\tilde m')=(u,m')$ for (R-forth$^2$). For every right choice node $t$ with $L(t)=\{\hat\gamma\}$ and every $m\in C(t)$, fix a choice function $c_m$ whose saturation outcome is $m$. Formulas tagged $L$ form the \emph{left side} of a label, formulas tagged $R$ the \emph{right side}; a choice function $c$ splits into its restrictions $c_L$ and $c_R$ to the two sides, and since no $E$-edge joins the two sides, $S_c(\Theta)$, $M_c(\Theta)$ and $R_c(\Theta)$ are the unions of those of the two sides, computed with $c_L$ and $c_R$. We fix a topological order of $E$ in which all left formulas precede all right formulas; it exists for the same reason.

By Lemma~\ref{lem:anf-properties}, every choice node of $\mathcal T_{\widehat\varphi}$ has a singleton label $\{\hat\gamma\}$ and every consistent $\nabla$-non-free modal node a label $\{\nabla\Phi\}\cup L$. In $\hat\gamma\in\ANF$ the only conjunctions outside the scope of $\nabla$ are of the form $\bigwedge L$ and $\nabla\Phi\land\bigwedge L$ (Definition~\ref{def:anf-set}), and decomposing these conjunctions yields only modal formulas. Hence the only conjunctions of $\sim\hat\gamma$ outside the scope of $\nabla$ are the duals $\sim D$ of the disjunctions $D$ of $\hat\gamma$, and saturating $\sim\hat\gamma$ along one conjunct of each of them reaches either a literal or $\sim(\nabla\Phi\land\bigwedge L)$, a disjunction whose disjuncts are the literals $\sim l$ ($l\in L$) and the disjuncts $\nabla\emptyset$, $\nabla\{\sim\psi\}$ ($\psi\in\Phi$) and $\nabla\{\bigwedge_{\psi\in\Phi}\sim\psi,\top\}$ of $\sim\nabla\Phi$ (Definition~\ref{def:negation}). The conjunction $\bigwedge_{\psi\in\Phi}\sim\psi$ lies under $\nabla$ and is not decomposed by saturation. Thus, once each $\sim D$ is thinned to one conjunct, the right side of the saturation is a single path, and resolving its disjunctions by $c_R$ yields exactly one right modal formula.

\emph{Shape of the refutation.} We build $\mathcal R$ bottom-up. Apart from auxiliary choice nodes whose modal children are all inconsistent leaves, each choice node of $\mathcal R$ is a \emph{pair node} or a \emph{diamond node}. A pair node is attached to a choice pair $(\tilde t,t)\in Z$ with $L(t)=\{\hat\gamma\}$ and labelled $(\{L\}\times L(\tilde t))\cup\{(R,\sim\hat\gamma)\}$, i.e.\ it represents the sequent $L(\tilde t)\vdash\hat\gamma$. Each modal node that is not an inconsistent leaf is attached to a modal pair $(\tilde m,m)\in Z$ and labelled $(\{L\}\times L(\tilde m))\cup\{(R,\delta)\}$ for one modal formula $\delta$. The root is the pair node of the root pair. Below we omit the tags when the side is clear.

\emph{Pair node for $(\tilde t,t)$.} For a choice function $c$, the left side is saturated canonically, giving the modal child $\tilde m$ of $\tilde t$ with outcome $c_L$; put $m:=\mathsf{ch}(\tilde t,t,\tilde m)$. On the right side, every conjunction $\sim D$ is thinned to the conjunct $\sim c_m(D)$; the remaining right disjunctions are resolved by $c_R$, yielding one modal formula $\delta$. The modal child for $c$ is labelled $L(\tilde m)\cup\{\delta\}$ and attached to $(\tilde m,m)\in Z$.

\emph{Modal node attached to $(\tilde m,m)$ with label $L(\tilde m)\cup\{\delta\}$.} If $L(\tilde m)$ is inconsistent, the node is a leaf. Otherwise $L(m)$ is consistent by (R-loc), and $\delta$ is one of the following.
\begin{itemize}
\item $\delta=\sim l$ for a literal $l\in L(m)$, or $\delta=\bot$. By (R-loc), $l\in L(\tilde m)$, so the label is inconsistent (consistency is computed on underlying formulas) and the node is a leaf. This covers all $\delta$ when $L(m)$ is $\nabla$-free.
\item $L(m)=\{\nabla\Phi\}\cup L$ is $\nabla$-non-free; then $L(\tilde m)=\{\nabla\Psi_1,\dots,\nabla\Psi_i\}\cup\Lit(L(\tilde m))$ is $\nabla$-non-free by (R-loc).
\begin{itemize}
\item $\delta=\nabla\emptyset$ ($\Phi\neq\emptyset$). Since $C(m)\neq\emptyset$, (R-back) gives $C(\tilde m)\neq\emptyset$, so some $\Psi_k\neq\emptyset$; the node has the $(k,\theta)$-successor for some $\theta\in\Psi_k$. Its label contains the box part $\bigvee\emptyset=\bot$ of $\delta$, so all its modal children are inconsistent leaves.
\item $\delta=\nabla\{\top\}$, a disjunct of $\sim\nabla\emptyset=\bot\lor\nabla\{\top\}$ ($\Phi=\emptyset$, so $\bigwedge_{\psi\in\Phi}\sim\psi=\top$ and the last disjunct of Definition~\ref{def:negation} is $\nabla\{\top,\top\}=\nabla\{\top\}$; the disjunct $\bot$ gives an inconsistent leaf). The node has the successor with principal formula $\delta$ and element $\top$, labelled $\{\top\}\cup B$ with $B$ as in Lemma~\ref{lem:diamond-macro-step}. For every $c$, its modal child contains $L(\tilde m')$ for a two-step successor $\tilde m'$ of $\tilde m$ (Lemma~\ref{lem:diamond-macro-step}, or $\bot$ if some $\Psi_n=\emptyset$), and every such $\tilde m'$ is inconsistent because (R-forth$^2$) would otherwise require a child of $m$, which has none. All modal children are leaves.
\item $\delta=\nabla\{\sim\psi\}$ for some $\psi\in\Phi$. Let $u\in C(m)$ be the child with $L(u)=\{\psi\}$ and $\tilde u:=\mathsf{bk}(\tilde m,m,u)$, say the $(k,\theta)$-child of $\tilde m$. The node has the $(k,\theta)$-successor, labelled $L(\tilde u)\cup\{\sim\psi\}$, where $\sim\psi=\bigvee\{\sim\psi\}$ is the box part of $\delta$: this is the pair node for $(\tilde u,u)\in Z$.
\item $\delta=\nabla\{\bigwedge_{\psi\in\Phi}\sim\psi,\top\}$ ($\Phi\neq\emptyset$). The node has the successor with principal formula $\delta$ and element $\bigwedge_{\psi\in\Phi}\sim\psi$, labelled $D:=B\cup\{\bigwedge_{\psi\in\Phi}\sim\psi\}$: the \emph{diamond node} attached to $(\tilde m,m)$.
\end{itemize}
\end{itemize}

\emph{Diamond node $D$ attached to $(\tilde m,m)$.} As above some $\Psi_k\neq\emptyset$; fix such a $k$, and fix once and for all some $\psi_0\in\Phi$, the child $u_0\in C(m)$ with $L(u_0)=\{\psi_0\}$, and some $m'_0\in C(u_0)$ (it exists since $\Out(\{\psi_0\})\neq\emptyset$). For a choice function $c$, Lemma~\ref{lem:diamond-macro-step} applied to $c_L$ yields the two-step successor $\tilde m'$ of $\tilde m$ (through the $(k,\theta_k)$-child $\tilde u$) with $M_{c_L}(B)=L(\tilde m')$.
\begin{itemize}
\item If $L(\tilde m')$ is consistent, let $(u,m'):=\mathsf{fo}(\tilde m,m,\tilde u,\tilde m')$, so $L(u)=\{\psi\}$ for some $\psi\in\Phi$ and $(\tilde m',m')\in Z$, and put $(\psi^*,m^*):=(\psi,m')$.
\item If $L(\tilde m')$ is inconsistent, put $(\psi^*,m^*):=(\psi_0,m'_0)$.
\end{itemize}
Thin the right conjunction $\bigwedge_{\psi\in\Phi}\sim\psi$ to $\sim\psi^*$, thin every conjunction $\sim D'$ of $\sim\psi^*$ to $\sim c_{m^*}(D')$, and resolve the remaining right disjunctions by $c_R$; as at a pair node this yields one right modal formula $\delta'$. The modal child of $D$ for $c$ is labelled $L(\tilde m')\cup\{\delta'\}$; in the first case it is attached to $(\tilde m',m')$, in the second it is an inconsistent leaf.

\emph{Local conditions.} Every modal node of $\mathcal R$ is inconsistent or has exactly one successor, every choice node has a child for every $c$, and every leaf is inconsistent; so $\mathcal R$ satisfies Definition~\ref{def:refutations}(1)--(2) and the leaf condition of (3).

\emph{Sequentiality.} The only thinnings are of right conjunctions, each to a single conjunct; left formulas are never thinned, so every left formula of $S_c(\Delta)$ belongs to $K_c$. At a pair node the conjunct is determined by $m$, hence by $\tilde m$, i.e.\ by $c_L$; at a diamond node it is determined by $c_L$ through $\tilde m'$ and the fixed witnesses. A right formula belongs to $K_c$ iff it lies on the single right path described in the conventions, which is determined by these conjuncts and by the values of $c_R$ at the right disjunctions of the path that precede it. All these disjunctions belong to $K_c$, and the left ones precede every right formula in the topological order; hence $\mathcal R$ is sequential with respect to this order.

\emph{Thinness.} Every right conjunction occurring in a thinning is thinned to one conjunct, including those inside the fixed conjunct $\sim\psi_0$ at an inconsistent diamond step, so no right conjunction keeps both conjuncts. (This matters: $\sim\widehat\varphi$ need not be aconjunctive.) A left conjunction lies in $\Sub(\alpha)$, and activity is computed in $\alpha$ (Definition~\ref{def:two-sided}); aconjunctivity of $\alpha$ excludes a $\mu$-variable active in both conjuncts. So $\mathcal R$ is thin.

\emph{Infinite branches.} The modal nodes along a branch of $\mathcal R$ are attached to modal pairs forming a $Z$-path: a pair node followed by a modal node is an (R-ch) witness, a modal node with $\delta=\nabla\{\sim\psi\}$ followed by its successor is a step by (R-back) and (R-ch), and the passage through a diamond node is a step by (R-forth$^2$). Let $\mathcal B$ be an infinite branch, with left projection $\tilde\xi$ and right projection $\xi$. The left side of $\mathcal B$ carries exactly the traces of $\tilde\xi$ with the same regeneration sets: at pair nodes and single-successor steps the left saturation is canonical; at a diamond node, the modal step with the right principal formula $\delta$ sends every $\nabla\Psi_n\in L(\tilde m)$ to $\bigvee\Psi_n\in B$ without regeneration, and Lemma~\ref{lem:diamond-macro-step} identifies the composite with the canonical trace relation from $\tilde m$ to $\tilde m'$. The right side of $\mathcal B$ carries exactly one infinite trace, the dual of the unique infinite trace $\hat\tau$ of $\xi$ (Lemma~\ref{lem:anf-properties}): the extra steps through thinned conjunctions and through the elements $\bigwedge_\psi\sim\psi$ regenerate nothing, and the regenerated variables are those of $\hat\tau$, with $\mu$ and $\nu$ interchanged. If $\tilde\xi$ is even, its $\mu$-trace is a $\mu$-trace of $\mathcal B$. Otherwise (R-par) forces $\xi$ to be odd, so the $\sqsubseteq$-least infinitely regenerated variable of $\hat\tau$ is a $\nu$-variable of $\widehat\varphi$, i.e.\ a $\mu$-variable of $\sim\widehat\varphi$, and the dual trace is a $\mu$-trace of $\mathcal B$. Hence every infinite branch is even.
\end{proof}

\begin{corollary}\label{cor:consequence-provable}
Under the hypotheses of Theorem~\ref{thm:transfer}, $\vdash \alpha \to \widehat{\varphi}$.
\end{corollary}

\begin{proof}
By Theorems~\ref{thm:transfer} and~\ref{thm:thin-refutation}, $\{(L,\alpha),(R,\sim\widehat\varphi)\}\vdash$, i.e.\ $\alpha\vdash\widehat\varphi$.
\end{proof}

\section{Positive-context lifting and completeness}\label{sec:completeness}
\subsection{Positive-context lifting and the least-fixpoint step}\label{sec:pcc-fixpoint-corollary}

\begin{theorem}[Positive-context lifting of tableau consequence]
\label{thm:positive-context-closure}
Let $A(X)$, $B$, and $C$ be formulas in $\nabla$-form satisfying the following
conditions:
\begin{enumerate}
\item $A(X)$, $B$, and $C$ are clean, guarded, and box-isolated
(Definitions~\ref{def:clean-guarded} and~\ref{def:box-isolated}).
\item The variable $X$ occurs only positively in $A(X)$ (i.e.\ the literal
$\neg X$ does not occur in $A(X)$) and does not occur in $B$ or $C$.
\item The sets $\Bound(A(X))$, $\Bound(B)$, and $\Bound(C)$ are pairwise
disjoint, and no bound variable of any one of the three formulas occurs
(free or bound) in either of the other two.
\item Both $B$ and $C$ have outermost fixpoint binders: $B=\eta_{z_B}z_B.\,B_0$ and $C=\eta_{z_C}z_C.\,C_0$ for some $\eta_{z_B},\eta_{z_C}\in\{\mu,\nu\}$, where $z_B\in\Bound(B)$ and $z_C\in\Bound(C)$ (so neither occurs in $A(X)$ by~(3)).
\item There exists a tableau consequence relation $W : \mathcal{T}_B \Rightarrow \mathcal{T}_C$
in the sense of Definition~\ref{def:consequence-relation}.
\end{enumerate}
Then $A(B)$ and $A(C)$ are clean, guarded, and box-isolated, and there exists a tableau consequence relation
\[
Z : \mathcal{T}_{A(B)} \Rightarrow \mathcal{T}_{A(C)}.
\]
Moreover, $Z$ may be chosen so that every modal pair in $Z$ lies on a finite
$Z$-path from the roots.
\end{theorem}
\begin{remark}[Scope of the lifting theorem]\label{rem:pcc-hypotheses}
Box isolation makes the box-chain supports independent (Lemma~\ref{lem:pcc-context-support}), and the outermost binders of Hypothesis~(4) make both substitutions injective (Lemma~\ref{lem:pcc-outer-inj}). These are hypotheses of the present proof, not claimed to be necessary; closure is asserted only for contexts satisfying all of them.
\end{remark}

\begin{lemma}[Unfolding consequence for a renamed fixpoint]
\label{lem:pcc-unfold}
Let $\alpha(x)$ be clean, guarded, and box-isolated, with $x \in \Free(\alpha)$
positive and guarded, and set $\psi := \mu x.\alpha(x)$. Let
$\psi' := \mu x'.\alpha'(x')$ be obtained from $\psi$ by renaming every bound
variable $z \in \Bound(\psi)$ to a fresh variable $z'$ not occurring in $\psi$.
Then $\alpha(\psi') := \alpha(x)[\psi'/x]$ is clean, guarded, and box-isolated, and
there exists a tableau consequence relation
\[
U : \mathcal{T}_{\alpha(\psi')} \Rightarrow \mathcal{T}_\psi
\]
every modal pair of which lies on a $U$-path.
\end{lemma}

The proofs of Theorem~\ref{thm:positive-context-closure} and Lemma~\ref{lem:pcc-unfold} are given in Section~\ref{sec:pcc}; they use no result of the present section and, from Section~\ref{sec:transfer}, only Lemma~\ref{lem:diamond-macro-step}, so there is no circularity.

\begin{proposition}[Fixpoint step; Claim~(g), Walukiewicz {\cite[Lemma~39]{Walukiewicz00}}]
\label{prop:fixpoint-step}
Let $\widehat\alpha(x) \in \ANF$ be box-isolated, with $x \in \Free(\widehat\alpha)$
positive and guarded, and set $\psi := \mu x.\widehat\alpha(x)$ and
$\widehat\varphi := \anf(\psi)$, with the bound variables of $\widehat\varphi$
distinct from those of $\widehat\alpha$. Then
$\mathcal{T}_{\widehat\alpha(\widehat\varphi)} \Rightarrow \mathcal{T}_{\widehat\varphi}$,
and consequently $\vdash \widehat\alpha(\widehat\varphi) \to \widehat\varphi$.
\end{proposition}

\begin{proof}
Let $\psi' := \mu x'.\widehat\alpha'(x')$ be obtained from $\psi$ by renaming every
bound variable $z \in \Bound(\psi)$ to a fresh variable $z'$ occurring nowhere in
$\widehat\alpha(x)$ or $\widehat\varphi$, and let $\iota : \mathcal{T}_\psi \to \mathcal{T}_{\psi'}$
be the canonical syntactic isomorphism induced by $z \mapsto z'$ (which is in
particular a bisimulation map). By Theorem~\ref{thm:anf}, $\widehat\varphi = \anf(\psi) \in \ANF$,
$\Free(\widehat\varphi) \subseteq \Free(\psi) = \Free(\widehat\alpha) \setminus \{x\}$,
and there is a bisimulation map $\zeta_{\anf} : \mathcal{T}_\psi \to \mathcal{T}_{\widehat\varphi}$.
By Lemma~\ref{lem:pcc-inv-bisim} and Lemma~\ref{lem:relation-map-composition},
\[
W := \Graph(\iota) \circ \Graph(\zeta_{\anf})^{-1} : \mathcal{T}_{\widehat\varphi} \Rightarrow \mathcal{T}_{\psi'}
\]
is a tableau consequence relation.

We now instantiate Theorem~\ref{thm:positive-context-closure} with the substitution
\[
\begin{aligned}
A(X)&:=\widehat\alpha(X)\quad(\text{with fresh }X\text{ replacing free }x),\\
B&:=\widehat\varphi,\qquad C:=\psi'=\mu x'.\widehat\alpha'(x').
\end{aligned}
\]
Hypotheses~(1)--(5) hold: (1) follows from the assumptions, Theorem~\ref{thm:anf} and Lemma~\ref{lem:box-isolation}(2),(3); positivity and freshness give~(2); the fresh bound variables, cleanness of $\widehat\alpha$ and $\Free(\widehat\varphi)\cup\Free(\psi')\subseteq\Free(\widehat\alpha)\setminus\{x\}$ give~(3); Definition~\ref{def:anf-construction} and the definition of $\psi'$ give the outermost binders in~(4); and $W$ witnesses~(5).
Theorem~\ref{thm:positive-context-closure} therefore yields a consequence relation
$Z_1 : \mathcal{T}_{\widehat\alpha(\widehat\varphi)} \Rightarrow \mathcal{T}_{\widehat\alpha(\psi')}$
every modal pair of which lies on a finite $Z_1$-path from the roots. Next, Lemma~\ref{lem:pcc-unfold} gives a
consequence relation $U : \mathcal{T}_{\widehat\alpha(\psi')} \Rightarrow \mathcal{T}_\psi$
every modal pair of which lies on a $U$-path, and $\Graph(\zeta_{\anf}) : \mathcal{T}_\psi \Rightarrow \mathcal{T}_{\widehat\varphi}$
likewise has every modal pair on a path. Two applications of
Lemma~\ref{lem:relation-map-composition} (composition of consequence relations)
yield
\[
Z := \Graph(\zeta_{\anf}) \circ U \circ Z_1 :
\mathcal{T}_{\widehat\alpha(\widehat\varphi)} \Rightarrow \mathcal{T}_{\widehat\varphi}.
\]
Finally, $\widehat\alpha(\widehat\varphi)$ is aconjunctive by Lemma~\ref{lem:anf-closure};
renaming the bound variables of $\widehat\varphi$ apart from those of
$\widehat\alpha(\widehat\varphi)$ on the right-hand side, applying
Corollary~\ref{cor:consequence-provable}, and cutting with Lemma~\ref{lem:standard}(6)
gives
$\vdash \widehat\alpha(\widehat\varphi) \to \widehat\varphi$.
\end{proof}

Applying Theorem~\ref{thm:positive-context-closure} in both directions gives context-stable ANF replacement.
\begin{corollary}[Context-stable normal form]
\label{cor:pcc-anf}
Let $A(X)$ and $\varphi$ be clean, guarded, box-isolated formulas in $\nabla$-form
such that:
\begin{enumerate}[(i)]
\item $X$ occurs only positively in $A(X)$ and does not occur in $\varphi$;
\item $\Bound(A(X))\cap\Bound(\varphi)=\emptyset$, and no bound variable of
      either formula occurs in the other;
\item $\varphi$ has an outermost fixpoint binder (in particular, this holds for
      every fixpoint formula $\varphi=\eta x.\,\varphi_0$).
\end{enumerate}
Put $\widehat\varphi := \anf(\varphi)$, with its fresh bound variables chosen
outside $A(X)$ and $\varphi$. Then $A(\varphi)$ and $A(\widehat\varphi)$ are
clean, guarded, and box-isolated, and there exist tableau consequence relations
\[
  \mathcal T_{A(\varphi)} \Rightarrow \mathcal T_{A(\widehat\varphi)}
  \qquad\text{and}\qquad
  \mathcal T_{A(\widehat\varphi)} \Rightarrow \mathcal T_{A(\varphi)},
\]
each having every modal pair on a finite path from the roots.
\end{corollary}

\begin{proof}
Applying Theorem~\ref{thm:anf} to $\Gamma=\{\varphi\}$ gives
$\widehat\varphi\in\ANF$, $\Free(\widehat\varphi)\subseteq\Free(\varphi)$, and a
bisimulation map $\zeta_{\anf}:\mathcal T_\varphi\to\mathcal T_{\widehat\varphi}$.
The formula $\widehat\varphi$ is clean and guarded
(Definition~\ref{def:anf-set}), box-isolated (Lemma~\ref{lem:box-isolation}(2)),
and has an outermost fixpoint binder $\eta_r x_r$ for the root $r$
(Definition~\ref{def:anf-construction}). We check Hypotheses~(1)--(5) of
Theorem~\ref{thm:positive-context-closure} for both triples
$(A(X),\varphi,\widehat\varphi)$ and $(A(X),\widehat\varphi,\varphi)$:
(1)~holds by assumption and the properties of $\widehat\varphi$ just noted;
(2)~follows from~(i), $\Free(\widehat\varphi)\subseteq\Free(\varphi)$, and the
freshness of $\Bound(\widehat\varphi)$;
(3)~follows from~(ii), the freshness of $\Bound(\widehat\varphi)$, and
$\Free(\widehat\varphi)\subseteq\Free(\varphi)$ together with cleanness of
$\varphi$;
(4)~holds by~(iii) and the outermost binder of $\widehat\varphi$; and
(5)~is witnessed by $\Graph(\zeta_{\anf}):\mathcal T_\varphi\Rightarrow\mathcal T_{\widehat\varphi}$
(Definition~\ref{def:tableau-map}) and
$\Graph(\zeta_{\anf})^{-1}:\mathcal T_{\widehat\varphi}\Rightarrow\mathcal T_\varphi$
(Lemma~\ref{lem:pcc-inv-bisim}). Two applications of
Theorem~\ref{thm:positive-context-closure} give all claims, including syntactic
preservation and the finite-path property.
\end{proof}

\begin{remark}\label{rem:pcc-anf}
\begin{enumerate}[(a)]
\item If $\varphi$ has no outermost fixpoint binder, apply Theorem~\ref{thm:positive-context-closure} to $\varphi^\circ:=\nu z.\varphi$ for a fresh $z$ ($\vdash\varphi^\circ\leftrightarrow\varphi$ by Lemma~\ref{lem:standard}(2)), using the root-unfolding bisimulation $\kappa:\mathcal T_{\varphi^\circ}\to\mathcal T_\varphi$ and $\zeta_{\anf}\circ\kappa$.
      The binder cannot in general be omitted: for $A(X)=(X\lor p)\land((q\lor r)\lor p)$ and $\varphi=q\lor r$, the substituted occurrence merges with an outer subformula, so $\mathcal T_{A(\varphi^\circ)}$ and $\mathcal T_{A(\varphi)}$ need not be bisimilar.
\item Corollary~\ref{cor:pcc-anf} is the tableau-theoretic analogue of the
      two-directional statement in~\cite[Theorem~4]{EnqvistSeifanVenema18}, with
      $\anf(\varphi)$ playing the role of $\operatorname{sim}(A)$; this is an
      analogy rather than a formal implication. Here the context $A(X)$ need be
      neither in ANF nor aconjunctive, and in the forward relation the inner
      left-hand formula $\varphi$ need not be in ANF.
\item Provability in $\Koz$ is not asserted here: Theorem~\ref{thm:transfer}
      requires an aconjunctive left-hand formula and an ANF right-hand
      formula, so $\vdash A(\varphi)\to A(\widehat\varphi)$ does not follow for
      an arbitrary positive context $A(X)$.
\end{enumerate}
\end{remark}

\subsection{The main inductive lemma and completeness}

\begin{lemma}[Main inductive lemma; Claim (d), Walukiewicz {\cite[Theorem~38]{Walukiewicz00}}]\label{lem:main-lemma}
For every clean guarded formula $\varphi$, there is a formula $\widehat{\varphi} \in \ANF$ such that $\varphi \equiv \widehat{\varphi}$, $\Free(\widehat{\varphi}) \subseteq \Free(\varphi)$, every free variable that is positive (guarded) in $\varphi$ is positive (guarded) in $\widehat{\varphi}$, and $\vdash \varphi \to \widehat{\varphi}$.
\end{lemma}

\begin{proof}
By induction on the structure of $\varphi$ (in the primitive syntax with $\Diamond, \Box$, translating $\nabla$-abbreviations via Lemma~\ref{lem:standard}(5)). In each case the induction hypothesis gives $\widehat\alpha,\widehat\beta\in\ANF$ with $\vdash\alpha\to\widehat\alpha$ and $\vdash\beta\to\widehat\beta$; bound variables are renamed apart where needed, which preserves these derivabilities by Lemma~\ref{lem:standard}(6) and $(\mathrm{Cut})$.
\begin{itemize}
\item \emph{Literals:} $\varphi \in \{\top, \bot, p, \neg p\}$ is already in $\ANF$ ($\bigwedge\{\varphi\}$ or $\bigwedge\emptyset$), so take $\widehat{\varphi} := \varphi$.
\item \emph{Disjunction $\varphi = \alpha \lor \beta$:} $\widehat{\varphi} := \widehat{\alpha} \lor \widehat{\beta} \in \ANF$ by Definition~\ref{def:anf-set}(2), and monotonicity (Lemma~\ref{lem:standard}(3)) gives $\vdash \alpha \lor \beta \to \widehat{\alpha} \lor \widehat{\beta}$.
\item \emph{Modalities $\varphi = \Diamond\alpha$ and $\varphi = \Box\alpha$:} By Definition~\ref{def:anf-set}, $\nabla\{\widehat{\alpha}, \top\} \in \ANF$ and $\nabla\emptyset \lor \nabla\{\widehat{\alpha}\} \in \ANF$. Monotonicity (Lemma~\ref{lem:standard}(3), whose modal cases are rule $(\Diamond)$) together with the equivalences of Lemma~\ref{lem:standard}(5) gives $\vdash \Diamond\alpha \to \Diamond\widehat{\alpha}$ and $\vdash \Diamond\widehat{\alpha} \leftrightarrow \nabla\{\widehat{\alpha}, \top\}$, hence $\vdash \Diamond\alpha \to \nabla\{\widehat{\alpha}, \top\}$ by $(\mathrm{Cut})$; likewise $\vdash \Box\alpha \to \Box\widehat{\alpha}$ and $\vdash \Box\widehat{\alpha} \leftrightarrow \nabla\emptyset \lor \nabla\{\widehat{\alpha}\}$ give $\vdash \Box\alpha \to \nabla\emptyset \lor \nabla\{\widehat{\alpha}\}$.
\item \emph{Conjunction $\varphi = \alpha \land \beta$ and greatest fixpoint $\varphi = \nu x.\alpha(x)$:} Let $\psi := \widehat{\alpha} \land \widehat{\beta}$, respectively $\psi := \nu x.\widehat{\alpha}(x)$; then $\vdash\varphi\to\psi$ by monotonicity (Lemma~\ref{lem:standard}(3)). By Lemma~\ref{lem:anf-closure}, $\psi$ is clean and aconjunctive, and it is guarded since the induction hypothesis preserves guardedness. Set $\widehat{\varphi} := \anf(\psi) \in \ANF$; its bound variables are fresh. Theorem~\ref{thm:anf} provides a bisimulation map (and hence a consequence relation) $\mathcal{T}_\psi \to \mathcal{T}_{\widehat{\varphi}}$, so Corollary~\ref{cor:consequence-provable} (Claims~(h) and (f)) yields $\vdash \psi \to \widehat{\varphi}$, whence $\vdash \varphi \to \widehat{\varphi}$ by $(\mathrm{Cut})$.
\item \emph{Least fixpoint $\varphi = \mu x.\alpha(x)$:} By Lemma~\ref{lem:box-isolation}(1), we may replace $\widehat\alpha(x)$ by a provably equivalent box-isolated ANF formula, preserving the induction hypothesis by $(\mathrm{Cut})$. Thus $\widehat\alpha(x)$ is box-isolated, with $x$ positive and guarded. Put $\psi := \mu x.\widehat{\alpha}(x)$. By the induction hypothesis and monotonicity, $\vdash \mu x.\alpha(x) \to \psi$, and the induction hypothesis also gives $\varphi\equiv\psi$. If $x \notin \Free(\widehat{\alpha})$, take $\widehat{\varphi} := \widehat{\alpha}$. Since $x$ is absent from $\widehat\alpha$, unfolding (Lemma~\ref{lem:standard}(2)) gives $\vdash\psi\leftrightarrow\widehat\alpha$, so $\vdash\varphi\to\widehat\varphi$ and $\varphi\equiv\widehat\varphi$. Otherwise put $\widehat{\varphi} := \anf(\psi) \in \ANF$; its bound variables are fresh, hence distinct from those of $\widehat{\alpha}$. Proposition~\ref{prop:fixpoint-step} gives $\vdash \widehat{\alpha}(\widehat{\varphi}) \to \widehat{\varphi}$, and Park induction yields $\vdash \psi \to \widehat{\varphi}$; hence $\vdash\varphi\to\widehat\varphi$ by $(\mathrm{Cut})$. In this case $\psi\equiv\widehat\varphi$ by Theorem~\ref{thm:anf}, so again $\varphi\equiv\widehat\varphi$.
\end{itemize}
In every case, the required conditions on free variables, positivity and guardedness follow from the induction hypothesis and the preservation properties of the constructions used above (in particular, Theorem~\ref{thm:anf} and Lemma~\ref{lem:box-isolation}(1)).
\end{proof}
\begin{theorem}[Completeness of $\Koz$; Claim (a), Walukiewicz {\cite[Theorem~37]{Walukiewicz00}}]\label{thm:completeness}
If a formula $\varphi$ of the modal $\mu$-calculus is unsatisfiable, then $\vdash \neg\varphi$. Consequently, $\models\varphi$ implies $\vdash\varphi$, and $\alpha\models\beta$ implies $\vdash\alpha\to\beta$.
\end{theorem}

\begin{proof}
Let $\varphi$ be unsatisfiable. By Lemma~\ref{lem:standard}(5), $\varphi$ is provably equivalent in $\Koz$ to a clean guarded formula $\varphi'$, which is also unsatisfiable. By Lemma~\ref{lem:main-lemma} (Claim~(d)), there is $\widehat{\varphi} \in \ANF$ such that $\varphi' \equiv \widehat{\varphi}$ and $\vdash \varphi' \to \widehat{\varphi}$. Since $\varphi'$ is unsatisfiable, so is $\widehat{\varphi}$. Taking $\Gamma_0 := \{\widehat{\varphi}\}$, Theorem~\ref{thm:unsat-refutation} (Claim~(e)) gives a sequential refutation of $\Gamma := \{\widehat{\varphi}\}$ in which all thinnings are full. By Lemma~\ref{lem:anf-properties}, $\widehat{\varphi}$ is aconjunctive, so by Lemma~\ref{lem:aconj-thin} this refutation is thin. By Theorem~\ref{thm:thin-refutation} (Claim~(f)), we obtain $\{\widehat{\varphi}\} \vdash$, i.e.\ $\vdash \neg\widehat{\varphi}$ (Claim~(c)). Two applications of $(\mathrm{Cut})$ to $\vdash \varphi \to \varphi'$, $\vdash \varphi' \to \widehat{\varphi}$ and $\widehat{\varphi}\vdash$ yield $\varphi\vdash$, i.e.\ $\vdash \neg\varphi$. The consequences follow by applying this to $\sim\varphi$ and to $\alpha\land\sim\beta$ (Section~\ref{sec:koz}).
\end{proof}

\section{Proofs of positive-context lifting and unfolding consequence}
\label{sec:pcc}

This section proves Theorem~\ref{thm:positive-context-closure} and Lemma~\ref{lem:pcc-unfold}, postponed from Section~\ref{sec:completeness}.
After syntactic preliminaries, we construct the older-wins relay, verify trace preservation, and specialise the mechanism to two lanes for unfolding.
\subsection{Full successor labels and syntactic preliminaries}
\label{sec:pcc-prelims}

By Definition~\ref{def:modal-successors}, the principal formula $\nabla\Psi_k$
of a modal step omits its own box part $\bigvee\Psi_k$ from the successor label.
Because the ambient target tableau may treat $\nabla\Psi_k$ as non-principal
where a copy treated it as principal (or vice versa), we adjoin the omitted box
part. For a consistent $\nabla$-non-free modal node $n$ of a canonical tableau
with $L(n)=\{\nabla\Psi_1,\dots,\nabla\Psi_i\}\cup\Lit(L(n))$ and its
$(k,\theta)$-child $u$, put
\[
L^+(u):=L(u)\cup\{\textstyle\bigvee\Psi_k\},
\]
and for the root $r$ put $L^+(r):=L(r)$. For a choice function $c$ realising an
edge $(u,m')$, let $c^+$ agree with $c$ except on the box chain of
$\nabla\Psi_k$, along which $c^+$ leads to $\theta$ (with $c^+:=c$ at the root).

\begin{lemma}[Full successor labels]\label{lem:full-successor}
Let $\Gamma_0=\{\chi\}$ with $\chi$ box-isolated, and let $n$, $u$, $m'\in C(u)$,
$c$ and $c^+$ be as above. Then:
\begin{enumerate}
\item $S_{c^+}(L^+(u))=S_c(L(u))\cup\mathcal C$, where $\mathcal C$ is the set
of disjunctions of the box chain of $\nabla\Psi_k$ on the path from
$\bigvee\Psi_k$ to $\theta$; in particular $M_{c^+}(L^+(u))=L(m')$;
\item $R(n,u)\circ R(u,m')=\{(\nabla\Psi_j,\varpi,V)\mid 1\le j\le i,\ (\bigvee\Psi_j,\varpi,V)\in R_{c^+}(L^+(u))\}$.
\end{enumerate}
\end{lemma}

\begin{proof}
If $|\Psi_k|=1$, then $\bigvee\Psi_k=\theta$, $L^+(u)=L(u)$ and $\mathcal C=\emptyset$.
Otherwise $L(u)=\{\theta\}\cup\{\bigvee\Psi_j\mid j\neq k\}$ contains no
disjunction of the box chain of $\nabla\Psi_k$: $\theta$ is a fixpoint formula,
and distinct $\nabla$-subformulas of a box-isolated formula have disjoint
element sets and hence disjoint box chains. By Lemma~\ref{lem:box-isolation}(4),
no $E_c$-path from $L(u)$ meets this box chain, so $E_{c^+}$-paths from $L(u)$
coincide with $E_c$-paths, while from $\bigvee\Psi_k$ the $E_{c^+}$-path runs
along $\mathcal C$ to $\theta$ without regeneration and continues as an
$E_c$-path from $\theta$. Since $\mathcal C$ contains no modal formula, (1) and
(2) follow from Definitions~\ref{def:saturation} and~\ref{def:modal-successors}.
\end{proof}

Throughout the rest of this section, we write $r_{\mathcal U}$ for the root of a canonical tableau $\mathcal U$, and for each choice-to-modal edge $(u,m')$ we fix a choice function $c_{(u,m')}$ realising $(u,m')$.

\begin{lemma}[Syntactic preservation]\label{lem:pcc-syntax}
Under Hypotheses~(1)--(4) of Theorem~\ref{thm:positive-context-closure}, both
$A(B)$ and $A(C)$ are clean, guarded, and box-isolated formulas in $\nabla$-form.
\end{lemma}

\begin{proof}
Let $D\in\{B,C\}$; positivity of $X$ ensures that $A(D)$ satisfies Definition~\ref{def:formulas}.
Hypothesis~(3) gives both $\Free(D)\cap\Bound(A(X))=\emptyset$ and $\Free(A(X))\cap\Bound(D)=\emptyset$; together with the disjoint bound-variable sets and Hypothesis~(1), this gives cleanness and preserves guardedness by the usual substitution argument (Remark~\ref{rem:sub-set}).
For box isolation, every $\nabla$-subformula of $A(D)$ is either $\nabla\Xi\in\Sub(D)$
or $\nabla(\Psi[D/X])$ for some $\nabla\Psi\in\Sub(A(X))$. For $\nabla\Xi\in\Sub(D)$,
each $\xi\in\Xi$ has the form $\eta_ww.\varphi_w$ with $w\in\Bound(D)$; since
$w$ does not occur in $A(X)$, $\xi$ cannot have the form $\gamma[D/X]$ for
$\gamma\in\Sub(A(X))\setminus\{X\}$, so every occurrence of $\xi$ in $A(D)$ lies
inside $D$, where it is an element of $\nabla\Xi$. For $\nabla\Psi\in\Sub(A(X))$,
each $\theta\in\Psi$ is $\eta_zz.\varphi_z$ with $z\in\Bound(A(X))$ (in particular
$\theta\neq X$). Since $z$ does not occur in $D$, $\theta[D/X]=\eta_zz.(\varphi_z[D/X])$
cannot occur inside $D$; any occurrence of $\theta[D/X]$ in $A(D)$ therefore
arises from some $\gamma\in\Sub(A(X))\setminus\{X\}$ with $\gamma[D/X]=\theta[D/X]$,
which by comparing outermost binders and cleanness of $A(X)$ forces $\gamma=\theta$,
an element of $\nabla\Psi$. Hence $A(D)$ is box-isolated.
\end{proof}

For $\gamma\in\Sub(A(X))$, set $\gamma^\flat:=\gamma[B/X]\in\Sub(A(B))$ and
$\gamma^\sharp:=\gamma[C/X]\in\Sub(A(C))$, so $X^\flat=B$ and $X^\sharp=C$. For
$\nabla\Psi\in\Sub(A(X))$, set $\Psi^\flat:=\{\theta^\flat\mid\theta\in\Psi\}$,
$\Psi^\sharp:=\{\theta^\sharp\mid\theta\in\Psi\}$,
$(\bigvee\Psi)^\flat:=\bigvee(\Psi^\flat)$, and
$(\bigvee\Psi)^\sharp:=\bigvee(\Psi^\sharp)$, and put
\[
\mathrm{Head}(A(X)):=\bigl\{\textstyle\bigvee\Psi\;\big|\;\nabla\Psi\in\Sub(A(X)),\;|\Psi|\ge 2\bigr\}.
\]
By Lemma~\ref{lem:box-isolation}(4)(a), $\mathrm{Head}(A(X))\cap\Sub(A(X))=\emptyset$.

\begin{lemma}[Injectivity and local commutation of outer substitution]
\label{lem:pcc-outer-inj}
Under Hypotheses~(1)--(4) of Theorem~\ref{thm:positive-context-closure}:
\begin{enumerate}[(a)]
\item both $\gamma\mapsto\gamma^\flat$ and $\gamma\mapsto\gamma^\sharp$ are injective on $\Sub(A(X))\cup\mathrm{Head}(A(X))$;
\item for $\gamma\in\Sub(A(X))\cup\mathrm{Head}(A(X))$, $\gamma^\flat=B$ implies $\gamma=X$, and $\gamma^\sharp=C$ implies $\gamma=X$; moreover, if $\gamma\in\Sub(A(X))\setminus\{X\}$ is non-modal
in $\Cl(A(X))$ and $\gamma^\sharp\in\Cl(C)$, then $X\notin\Free(\gamma)$
and $\gamma^\flat=\gamma=\gamma^\sharp$;
\item for $\gamma\in\Sub(A(X))\setminus\{X\}$: $\gamma$ is a literal (distinct
from $X,\neg X$) iff $\gamma^\sharp$ is a literal of $\Cl(A(C))$ iff
$\gamma^\flat$ is a literal of $\Cl(A(B))$, in which case
$\gamma^\flat=\gamma=\gamma^\sharp$; $\gamma=\nabla\Psi$ iff
$\gamma^\sharp=\nabla(\Psi^\sharp)$ iff $\gamma^\flat=\nabla(\Psi^\flat)$;
and if $\gamma$ is non-modal ($\alpha_1\land\alpha_2$, $\alpha_1\lor\alpha_2$,
$\eta_yy.\varphi_y$, or $y\in\Bound(A(X))$), its immediate reducts under
$(\cdot)^\flat$ in $\Cl(A(B))$ and under $(\cdot)^\sharp$ in $\Cl(A(C))$
are the $(\cdot)^\flat$- and $(\cdot)^\sharp$-images of those of $\gamma$
in $\Cl(A(X))$, with identical regeneration sets ($\{y\}$ when
$\gamma=y\in\Bound(A(X))$, and $\emptyset$ otherwise).
\end{enumerate}
\end{lemma}

\begin{proof}
View each formula $\chi$ as a finite ordered syntax tree $T(\chi)$ with internal
nodes labelled by $\land$, $\lor$, $\eta_ww$, or $\nabla$, and leaves by
literals. For either $D\in\{B,C\}$, Hypothesis~(4) gives an outermost binder $\eta_{z_D}z_D$ with $z_D$ absent from $A(X)$. In the ordered syntax tree of $\gamma[D/X]$, the roots of substituted copies of $D$ are exactly the nodes bearing that binder label: cleanness makes its defining binder unique within $D$, and separation excludes it from the outer context. Collapsing those maximal subtrees to $X$ recovers $\gamma$. This proves injectivity of both translations on $\Sub(A(X))$ and shows that $\gamma[D/X]=D$ implies $\gamma=X$.
If $\gamma\in\Sub(A(X))\setminus\{X\}$ is non-modal, $\gamma^\sharp\in\Cl(C)$, and $X\in\Free(\gamma)$, then $C$ would be a proper subformula of an element of $\Cl(C)$. This contradicts the uniqueness of its outermost defining binder, including for subformulas reached through a box chain. Thus $X\notin\Free(\gamma)$ and $\gamma^\flat=\gamma=\gamma^\sharp$, proving~(b).
For the remaining part of~(a), box isolation implies that translated box-chain heads do not belong to the corresponding substituted subformula sets. Each head is a disjunction of fixpoint formulas, and its elements can be recovered from its disjuncts. Injectivity on subformulas therefore extends to the heads, for both $B$ and $C$.
Part~(c) holds by
direct inspection of the root of $\gamma\neq X,\neg X$, using cleanness of
$A(B)$ and $A(C)$ (Lemma~\ref{lem:pcc-syntax}) for regenerations $y\to\varphi_y^\flat$
and $y\to\varphi_y^\sharp$.
\end{proof}

\begin{lemma}[Backward extension of trace tails]\label{lem:pcc-trace-extension}
Let $\xi=t_0t_1\cdots$ be an infinite branch of a canonical tableau $\mathcal T_\Gamma$
($\Gamma_0:=\Gamma$ clean and guarded), and suppose $\gamma_n\in L(t_n)$ and
$V_n\subseteq\Bound(\Gamma_0)$ satisfy $(\gamma_n,\gamma_{n+1},V_n)\in R(t_n,t_{n+1})$
for all $n\ge N$. Then there exist $\gamma_n\in L(t_n)$ and $V_n\subseteq\Bound(\Gamma_0)$
for $0\le n<N$ making $\tau=(\gamma_n,V_n)_{n\ge 0}$ an infinite trace on $\xi$
from $t_0$, with $\Reg^\infty(\tau)=\{x\in\Bound(\Gamma_0)\mid x\in V_n\text{ for infinitely many }n\ge N\}$.
\end{lemma}

\begin{proof}
Every formula in $L(t_n)$ ($n\ge 1$) has a predecessor in $L(t_{n-1})$ under
$R(t_{n-1},t_n)$: at a choice-to-modal edge by Definition~\ref{def:saturation}
(reachability from $L(t_{n-1})$ in $E_c$), and at a modal-to-choice edge by
Definition~\ref{def:modal-successors} ($\nabla\Psi_k\to\psi$ and
$\nabla\Psi_j\to\bigvee\Psi_j$). Stepping backwards $N$ times gives $\tau$, and
finitely many initial sets $V_0,\dots,V_{N-1}$ do not affect $\Reg^\infty(\tau)$.
\end{proof}

\subsection{Positions, one-round construction, and older-wins merging}
\label{sec:pcc-rounds}
The birth-time ordering is related to the ordering of trace histories in the proof of~\cite[Proposition~7.4]{EnqvistSeifanVenema18}. Here the ordered objects are instances of an arbitrary consequence relation $W$, rather than histories leading into a disjunctive companion, and ownership is assigned formula by formula when supports merge.

Throughout Sections~\ref{sec:pcc-rounds}--\ref{sec:pcc-completion}, fix
Hypotheses~(1)--(5) of Theorem~\ref{thm:positive-context-closure} and write
$\widetilde{\mathcal T}:=\mathcal T_{A(B)}$ and $\mathcal T:=\mathcal T_{A(C)}$.
By closing the root pair $(r_{\mathcal T_B},r_{\mathcal T_C})\in W$ inductively
under fixed witnesses for~(R-ch), (R-back) followed by~(R-ch), and~(R-forth$^2$)
(and retaining only those pairs), we may assume without loss of generality that
every modal pair in $W$ lies on a finite $W$-path from
$(r_{\mathcal T_B},r_{\mathcal T_C})$---unique because $\mathcal T_B$ and
$\mathcal T_C$ are trees---since (R-loc) and~(R-par) are inherited by any
subrelation of $W$.

\begin{definition}[Positions and ownership]\label{def:pcc-position}
A \emph{position} at $(\tilde m,m)\in\Mod(\widetilde{\mathcal T})\times\Mod(\mathcal T)$
consists of a finite set $\mathcal H\subseteq\mathbb N$ of birth times and an
assignment of a modal pair $(b_h,c_h)\in W$ to each $h\in\mathcal H$ such that
\begin{align}
L(b_h)&\subseteq L(\tilde m)\qquad\text{for every }h\in\mathcal H,\label{eq:pcc-I1}\\
L(m)&\subseteq\bigcup_{h\in\mathcal H}L(c_h)\;\cup\;
\bigl\{\gamma^\sharp\;\big|\;\gamma\in\Sub(A(X))\setminus\{X\},\;
\gamma^\sharp\text{ modal in }\Cl(A(C)),\;\gamma^\flat\in L(\tilde m)\bigr\}.\label{eq:pcc-I2}
\end{align}
The \emph{owner} of $\varpi\in L(m)$ is
\[
\own(\varpi):=\min\Bigl(\{h\in\mathcal H\mid\varpi\in L(c_h)\}\cup
\bigl\{\infty\;\big|\;\exists\gamma\in\Sub(A(X))\setminus\{X\}\;(\gamma^\sharp=\varpi\land\gamma^\flat\in L(\tilde m))\bigr\}\Bigr),
\]
which is well-defined by~\eqref{eq:pcc-I2}; when $\own(\varpi)=\infty$, $\gamma$
is unique by Lemma~\ref{lem:pcc-outer-inj}(a).
\end{definition}

\begin{lemma}[Local conditions]\label{lem:pcc-local}
Every modal pair $(\tilde m,m)$ carrying a position satisfies~(R-loc).
\end{lemma}

\begin{proof}
Assume $L(\tilde m)$ is consistent. By~\eqref{eq:pcc-I1}, each $L(b_h)$ is
consistent, so (R-loc) for $W$ at $(b_h,c_h)$ gives consistency of $L(c_h)$,
$\Lit(L(c_h))\subseteq\Lit(L(b_h))\subseteq\Lit(L(\tilde m))$, and that $L(b_h)$
is $\nabla$-non-free if $L(c_h)$ is. For $\varpi\in L(m)$ with $\own(\varpi)=\infty$
and witness $\gamma\in\Sub(A(X))\setminus\{X\}$, Lemma~\ref{lem:pcc-outer-inj}(c)
gives $\gamma^\flat=\varpi\in\Lit(L(\tilde m))$ if $\varpi$ is a literal, and
$\gamma^\flat=\nabla(\Psi^\flat)\in L(\tilde m)$ if $\varpi=\nabla(\Psi^\sharp)$.
Hence $\Lit(L(m))\subseteq\Lit(L(\tilde m))$, $L(m)$ is consistent, and
$L(\tilde m)$ is $\nabla$-non-free whenever $L(m)$ is.
\end{proof}

\paragraph{Invariant guide.}
The relay maintains three facts, each matched to a consequence clause:
\begin{itemize}
\item \emph{Left support:} every active $h$ has $(b_h,c_h)\in W$ and $L(b_h)\subseteq L(\tilde m)$ (\eqref{eq:pcc-I1}). This makes the $W$-witnesses available for (R-back) and (R-forth$^2$).
\item \emph{Right coverage:} $L(m)$ is covered by the copy labels and translated outer modal formulas (\eqref{eq:pcc-I2}); the minimum supporting birth time defines $\own$. Together with left support and (R-loc) for $W$, this proves (R-loc) for $Z$.
\item \emph{Trace monotonicity:} each round satisfies $\own_{\mathrm{next}}(\varpi)\le\own(\varepsilon)$ for a target trace step, in the order $0<1<2<\cdots<\infty$ (\eqref{eq:pcc-own-decr}). Stabilisation yields (R-par) by transferring a trace through the outer context or a persistent copy.
\end{itemize}
Let $N_{\mathrm{born}}\in\mathbb N$ denote the total number of copies born up to
$(\tilde m,m)$, so $\mathcal H\subseteq\{0,\dots,N_{\mathrm{born}}-1\}$.

\begin{figure}[htbp]
\centering
\includegraphics[width=0.85\textwidth]{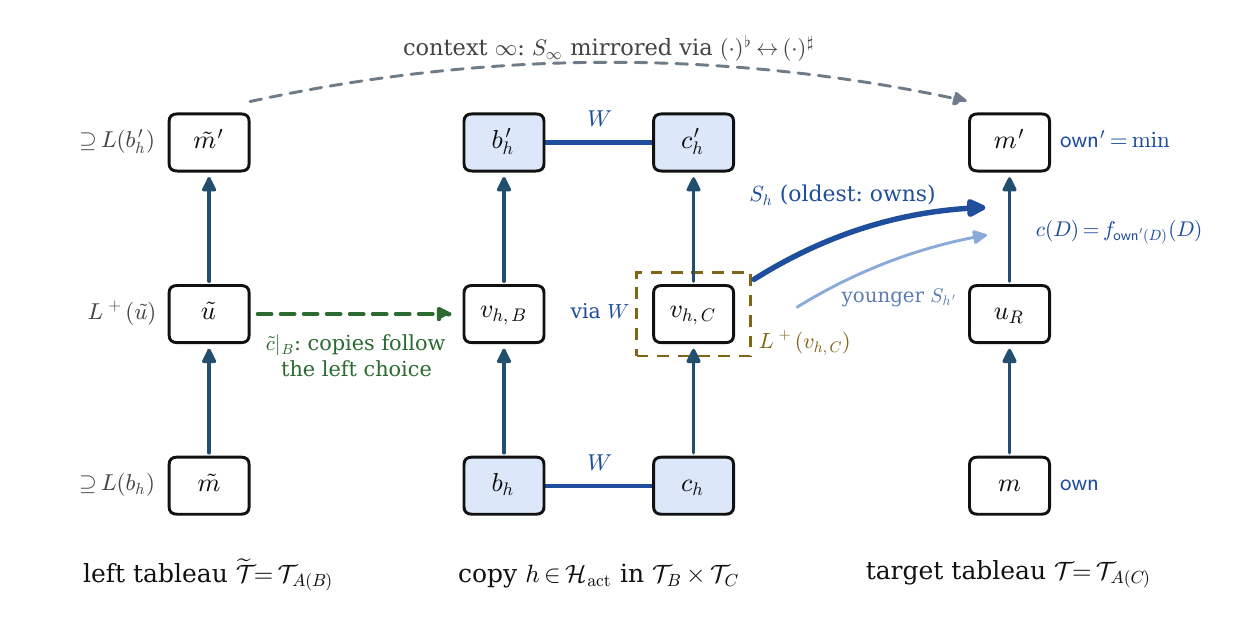}
\caption{One round of the older-wins construction. The left tableau supplies the saturation choice; an active copy advances through $W$ and supplies a target support $S_h$. At a shared formula, the oldest supporting copy determines the target choice; the outer context supplies $S_\infty$.}
\label{fig:pcc-round}
\end{figure}

Figure~\ref{fig:pcc-round} illustrates how an active copy follows the left saturation choice in Construction~\ref{con:pcc-round}, and how its support is combined with younger-copy and context supports by the older-wins rule.

\begin{construction}[One round from a position]\label{con:pcc-round}
Let $(\tilde m,m)$ carry a position $(\mathcal H,\{(b_h,c_h)\}_{h\in\mathcal H})$
with $L(\tilde m)$ consistent and $L(m)$ $\nabla$-non-free (hence consistent by
Lemma~\ref{lem:pcc-local}).

\smallskip\noindent\textbf{Step 1 (Discarding).} Put
$\mathcal H_{\mathrm{act}}:=\{h\in\mathcal H\mid L(c_h)\text{ is $\nabla$-non-free}\}$.
Dropping $\mathcal H\setminus\mathcal H_{\mathrm{act}}$ leaves $\own(\nabla\Xi)$
unchanged for every $\nabla\Xi\in L(m)$. By (R-loc) for $W$, both $L(c_h)$ and
$L(b_h)$ are consistent and $\nabla$-non-free for all $h\in\mathcal H_{\mathrm{act}}$.

\smallskip\noindent\textbf{Step 2a ((R-back) entry).} Given $u_R\in C(m)$, say
the $(k,\theta)$-successor for $\nabla\Xi_k\in L(m)$ and $\theta\in\Xi_k$, let
$\kappa_0:=\own(\nabla\Xi_k)\in\mathcal H_{\mathrm{act}}\cup\{\infty\}$.
\begin{itemize}
\item If $\kappa_0=h_0\in\mathcal H_{\mathrm{act}}$, let $v^0_C\in C(c_{h_0})$ have
principal formula $\nabla\Xi_k$ and element $\theta$. By (R-back) for $W$ at
$(b_{h_0},c_{h_0})$, pick $v^0_B\in C(b_{h_0})$ with $(v^0_B,v^0_C)\in W$;
say $v^0_B$ has principal formula $\nabla\Theta_*\in L(b_{h_0})\subseteq L(\tilde m)$
and element $\vartheta_*\in\Theta_*$. Let $\tilde u\in C(\tilde m)$ be the
successor of $L(\tilde m)$ with principal formula $\nabla\Theta_*$ and
element $\vartheta_*$.
\item If $\kappa_0=\infty$, let $\nabla\Psi\in\Sub(A(X))$ and $\theta_0\in\Psi$
be the unique formulas with $\nabla(\Psi^\sharp)=\nabla\Xi_k$,
$\nabla(\Psi^\flat)\in L(\tilde m)$ and $\theta_0^\sharp=\theta$
(Lemma~\ref{lem:pcc-outer-inj}(a),(c)), and let $\tilde u\in C(\tilde m)$
have principal formula $\nabla(\Psi^\flat)$ and element $\theta_0^\flat$.
\end{itemize}
For each $\tilde m'\in C(\tilde u)$: if $L(\tilde m')$ is inconsistent, pair
$\tilde m'$ with a fixed $m'_{\mathrm{inc}}(u_R)\in C(u_R)$ and terminate the
round without a position; if $L(\tilde m')$ is consistent, proceed to Step~3.

\smallskip\noindent\textbf{Step 2b ((R-forth$^2$) entry).} Given $\tilde u\in C(\tilde m)$
and $\tilde m'\in C(\tilde u)$ with $L(\tilde m')$ consistent, execute Step~3
first and then select $u_R\in C(m)$ in Step~4.

\smallskip\noindent\textbf{Step 3 (Advancing finite copies and newborn copy).}
Let $\tilde c_0:=c_{(\tilde u,\tilde m')}$ and $\tilde c:=\tilde c_0^+$, so
$M_{\tilde c}(L^+(\tilde u))=L(\tilde m')$ by Lemma~\ref{lem:full-successor}(1).
Since $L(\tilde m')$ is consistent, $\nabla\emptyset\notin L(\tilde m)$ (otherwise
$\bigvee\emptyset=\bot$ in $L(\tilde u)$; since $\bot$ is modal, it is retained in $L(\tilde m')$).
\begin{itemize}
\item \emph{Finite copies $h\in\mathcal H_{\mathrm{act}}$:} Write the
$\nabla$-formulas of $L(b_h)\subseteq L(\tilde m)$ as
$\nabla\Theta_{h,1},\dots,\nabla\Theta_{h,r_h}$ ($\Theta_{h,j}\neq\emptyset$).
Then $B_{b_h}:=\{\bigvee\Theta_{h,1},\dots,\bigvee\Theta_{h,r_h}\}\subseteq L^+(\tilde u)$
(including when some $\nabla\Theta_{h,j}$ is principal for $\tilde u$, by
definition of $L^+(\tilde u)$). Applying Lemma~\ref{lem:diamond-macro-step}
at $b_h$ to $\tilde c|_B:=\tilde c|_{\Cl(B)}$, let $\vartheta_{h,j}\in\Theta_{h,j}$
be the element reached from $\bigvee\Theta_{h,j}$.
In the (R-back) mode with $\kappa_0=h_0$ and $h=h_0$, $\tilde c=\tilde c_0^+$
leads from $\bigvee\Theta_*$ to $\vartheta_*$, so choosing $\nabla\Theta_*$
gives $v_{h_0,B}:=v^0_B$ and modal child $b'_{h_0}\in C(v^0_B)$ with outcome
$\tilde c|_B$; (R-ch) for $W$ at $(v^0_B,v^0_C)$ yields $c'_{h_0}\in C(v^0_C)$
with $(b'_{h_0},c'_{h_0})\in W$, and we put $v_{h_0,C}:=v^0_C$.
For all other $h\in\mathcal H_{\mathrm{act}}$, let $v_{h,B}\in C(b_h)$ be
the $(1,\vartheta_{h,1})$-child and $b'_h\in C(v_{h,B})$ its child with
outcome $\tilde c|_B$. Since $L(b'_h)=M_{\tilde c|_B}(B_{b_h})\subseteq L(\tilde m')$
is consistent, (R-forth$^2$) for $W$ at $(b_h,c_h)$ yields $v_{h,C}\in C(c_h)$
and $c'_h\in C(v_{h,C})$ with $(b'_h,c'_h)\in W$. In both cases put
$f_h:=c_{(v_{h,C},c'_h)}^+$ and $S_h:=S_{f_h}(L^+(v_{h,C}))$.
\item \emph{Newborn copy:} By Lemmas~\ref{lem:box-isolation}(4) and~\ref{lem:full-successor}(1),
$B\in S_{\tilde c}(L^+(\tilde u))$ iff $B\in S_{\tilde c_0}(L(\tilde u))$,
and no $E$-path from $B$ enters any box chain. If $B\in S_{\tilde c}(L^+(\tilde u))$,
let $b'_{h_{\mathrm{new}}}\in C(r_{\mathcal T_B})$ have outcome
$\tilde c_0|_{\Cl(B)}$, so $L(b'_{h_{\mathrm{new}}})=M_{\tilde c}(\{B\})\subseteq L(\tilde m')$
is consistent; (R-ch) for $W$ at $(r_{\mathcal T_B},r_{\mathcal T_C})$
gives $c'_{h_{\mathrm{new}}}\in C(r_{\mathcal T_C})$ with
$(b'_{h_{\mathrm{new}}},c'_{h_{\mathrm{new}}})\in W$. Put
$h_{\mathrm{new}}:=N_{\mathrm{born}}$,
$\mathcal H':=\mathcal H_{\mathrm{act}}\cup\{h_{\mathrm{new}}\}$,
$N'_{\mathrm{born}}:=N_{\mathrm{born}}+1$,
$f_{h_{\mathrm{new}}}:=c_{(r_{\mathcal T_C},c'_{h_{\mathrm{new}}})}$ and
$S_{h_{\mathrm{new}}}:=S_{f_{h_{\mathrm{new}}}}(\{C\})$. Otherwise put
$\mathcal H':=\mathcal H_{\mathrm{act}}$ and $N'_{\mathrm{born}}:=N_{\mathrm{born}}$.
\end{itemize}

\smallskip\noindent\textbf{Step 4 (Choosing $u_R$ in the (R-forth$^2$) mode).}
Note that $\nabla\emptyset\notin L(m)$: an owner $\infty$ would give
$\nabla\emptyset\in L(\tilde m)$, while an owner $h\in\mathcal H_{\mathrm{act}}$
would put $\bigvee\emptyset=\bot$ in $L(v_{h,C})$; since $\bot$ is modal, it is retained in $L(c'_h)$, violating
(R-loc) for $W$ at $(b'_h,c'_h)$. In the (R-forth$^2$) mode, pick any
$\nabla\Xi_k\in L(m)$ ($\Xi_k\neq\emptyset$) and let $\kappa_0:=\own(\nabla\Xi_k)$.
If $\kappa_0=h_0\in\mathcal H_{\mathrm{act}}$, let $\theta\in\Xi_k$ be reached by
$f_{h_0}$ from $\bigvee\Xi_k\in L^+(v_{h_0,C})$; if $\kappa_0=\infty$, with
$\nabla(\Psi^\sharp)=\nabla\Xi_k$, let $\theta:=\theta_0^\sharp\in\Xi_k$ where
$\theta_0^\flat\in\Psi^\flat$ is reached by $\tilde c$ from
$(\bigvee\Psi)^\flat\in L^+(\tilde u)$. Let $u_R\in C(m)$ be the $(k,\theta)$-successor.

\smallskip\noindent\textbf{Step 5 (Context support $S_\infty$ and older-wins choice $c$).}
Put $\Lambda_\infty:=\{\gamma\in(\Sub(A(X))\setminus\{X\})\cup\mathrm{Head}(A(X))\mid\gamma^\sharp\in L(u_R)\land\gamma^\flat\in S_{\tilde c}(L^+(\tilde u))\}$.
The exclusion of $X$ is explicit: elements of $L(u_R)$ are either fixpoint
formulas $\theta_0^\sharp$ with $\theta_0=\eta_zz.\chi\neq X$ (by box isolation)
or disjunctions $\bigvee\Xi_j$ ($|\Xi_j|\ge 2$), whereas $X^\sharp=C=\eta_{z_C}z_C.\,C_0$,
so $\gamma^\sharp\neq C$ by Lemma~\ref{lem:pcc-outer-inj}(b). Define $f_\infty$ on
$\Cl(A(C))$ by:
(i)~for $D_0=\alpha_1\lor\alpha_2\in\Sub(A(X))$ with $D_0^\flat\in S_{\tilde c}(L^+(\tilde u))$,
$f_\infty(D_0^\sharp):=\alpha_i^\sharp$ where $\tilde c(D_0^\flat)=\alpha_i^\flat$;
(ii)~for $\bigvee\Psi\in\Lambda_\infty\cap\mathrm{Head}(A(X))$, $f_\infty$ leads
along the box chain of $\nabla(\Psi^\sharp)$ to $\theta_\Psi^\sharp$, where
$\tilde c$ leads from $(\bigvee\Psi)^\flat$ to $\theta_\Psi^\flat\in\Psi^\flat$;
(iii)~left disjuncts elsewhere. Let $S_\infty\subseteq\Cl(A(C))\setminus\{C\}$ be
the set of formulas reachable from $\Lambda_\infty^\sharp$ by $C$-avoiding
$E_{f_\infty}$-paths. For $\delta\in\mathcal S:=\bigcup_{\kappa\in\mathcal H'\cup\{\infty\}}S_\kappa$,
put $\own'(\delta):=\min\{\kappa\mid\delta\in S_\kappa\}$, define $c(D):=f_{\own'(D)}(D)$
for $D\in\mathcal S$ (left disjunct outside $\mathcal S$), and let $m'\in C(u_R)$
be the child with outcome $(M_c(L(u_R)),R_c(L(u_R)))$.
\end{construction}

\begin{lemma}[Individual copy and context supports]
\label{lem:pcc-copy-supports}\label{lem:pcc-context-support}
In Construction~\ref{con:pcc-round} (with $L(\tilde m')$ consistent):
\begin{enumerate}[(a)]
\item For each $h\in\mathcal H'$, $S_h\subseteq\Cl(C)$ is closed under
$E_{f_h}$-edges in $\Cl(A(C))$, $L(b'_h)\subseteq L(\tilde m')$, and a
formula $\varpi\in S_h$ is modal in $\Cl(A(C))$ iff $\varpi\in L(c'_h)$. If
$h_{\mathrm{new}}\in\mathcal H'$, then $C\in S_{h_{\mathrm{new}}}$.
\item $f_\infty$ is well-defined; every $\delta\in S_\infty$ is either a
box-chain disjunction on the $E_{f_\infty}$-path from $(\bigvee\Psi)^\sharp$
to $\theta_\Psi^\sharp$ ($\bigvee\Psi\in\Lambda_\infty\cap\mathrm{Head}(A(X))$),
or $\gamma^\sharp$ for a unique $\gamma\in\Sub(A(X))\setminus\{X\}$ with
$\gamma^\flat\in S_{\tilde c}(L^+(\tilde u))$; and if $\varpi\in S_\infty$
is modal in $\Cl(A(C))$, then $\varpi=\gamma^\sharp$ with
$\gamma^\flat\in L(\tilde m')$.
\item If $\delta\in S_\infty$ and $(\delta,\delta')\in E_{f_\infty}$, then either
$\delta'\neq C$ and $\delta'\in S_\infty$, or $\delta'=C$, in which case
$B\in S_{\tilde c}(L^+(\tilde u))$, $h_{\mathrm{new}}\in\mathcal H'$ exists,
and $C\in S_{h_{\mathrm{new}}}$.
\item If $(\varepsilon,\delta_0,\emptyset)\in R(m,u_R)$ for $\varepsilon\in L(m)$,
then $\delta_0\in S_{\own(\varepsilon)}$.
\end{enumerate}
\end{lemma}

\begin{proof}
(a) By Hypothesis~(3) and Lemma~\ref{lem:pcc-syntax}, outgoing decomposition
edges and modality on $\Cl(C)$ coincide in $\Cl(C)$ and $\Cl(A(C))$, so $S_h$ is
$E_{f_h}$-closed; $L(b'_h)\subseteq L(\tilde m')$ was shown in Step~3, and
$C\in\{C\}\subseteq S_{h_{\mathrm{new}}}$. Since $C$ is box-isolated,
Lemma~\ref{lem:full-successor}(1) applied to $\mathcal T_C$ gives
$M_{f_h}(L^+(v_{h,C}))=L(c'_h)$ for $h\in\mathcal H_{\mathrm{act}}$, while
$M_{f_{h_{\mathrm{new}}}}(\{C\})=L(c'_{h_{\mathrm{new}}})$ by definition.

(b) In defining $f_\infty$, clause~(i) is conflict-free by injectivity of
$(\cdot)^\sharp$ and $(\cdot)^\flat$ on $\Sub(A(X))$ (Lemma~\ref{lem:pcc-outer-inj}(a)); by
box isolation of $A(C)$ and Lemma~\ref{lem:box-isolation}(4)(a), box-chain
disjunctions in clause~(ii) are disjoint from $\Sub(A(C))$ and pairwise disjoint
across distinct $\nabla(\Psi^\sharp)$, so $f_\infty$ is well-defined. Induction
on $C$-avoiding $E_{f_\infty}$-paths from $\Lambda_\infty^\sharp$, using
Lemma~\ref{lem:pcc-outer-inj}(b),(c) and clause~(i) of $f_\infty$, establishes
the characterisation of $\delta\in S_\infty$. A modal $\varpi\in S_\infty$ is
not a disjunction, so $\varpi=\gamma^\sharp$ with $\gamma^\flat$ modal in
$S_{\tilde c}(L^+(\tilde u))$, i.e.\ $\gamma^\flat\in L(\tilde m')$.

(c) If $\delta'\neq C$, appending $(\delta,\delta')$ preserves $C$-avoidance, so
$\delta'\in S_\infty$. Suppose $\delta'=C$. By box isolation of $A(X)$, no
$\theta_\Psi\in\Psi$ equals $X$, so Lemma~\ref{lem:pcc-outer-inj}(b) rules out
$\delta'=\theta_\Psi^\sharp$; hence $\delta=\gamma^\sharp$ with
$\gamma\in\Sub(A(X))\setminus\{X\}$ and $\gamma^\flat\in S_{\tilde c}(L^+(\tilde u))$.
By Lemma~\ref{lem:pcc-outer-inj}(c) and clause~(i) of $f_\infty$,
$\delta'=(\gamma')^\sharp=C$ for some $\gamma'\in\Sub(A(X))$ with
$(\gamma^\flat,(\gamma')^\flat)\in E_{\tilde c}$. Lemma~\ref{lem:pcc-outer-inj}(b)
gives $\gamma'=X$, so $B=X^\flat\in S_{\tilde c}(L^+(\tilde u))$, spawning
$h_{\mathrm{new}}\in\mathcal H'$ with $C\in S_{h_{\mathrm{new}}}$.

(d) Write $\varepsilon=\nabla\Xi\in L(m)$ and $\kappa:=\own(\varepsilon)$.
Suppose first $\kappa=h\in\mathcal H_{\mathrm{act}}$, so $\nabla\Xi\in L(c_h)$.
If $\varepsilon$ is non-principal for $(m,u_R)$, then $\delta_0=\bigvee\Xi$;
even if $\nabla\Xi$ is principal for copy $h$'s step $(c_h,v_{h,C})$, the full
successor label $L^+(v_{h,C})=L(v_{h,C})\cup\{\bigvee\Xi\}$ retains $\bigvee\Xi$,
so $\delta_0=\bigvee\Xi\in L^+(v_{h,C})\subseteq S_h$. If $\varepsilon=\nabla\Xi_k$
is principal for $(m,u_R)$, then $h=h_0$ and $\delta_0=\theta$; in Step~2a,
$\theta\in L(v^0_C)\subseteq S_{h_0}$, while in Step~4, $f_{h_0}$ reaches
$\theta$ from $\bigvee\Xi_k\in L^+(v_{h_0,C})$, so $\theta\in S_{h_0}$.
Now suppose $\kappa=\infty$, with $\nabla(\Psi^\sharp)=\nabla\Xi$ and
$\nabla(\Psi^\flat)\in L(\tilde m)$. If $\varepsilon$ is non-principal,
$(\bigvee\Psi)^\flat\in L^+(\tilde u)\subseteq S_{\tilde c}(L^+(\tilde u))$, so
$\bigvee\Psi\in\Lambda_\infty$ and $\delta_0=(\bigvee\Psi)^\sharp\in S_\infty$;
if $\varepsilon$ is principal, $\delta_0=\theta_0^\sharp$ ($\theta_0\in\Psi$,
$\theta_0\neq X$) with $\theta_0^\flat\in S_{\tilde c}(L^+(\tilde u))$ by
Step~2a or Step~4, so $\theta_0\in\Lambda_\infty$ and $\delta_0\in S_\infty$.
\end{proof}

\begin{lemma}[Older-wins merging and position preservation]
\label{lem:pcc-merge}
In Construction~\ref{con:pcc-round} (with $L(\tilde m')$ consistent), put
$\mathcal S:=\bigcup_{\kappa\in\mathcal H'\cup\{\infty\}}S_\kappa$. Then:
\begin{enumerate}[(a)]
\item $L(u_R)\subseteq\mathcal S$ and $S_c(L(u_R))\subseteq\mathcal S$;
\item along every edge $(\delta,\delta')\in E_c$ with $\delta\in S_c(L(u_R))$,
$\own'(\delta')\le\own'(\delta)$;
\item any $E_c$-path $\pi=\delta_0\cdots\delta_l$ in $S_c(L(u_R))$ with
$\own'(\delta_0)=\own'(\delta_l)=\kappa$ has constant owner $\kappa$ and is
an $E_{f_\kappa}$-path inside $S_\kappa$;
\item $(\tilde m',m')$ equipped with $\mathcal H'$ and $\{(b'_h,c'_h)\}_{h\in\mathcal H'}$
is a position whose owner function $\own_{\mathrm{next}}$ equals
$\own'|_{L(m')}$.
\end{enumerate}
\end{lemma}

\begin{proof}
By Lemma~\ref{lem:pcc-copy-supports}(d), $L(u_R)\subseteq\mathcal S$. Let
$\delta\in\mathcal S$ and $(\delta,\delta')\in E_c$, with $\kappa:=\own'(\delta)$
(so $\delta\in S_\kappa$). We check the following three cases:
\begin{itemize}
\item \emph{Case A ($\kappa=h\in\mathcal H'$):} Either $\delta$ is non-disjunctive
or $\delta'=c(\delta)=f_h(\delta)$; in both cases $(\delta,\delta')\in E_{f_h}$,
so $\delta'\in S_h\subseteq\mathcal S$ by Lemma~\ref{lem:pcc-copy-supports}(a)
and $\own'(\delta')\le h=\own'(\delta)$.
\item \emph{Case B ($\kappa=\infty$ and $\delta'\neq C$):} Similarly
$(\delta,\delta')\in E_{f_\infty}$, so Lemma~\ref{lem:pcc-copy-supports}(c)
gives $\delta'\in S_\infty\subseteq\mathcal S$ and $\own'(\delta')\le\infty=\own'(\delta)$.
\item \emph{Case C ($\kappa=\infty$ and $\delta'=C$):} Here $(\delta,C)\in E_{f_\infty}$,
so by Lemma~\ref{lem:pcc-copy-supports}(c) the newborn copy
$h_{\mathrm{new}}\in\mathcal H'$ exists and $C\in S_{h_{\mathrm{new}}}\subseteq\mathcal S$,
giving $\own'(C)\le h_{\mathrm{new}}<\infty=\own'(\delta)$.
\end{itemize}
\smallskip\noindent\emph{Choice conflicts.} If a disjunction $\delta=D$ lies in multiple supports $S_\lambda$ with conflicting choices, put $\kappa=\own'(D)=\min\{\lambda\mid D\in S_\lambda\}$
(necessarily finite when $|\{\lambda\mid D\in S_\lambda\}|\ge 2$). In $E_c$
the unique outgoing edge from $D$ is $(D,f_\kappa(D))$; unchosen disjuncts
$f_\lambda(D)$ ($\lambda>\kappa$) have no incoming edge from $D$ in $E_c$
and require no continuation, while Case~A keeps $\delta'=f_\kappa(D)\in S_\kappa\subseteq\mathcal S$
with $\own'(\delta')\le\kappa=\own'(D)$.
Induction along $E_c$-paths from $L(u_R)$ yields~(a) and~(b). For~(c),~(b) forces
$\kappa=\own'(\delta_l)\le\dots\le\own'(\delta_0)=\kappa$, so $c(\delta_i)=f_\kappa(\delta_i)$
at every disjunction on $\pi$, making $\pi$ an $E_{f_\kappa}$-path in $S_\kappa$.

For~(d), \eqref{eq:pcc-I1} holds by Lemma~\ref{lem:pcc-copy-supports}(a). For
$\delta\in L(m')=M_c(L(u_R))\subseteq\mathcal S$, if $\own'(\delta)=h\in\mathcal H'$
then $\delta\in L(c'_h)$ by Lemma~\ref{lem:pcc-copy-supports}(a), while if
$\own'(\delta)=\infty$ then $\delta=\gamma^\sharp$ with $\gamma^\flat\in L(\tilde m')$
by Lemma~\ref{lem:pcc-copy-supports}(b), proving~\eqref{eq:pcc-I2}. Since for
modal $\delta\in L(m')$ we have $\delta\in S_h\iff\delta\in L(c'_h)$ for every
$h\in\mathcal H'$ (Lemma~\ref{lem:pcc-copy-supports}(a)), $\own_{\mathrm{next}}(\delta)$
and $\own'(\delta)$ take the minimum over the same subset of $\mathcal H'$ (and
equal $\infty$ iff that subset is empty).
\end{proof}

\paragraph{The initial round.}
From $(\tilde t_0,t_0):=(r_{\widetilde{\mathcal T}},r_{\mathcal T})=(\{A(B)\},\{A(C)\})$,
pair any inconsistent $\tilde m_0\in C(\tilde t_0)$ with a fixed
$m_{0,\mathrm{inc}}\in C(t_0)$. For consistent $\tilde m_0\in C(\tilde t_0)$ with
$\tilde c_0:=c_{(\tilde t_0,\tilde m_0)}$, run Steps~3 and~5 with
$\mathcal H_{\mathrm{act}}:=\emptyset$, $N_{\mathrm{born}}:=0$, $\tilde u:=\tilde t_0$
and $u_R:=t_0$: if $A(X)=X$, newborn copy $0$ is created ($\mathcal H'=\{0\}$,
$S_\infty=\emptyset$, $C\in S_0$); if $A(X)\neq X$, $\Lambda_\infty=\{A(X)\}$
($A(C)\in S_\infty$ by Lemma~\ref{lem:pcc-outer-inj}(b)) and newborn copy $0$ is
created iff $B\in S_{\tilde c_0}(\{A(B)\})$. Lemmas~\ref{lem:pcc-copy-supports}(a)--(c)
and~\ref{lem:pcc-merge}(b)--(d) apply verbatim, giving a position at
$(\tilde m_0,m_0)$ with $\own_0=\own'|_{L(m_0)}$.

\subsection{Trace preservation and completion of the proof}
\label{sec:pcc-completion}

\begin{definition}[The relation $Z$]\label{def:pcc-Z}
Define strata $Z^{(d)}$ ($d\ge 0$), attaching a position and $N_{\mathrm{born}}$
to every modal pair with consistent left label, by recursion on $d$:
\begin{itemize}
\item $Z^{(0)}$ contains $(r_{\widetilde{\mathcal T}},r_{\mathcal T})$ and all
initial-round modal pairs $(\tilde m_0,m_0)$ ($\tilde m_0\in C(r_{\widetilde{\mathcal T}})$).
\item For each modal pair $(\tilde m,m)\in Z^{(d)}$ with $L(\tilde m)$ consistent
and $L(m)$ $\nabla$-non-free:
First add all pairs from clause~(a): for every $u_R\in C(m)$,
$Z^{(d+1)}$ contains $(\tilde u,u_R)$ from Step~2a and, for all
$\tilde m'\in C(\tilde u)$, the resulting modal pair $(\tilde m',m')$.
Then apply clause~(b) only for each $\tilde u\in C(\tilde m)$ that was not paired with any $u_R\in C(m)$ in clause~(a), and for every consistent $\tilde m'\in C(\tilde u)$. Add the modal pair produced by Steps~2b--5. No pair constructed in clause~(a) is revisited.
\end{itemize}
Put $Z:=\bigcup_{d\ge 0}Z^{(d)}$.
\end{definition}

\begin{lemma}[Uniqueness of positions and local clauses]\label{lem:pcc-Z-local}
Every modal pair $(\tilde m,m)\in Z$ lies at depth $2d+1$ for a unique $d\ge 0$,
lies on a unique finite $Z$-path, and---when $L(\tilde m)$ is consistent---carries
a unique position. Moreover, $Z$ relates the roots and satisfies~(R-ch), (R-loc),
(R-back), and~(R-forth$^2$).
\end{lemma}

\begin{proof}
Choice and modal pairs in $Z^{(d)}$ have components at depths $2d$ and $2d+1$, respectively. Each modal node in either tableau has a unique parent choice node and a unique grandparent modal node (except at the initial depth). Clause~(a) processes all children of every selected left choice node; clause~(b) processes only left choice nodes not selected in~(a). These sets are disjoint. Within clause~(a), distinct right children $u_R$ have disjoint child sets $C(u_R)$, so even if the same left choice node is selected more than once, the resulting modal pairs have different right components. Hence each modal pair at depth $2d+3$ is generated by exactly one round from its uniquely determined grandparent pair in $Z^{(d)}$ and has a unique position when its left label is consistent. The grandparent pair necessarily belongs to $Z^{(d)}$; induction on $d$ therefore yields a unique finite $Z$-path for every modal pair. The construction directly supplies (R-ch), (R-back), and (R-forth$^2$), and Lemma~\ref{lem:pcc-local} supplies (R-loc).
\end{proof}

\begin{lemma}[Round-step trace preservation]\label{lem:pcc-trace-step}
\begin{enumerate}[(A)]
\item Let $(\tilde m',m')\in Z^{(d+1)}$ ($L(\tilde m')$ consistent) be generated
from $(\tilde m,m)\in Z^{(d)}$ through $(\tilde u,u_R)$. For every
$(\varepsilon,\varpi,V)\in R(m,u_R)\circ R(u_R,m')$,
\begin{equation}\label{eq:pcc-own-decr}
\own_{\mathrm{next}}(\varpi)\le\own(\varepsilon).
\end{equation}
If equality $\own_{\mathrm{next}}(\varpi)=\own(\varepsilon)=\kappa$ holds:
\begin{enumerate}[(i)]
\item if $\kappa=h\in\mathcal H_{\mathrm{act}}$, then $V\subseteq\Bound(C)$
and $(\varepsilon,\varpi,V)\in R(c_h,v_{h,C})\circ R(v_{h,C},c'_h)$
in $\mathcal T_C$;
\item if $\kappa=\infty$, then for the unique $\gamma,\tilde\gamma\in\Sub(A(X))\setminus\{X\}$
with $\gamma^\sharp=\varepsilon$ and $\tilde\gamma^\sharp=\varpi$, we
have $V\subseteq\Bound(A(X))$ and
$(\gamma^\flat,\tilde\gamma^\flat,V)\in R(\tilde m,\tilde u)\circ R(\tilde u,\tilde m')$
in $\widetilde{\mathcal T}$.
\end{enumerate}
\item In the initial round with $L(\tilde m_0)$ consistent, if
$(A(C),\varpi_0,V_0)\in R(r_{\mathcal T},m_0)$ and $\own_0(\varpi_0)=\infty$,
then $V_0\subseteq\Bound(A(X))$ and
$(A(B),\gamma_0^\flat,V_0)\in R(r_{\widetilde{\mathcal T}},\tilde m_0)$ for
$\gamma_0^\sharp=\varpi_0$.
\end{enumerate}
\end{lemma}

\begin{proof}
(A) Pick $\delta_0\in L(u_R)$ and an $E_c$-path $\pi$ from $\delta_0$ to $\varpi$
with $\Reg(\pi)=V$ witnessing $(\varepsilon,\delta_0,\emptyset)\in R(m,u_R)$ and
$(\delta_0,\varpi,V)\in R(u_R,m')$. Lemmas~\ref{lem:pcc-copy-supports}(d)
and~\ref{lem:pcc-merge}(b),(d) give
$\own_{\mathrm{next}}(\varpi)=\own'(\varpi)\le\own'(\delta_0)\le\own(\varepsilon)$,
proving~\eqref{eq:pcc-own-decr}. When equality holds with value $\kappa$,
Lemma~\ref{lem:pcc-merge}(c) makes $\pi$ an $E_{f_\kappa}$-path inside $S_\kappa$.

For~(i), $\kappa=h\in\mathcal H_{\mathrm{act}}$ and $\varepsilon=\nabla\Xi\in L(c_h)$.
If $\delta_0=\bigvee\Xi$, then $\bigvee\Xi\in L^+(v_{h,C})$; if $\delta_0=\theta\in\Xi_k$
($\varepsilon$ principal, so $h=h_0$), then by Step~2a or Step~4, $f_{h_0}$ leads
along the box chain of $\bigvee\Xi_k\in L^+(v_{h_0,C})$ to $\theta$ without
regeneration (Lemma~\ref{lem:box-isolation}(4)(c)). Prefixing this box chain when
needed gives $(\bigvee\Xi,\varpi,V)\in R_{f_h}(L^+(v_{h,C}))$, and
Lemma~\ref{lem:full-successor}(2) on $\mathcal T_C$ yields~(i).

For~(ii), $\kappa=\infty$ and $\varepsilon=\nabla(\Psi^\sharp)$ with
$\nabla(\Psi^\flat)\in L(\tilde m)$. Whether $\delta_0=(\bigvee\Psi)^\sharp$
(using clause~(ii) of $f_\infty$) or $\delta_0=\theta_0^\sharp$ ($\varepsilon$
principal, using Step~2a or Step~4), $\tilde c$ leads without regeneration along
the box chain from $(\bigvee\Psi)^\flat\in L^+(\tilde u)$ to $\theta_*^\flat\in\Psi^\flat$,
where $\theta_*^\sharp$ is the element of $\Psi^\sharp$ entered by $\pi$. From
$\theta_*^\sharp$ to $\tilde\gamma^\sharp$, $\pi$ lies in $S_\infty\cap\Sub(A(C))$
(avoiding $C$) and translates edge by edge via Lemma~\ref{lem:pcc-outer-inj}(c)
and clause~(i) of $f_\infty$ to an $E_{\tilde c}$-path from $\theta_*^\flat$ to
$\tilde\gamma^\flat$ with the same regeneration set $V\subseteq\Bound(A(X))$.
Thus $((\bigvee\Psi)^\flat,\tilde\gamma^\flat,V)\in R_{\tilde c}(L^+(\tilde u))$,
and Lemma~\ref{lem:full-successor}(2) on $\widetilde{\mathcal T}$ yields~(ii).

Part~(B) is the same edge-by-edge translation along the $C$-avoiding
$E_{f_\infty}$-path from $A(X)^\sharp=A(C)$ to $\gamma_0^\sharp=\varpi_0$.
\end{proof}

\begin{figure}[t]
\centering
\includegraphics[width=0.9\textwidth]{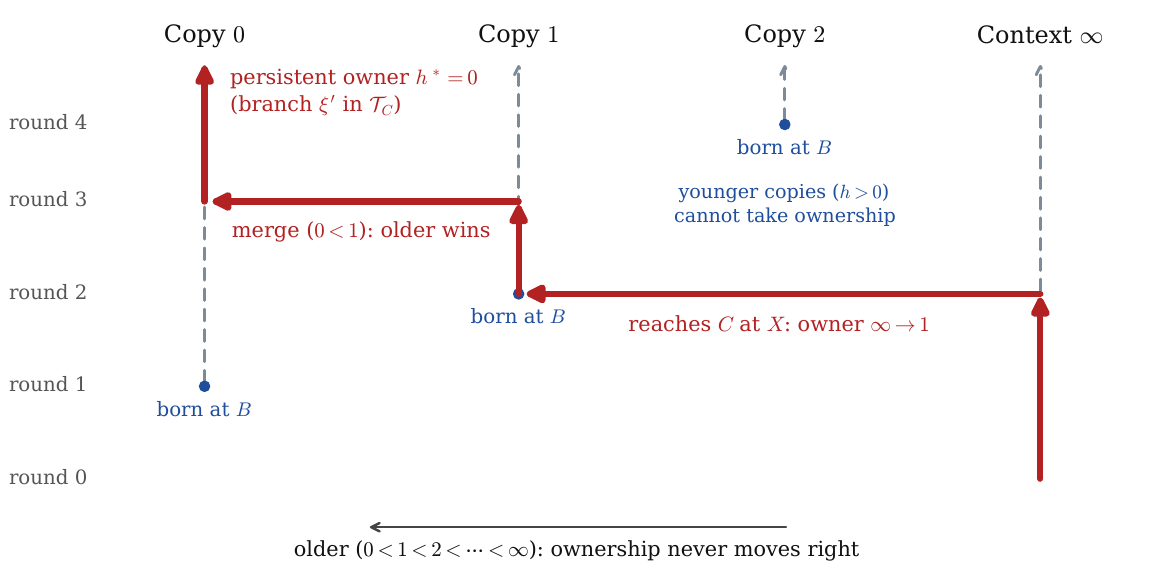}
\caption{Schematic older-wins relay. Rounds run upwards; a target trace can change owner only towards an older copy, and ultimately follows one persistent lane. Dashed lines are copy lanes, not tableau edges.}
\label{fig:relay}
\end{figure}
\begin{proof}[Proof of Theorem~\ref{thm:positive-context-closure}]
Syntactic preservation is Lemma~\ref{lem:pcc-syntax}, and the local clauses of
$Z$ (together with the property that every modal pair in $Z$ lies on a finite
$Z$-path from the roots) are Lemma~\ref{lem:pcc-Z-local}. For~(R-par), let $(\tilde m_j,m_j)_{j\ge 0}$
be an infinite $Z$-path with left projection
$\tilde\xi=\tilde t_0\tilde m_0\tilde u_0\tilde m_1\cdots$ and even right
projection $\xi=t_0m_0u_0m_1\cdots$. Since the path is infinite, every
$L(\tilde m_j)$ is consistent and every $L(m_j)$ is $\nabla$-non-free, so each
$(\tilde m_j,m_j)$ carries its unique position $(\mathcal H_j,\{(b_{j,h},c_{j,h})\})$
with owner function $\own_j$. Let $\tau$ be a $\mu$-trace on $\xi$, with
$(A(C),\varepsilon_0,V_{-1})\in R(t_0,m_0)$ and
$(\varepsilon_j,\varepsilon_{j+1},V_j)\in R(m_j,u_j)\circ R(u_j,m_{j+1})$
($j\ge 0$), whose $\sqsubseteq$-least infinitely regenerated variable
$z^*\in\Reg^\infty(\tau)$ has $\eta_{z^*}=\mu$ in $A(C)$. Each $\varepsilon_j$ is
a $\nabla$-formula. By~\eqref{eq:pcc-own-decr}, $o_j:=\own_j(\varepsilon_j)$ is
non-increasing in $\mathbb N\cup\{\infty\}$, so there exist $J\ge 0$ and
$\kappa^*\in\mathbb N\cup\{\infty\}$ with $o_j=\kappa^*$ for all $j\ge J$.

\smallskip\noindent\emph{Case 1: $\kappa^*=\infty$.} Then $o_j=\infty$ for all
$j\ge 0$. Writing $\varepsilon_j=\gamma_j^\sharp$ ($\gamma_j\in\Sub(A(X))\setminus\{X\}$),
Lemma~\ref{lem:pcc-trace-step}(B) and~(A)(ii) yield an infinite trace $\tilde\tau$
on $\tilde\xi$ through $\gamma_j^\flat\in L(\tilde m_j)$ with identical
regeneration sets, so $\Reg^\infty(\tilde\tau)=\Reg^\infty(\tau)\subseteq\Bound(A(X))$.
By Hypothesis~(3) and Lemma~\ref{lem:pcc-syntax}, binder types and $\sqsubseteq$
on $\Bound(A(X))$ agree in $A(C)$ and $A(B)$, so $z^*$ is the $\sqsubseteq$-least
element of $\Reg^\infty(\tilde\tau)$ in $A(B)$ and $\tilde\xi$ is even.

\smallskip\noindent\emph{Case 2: $\kappa^*=h\in\mathbb N$.} Let $r_0\le J$ be the
index at which copy $h$ first appears in $\mathcal H_{r_0}$ ($r_0=0$ if born at
the root, $r_0=b+1$ if born at $(\tilde u_b,\tilde m_{b+1})$), so
$b_{r_0,h}\in C(r_{\mathcal T_B})$ and $c_{r_0,h}\in C(r_{\mathcal T_C})$. For
$j\ge J$, $\varepsilon_j\in L(c_{j,h})$ is a $\nabla$-formula, so
$h\in\mathcal H_{j,\mathrm{act}}$; since newly born copies at $j\ge r_0$ receive
birth times $N_{\mathrm{born}}>h$, copy $h$ could never have re-entered after
being discarded, whence $h\in\mathcal H_{j,\mathrm{act}}$ for \emph{all}
$j\ge r_0$. Thus $(b_{j,h},c_{j,h})_{j\ge r_0}$ is an infinite $W$-path from
$(r_{\mathcal T_B},r_{\mathcal T_C})$, with projections
$\xi_B=r_{\mathcal T_B}b_{r_0,h}v_{r_0,h,B}\cdots$ in $\mathcal T_B$ and
$\xi_C=r_{\mathcal T_C}c_{r_0,h}v_{r_0,h,C}\cdots$ in $\mathcal T_C$.

For all $j\ge J$, Lemma~\ref{lem:pcc-trace-step}(A)(i) gives
$(\varepsilon_j,\varepsilon_{j+1},V_j)\in R(c_{j,h},v_{j,h,C})\circ R(v_{j,h,C},c_{j+1,h})$
with $V_j\subseteq\Bound(C)$. By Lemma~\ref{lem:pcc-trace-extension}, this tail
extends backwards to an infinite trace $\tau_C$ on $\xi_C$ from $r_{\mathcal T_C}$
with $\Reg^\infty(\tau_C)=\Reg^\infty(\tau)\subseteq\Bound(C)$. Since defining
subformulas, $\sqsubseteq$, and $\mu/\nu$-types on $\Bound(C)$ coincide in $A(C)$
and $C$ (Hypothesis~(3)), $\tau_C$ is a $\mu$-trace and $\xi_C$ is even. By~(R-par)
for $W$, $\xi_B$ is even and carries a $\mu$-trace $\tau_B$ with steps
$(\beta_j,\beta_{j+1},U_j)\in R(b_{j,h},v_{j,h,B})\circ R(v_{j,h,B},b_{j+1,h})$
($j\ge r_0$, $\beta_j=\nabla\Theta_j\in L(b_{j,h})\subseteq L(\tilde m_j)$,
$U_j\subseteq\Bound(B)$) and $\sqsubseteq$-least infinitely regenerated
$\mu$-variable $w^*\in\Bound(B)$. For every $j\ge r_0$, with
$\tilde c_j:=c_{(\tilde u_j,\tilde m_{j+1})}^+$, the trace identity of
Lemma~\ref{lem:diamond-macro-step} (used in Step~3) gives
$(\bigvee\Theta_j,\beta_{j+1},U_j)\in R_{\tilde c_j|_B}(B_{b_{j,h}})\subseteq R_{\tilde c_j}(L^+(\tilde u_j))$
(since $B_{b_{j,h}}\subseteq L^+(\tilde u_j)$ and $\Bound(B)\cap\Bound(A(X))=\emptyset$),
and Lemma~\ref{lem:full-successor}(2) on $\widetilde{\mathcal T}$ yields
$(\beta_j,\beta_{j+1},U_j)\in R(\tilde m_j,\tilde u_j)\circ R(\tilde u_j,\tilde m_{j+1})$.
Extending $(\beta_j)_{j\ge r_0}$ backwards from $\tilde m_{r_0}$ to $\tilde t_0$
by Lemma~\ref{lem:pcc-trace-extension} gives an infinite trace $\tilde\tau$ on
$\tilde\xi$ with $\Reg^\infty(\tilde\tau)=\Reg^\infty(\tau_B)\subseteq\Bound(B)$.
Again $\sqsubseteq$ and $\mu/\nu$-types on $\Bound(B)$ coincide in $B$ and
$A(B)$, so $w^*$ witnesses that $\tilde\tau$ is a $\mu$-trace and $\tilde\xi$ is
even. Thus $\mathcal T_{A(B)}\Rightarrow_Z\mathcal T_{A(C)}$.
\end{proof}

\begin{example}[Two aged copies sharing a target formula]\label{ex:pcc-older-wins}
Here is a local support calculation, illustrating Step~5 rather than specifying an entire tableau. Suppose Copy~$0$ was born in an earlier round and Copy~$1$ in the next, and both are active. Let $D=\alpha\lor\beta$ be a target disjunction, with $D\in S_0\cap S_1$, $f_0(D)=\alpha$ and $f_1(D)=\beta$. Then
\[
\own'(D)=\min\{0,1\}=0,\qquad c(D)=f_0(D)=\alpha.
\]
Thus the target follows $D\to\alpha$ even though the younger copy would choose $\beta$. If $\alpha\in S_0$, then $\own'(\alpha)\le0$, hence $\own'(\alpha)=0$. Both copies retain their original birth times and their respective $W$-pairs; only the formula's owner is selected. In the next round the same calculation applies whenever both supports again contain $D$. This local example isolates the merge mechanism without asserting that every pair of conflicting supports must arise on one prescribed branch.
\end{example}
\subsection{Proof of the unfolding consequence lemma}
\label{sec:pcc-unfold-proof}

\begin{proof}[Proof of Lemma~\ref{lem:pcc-unfold}]
By Lemma~\ref{lem:box-isolation}(3), $\psi$ and its renamed copy $\psi'$ are
clean, guarded, and box-isolated, and hence so is $\alpha(\psi')$ by
Lemma~\ref{lem:pcc-syntax} (with $A(X):=\alpha(X)$ and $D:=\psi'$). Write
$\widetilde{\mathcal U}:=\mathcal T_{\alpha(\psi')}$ and $\mathcal T:=\mathcal T_\psi$.
Every $\gamma\in\Cl(\psi)$ has an isomorphic copy $\gamma'\in\Cl(\psi')\subseteq\Cl(\alpha(\psi'))$
via $z\mapsto z'$ ($z\in\Bound(\psi)$), matching box chains as macro steps
$\bigvee\Psi\to\theta$ and $\bigvee(\Psi')\to\theta'$. For
$\gamma\in\Sub(\alpha(x))\cup\mathrm{Head}(\alpha(x))$, set
$\gamma^\flat:=\gamma[\psi'/x]\in\Cl(\alpha(\psi'))$ and $\psi^\flat:=\alpha(\psi')$,
so $x^\flat=\psi'$, while on the right side $\Sub(\alpha(x))$ embeds into
$\Cl(\psi)$ by the identity map (playing the role of $(\cdot)^\sharp$ in
Construction~\ref{con:pcc-round}); note that $\gamma\mapsto\gamma^\flat$ is
injective on $\Sub(\alpha(x))\cup\mathrm{Head}(\alpha(x))$ by
Lemma~\ref{lem:pcc-outer-inj}(a) applied with $C=\psi'$.

We use two owner lanes $0<\infty$: a modal pair $(\tilde m,m)\in\Mod(\widetilde{\mathcal U})\times\Mod(\mathcal T)$
\emph{carries a two-lane position} if
\begin{equation}\label{eq:pcc-2lane}
L(m)\subseteq\{\varpi\in\Cl(\psi)\mid\varpi'\in L(\tilde m)\}\;\cup\;
\{\varpi\in\Sub(\alpha(x))\setminus\{x\}\mid\varpi^\flat\in L(\tilde m)\},
\end{equation}
with $\own(\varpi):=0$ if $\varpi'\in L(\tilde m)$ and $\own(\varpi):=\infty$
otherwise. For consistent $L(\tilde m)$,~\eqref{eq:pcc-2lane} implies~(R-loc) at
once since $l'=l=l^\flat$ on literals and $(\nabla\Xi)',(\nabla\Xi)^\flat$ are
$\nabla$-formulas.

\smallskip\noindent\emph{One round from $(\tilde m,m)$ ($L(\tilde m)$ consistent,
$L(m)$ $\nabla$-non-free).}
In the (R-back) mode, given the $(k,\theta)$-successor $u_R\in C(m)$ for
$\nabla\Xi_k\in L(m)$, let $\tilde u\in C(\tilde m)$ have principal formula
$\nabla(\Xi'_k)$ and element $\theta'$ if $\own(\nabla\Xi_k)=0$, and principal
formula $\nabla(\Xi_k^\flat)$ and element $\theta^\flat$ if $\own(\nabla\Xi_k)=\infty$;
pair inconsistent $\tilde m'\in C(\tilde u)$ with a fixed $m'_{\mathrm{inc}}(u_R)\in C(u_R)$.
Given $\tilde u\in C(\tilde m)$ and consistent $\tilde m'\in C(\tilde u)$, put
$\tilde c_0:=c_{(\tilde u,\tilde m')}$ and $\tilde c:=\tilde c_0^+$; then
$\nabla\emptyset\notin L(\tilde m)$ and hence $\nabla\emptyset\notin L(m)$
by~\eqref{eq:pcc-2lane}. In the (R-forth$^2$) mode, pick any $\nabla\Xi_k\in L(m)$
and let $u_R\in C(m)$ be the $(k,\theta)$-successor where $\tilde c$ leads to
$\theta'$ from $\bigvee(\Xi'_k)\in L^+(\tilde u)$ (if $\own(\nabla\Xi_k)=0$) or
to $\theta^\flat$ from $\bigvee(\Xi_k^\flat)\in L^+(\tilde u)$ (if $\own(\nabla\Xi_k)=\infty$).

Define choice functions $f_0,f_\infty$ on $\Cl(\psi)$: $f_0(D)$ is the inverse
$'$-image of $\tilde c(D')$ for $D\in\Sub(\psi)$, and on the box chain of
$\nabla\Xi\in\Sub(\psi)$ with $\bigvee(\Xi')\in S_{\tilde c}(L^+(\tilde u))$ it
leads to the $\theta\in\Xi$ with $\theta'$ reached by $\tilde c$ from
$\bigvee(\Xi')$; let $S_0\subseteq\Cl(\psi)$ be the $E_{f_0}$-closure of
$\{\delta\in\Sub(\psi)\cup\mathrm{Head}(\psi)\mid\delta'\in S_{\tilde c}(L^+(\tilde u))\}$,
so modal $\varpi\in S_0$ satisfy $\varpi'\in L(\tilde m')$. Define $f_\infty$ to mirror $\tilde c$ through $(\cdot)^\flat$ on the outer context, and let $S_\infty$ be the set of formulas reachable from
$\{\gamma\in L(u_R)\mid\gamma\in\Sub(\alpha(x))\cup\mathrm{Head}(\alpha(x)),\;\gamma^\flat\in S_{\tilde c}(L^+(\tilde u))\}$
by $E_{f_\infty}$-paths which do not traverse $x\to\alpha(x)$.
Put $\own'(\delta):=0$ on $S_0$ and $\own'(\delta):=\infty$ on $S_\infty\setminus S_0$, set $c(D):=f_{\own'(D)}(D)$ for $D\in S_0\cup S_\infty$ (left disjunct outside $S_0\cup S_\infty$), and let $m'\in C(u_R)$ be the modal child with outcome $c$.

\begin{claim}[Two-lane support and position preservation]\label{clm:pcc-two-lane}
With the choices above:
\begin{enumerate}[(i)]
\item $L(u_R)\subseteq S_0\cup S_\infty$;
\item every $E_c$-edge from $S_0\cup S_\infty$ stays in this union and does not increase its owner in $0<\infty$;
\item the modal outcome $(\tilde m',m')$ again satisfies~\eqref{eq:pcc-2lane}, with $\own_{\mathrm{next}}=\own'|_{L(m')}$.
\end{enumerate}
\end{claim}
\begin{proof}
For (i), if a formula of $L(u_R)$ is generated by a modal predecessor owned by lane $0$, it belongs to the full successor support $S_0$; if its predecessor is owned by $\infty$, the corresponding outer formula is a generator of $S_\infty$.

For (ii), $S_0$ is $E_{f_0}$-closed. The set $S_\infty$ is $E_{f_\infty}$-closed until it reaches $x$. If $x\in S_\infty$, then $x^\flat=\psi'\in S_{\tilde c}(L^+(\tilde u))$, and the unfolding edge $\psi'\to\alpha'(x')=(\alpha(x))'$ places $\alpha(x)$ in $S_0$. Every selected successor remains in the union, and its owner is at most that of its predecessor. It may decrease from $\infty$ to $0$ whenever the successor is also supplied by $S_0$; the transition at $x$ is the point at which the inner lane is necessarily entered.

For (iii), if $\varpi\in L(m')$, then $\varpi\in S_0\cup S_\infty$. For every modal formula $\varpi\in\Cl(\psi)$, $\varpi\in S_0$ if and only if $\varpi'\in L(\tilde m')$: both sides describe the modal outcomes of the same mirrored saturation. This also holds for literals, since $l'=l$. A modal formula in $S_\infty\setminus S_0$ has the form $\gamma\in\Sub(\alpha(x))\setminus\{x\}$ with $\gamma^\flat\in L(\tilde m')$. These are exactly the two alternatives in~\eqref{eq:pcc-2lane}; the equivalence for lane $0$, including literals, shows that $\own_{\mathrm{next}}=\own'|_{L(m')}$.
\end{proof}

\paragraph{The formula of~\cite[Example~5.20]{EnqvistSeifanVenema18}.}
Let $\alpha(x)=\Diamond((x\land p)\lor q)$. In the tableau construction, the disjunction $(x\land p)\lor q$ is resolved by the left and right tableaux in the same saturation round. If $q$ is selected, the outer lane remains active; if $x\land p$ is selected and the trace reaches $x$, control passes to the inner lane following $\psi'$. The failure described in~\cite[Example~5.20]{EnqvistSeifanVenema18} has a different source: in the automaton $\mu x.A$, the state is split at a disjunction and the right-hand choice is thereby fixed one round earlier. No such premature commitment occurs here, because both tableaux resolve the displayed disjunction in the same saturation step.

In the initial round from $(\{\alpha(\psi')\},\{\psi\})$, for consistent
$\tilde m_0\in C(r_{\widetilde{\mathcal U}})$ with $\tilde c_0:=c_{(r_{\widetilde{\mathcal U}},\tilde m_0)}$,
define $S_0$ from $S_{\tilde c_0}(\{\alpha(\psi')\})$ and start $S_\infty$ from
$\{\psi,\alpha(x)\}$ (as $\psi\to\alpha(x)$ has empty regeneration and
$(\alpha(x))^\flat=\alpha(\psi')$), obtaining $(\tilde m_0,m_0)$ satisfying~\eqref{eq:pcc-2lane}.

Define $U:=\bigcup_{d\ge 0}U^{(d)}$ recursively as in Definition~\ref{def:pcc-Z},
restricting clause~(b) to left choice nodes $\tilde u$ not paired with any right choice node in clause~(a); then every modal pair of $U$ lies on a unique
finite $U$-path and carries a unique two-lane position when $L(\tilde m)$ is
consistent, and $U$ satisfies~(R-ch), (R-loc), (R-back), and~(R-forth$^2$). For~(R-par),
let $\tau$ be a $\mu$-trace on the right projection $\xi$ of an infinite
$U$-path, with $(\psi,\varepsilon_0,V_{-1})\in R(r_{\mathcal T},m_0)$ and
$(\varepsilon_j,\varepsilon_{j+1},V_j)\in R(m_j,u_j)\circ R(u_j,m_{j+1})$. Since
$\own_{j+1}(\varepsilon_{j+1})\le\own_j(\varepsilon_j)\in\{0,\infty\}$:
\begin{itemize}
\item If $\own_j(\varepsilon_j)=\infty$ for all $j\ge 0$, no step traverses
$x\to\alpha(x)$. In the initial round, the $E_{f_\infty}$-path from $\psi$
to $\varepsilon_0$ begins with the deterministic unfolding $\psi\to\alpha(x)$
(empty regeneration) and avoids $x\to\alpha(x)$ thereafter, so its
$(\cdot)^\flat$-translation gives
$(\alpha(\psi'),\varepsilon_0^\flat,V_{-1})\in R(r_{\widetilde{\mathcal U}},\tilde m_0)$
with $V_{-1}\subseteq\Bound(\alpha)\setminus\{x\}$; for each $j\ge 0$,
Lemma~\ref{lem:full-successor}(2) gives
$(\varepsilon_j^\flat,\varepsilon_{j+1}^\flat,V_j)\in R(\tilde m_j,\tilde u_j)\circ R(\tilde u_j,\tilde m_{j+1})$.
Thus $\tau^\flat$ is a $\mu$-trace on $\tilde\xi$.
\item Otherwise $\own_j(\varepsilon_j)=0$ for all $j\ge J$ for some $J\ge 0$. For
each $j\ge J$, the constant-$0$ $E_c$-path from $L(u_j)$ to $\varepsilon_{j+1}$
is an $E_{f_0}$-path, whose $'$-image prefixes via the box chain of
$\varepsilon'_j\in L(\tilde m_j)$ to an $E_{\tilde c}$-path from
$(\bigvee\Xi_j)'\in L^+(\tilde u_j)$ to $\varepsilon'_{j+1}\in L(\tilde m_{j+1})$
with regenerated set $V'_j:=\{z'\mid z\in V_j\}$. Lemma~\ref{lem:full-successor}(2)
gives $(\varepsilon'_j,\varepsilon'_{j+1},V'_j)\in R(\tilde m_j,\tilde u_j)\circ R(\tilde u_j,\tilde m_{j+1})$
for all $j\ge J$, and Lemma~\ref{lem:pcc-trace-extension} extends this tail
to an infinite trace $\tilde\tau$ on $\tilde\xi$ with
$\Reg^\infty(\tilde\tau)=\{z'\mid z\in\Reg^\infty(\tau)\}\subseteq\Bound(\psi')$,
whose $\sqsubseteq$-least element is the $\mu$-variable $(z^*)'$.
\end{itemize}
Hence $\tilde\xi$ is even in both cases, establishing $U:\mathcal T_{\alpha(\psi')}\Rightarrow\mathcal T_\psi$.
\end{proof}

\section{Concluding remarks}\label{sec:conclusion}
Compared with~\cite[Theorem~4]{EnqvistSeifanVenema18}, our lifting theorem isolates a structural property of arbitrary tableau consequence, relaying $\mathcal T_B\Rightarrow\mathcal T_C$ through suitably separated positive contexts.

The deterministic parity monitor of Section~\ref{sec:anf} serves both the finite-game proof of Claim~(e) and the construction of automaton normal form; two directions suggest themselves.
\begin{enumerate}
\item \emph{Formalisation.} Canonical tableaux, decomposition graphs, consequence relations, tokens and aged copies are explicit finite or inductive data, which should make mechanisation in a proof assistant feasible.
\item \emph{Proof-theoretic translations.} After Kloibhofer's counterexample~\cite{Kloibhofer23}, it is open whether completeness can be obtained directly from an annotated or reset system~\cite{Jungteerapanich10,Stirling14,DekkerKloibhoferMartiVenema23,LeighWehr24}. The older-wins and reset mechanisms of Appendix~\ref{app:thin} are justified in $\Koz$ by the $(\mathrm{Record})$ rule of Lemma~\ref{lem:kozen} and the monotonicity Lemma~\ref{lem:app-decoration}; whether this interpretation extends to conjunctive branching is a natural question.
\end{enumerate}

\appendix
\section{Proof of the thin-refutation theorem (after Walukiewicz)}\label{app:thin}

We prove Theorem~\ref{thm:thin-refutation} by restating the argument of~\cite[Theorem~31]{Walukiewicz00}, which generalises Kozen's proof for aconjunctive formulas~\cite{Kozen83}, for the single-rule expansion $\mathcal R^+$ of Construction~\ref{con:expansion} and for tagged sets: a context is recorded at every regeneration of a $\mu$-variable; on every infinite branch some recorded context recurs, and the branch can be cut there. Walukiewicz's token lists become tokens (variable, birth time) carried by formula occurrences; the correspondence is listed in Remark~\ref{rem:app-correspondence}.

Throughout, $\Gamma_0,\Delta_0$ are fixed as in Definition~\ref{def:two-sided}, $\mathcal R$ is a thin sequential refutation of a finite $\Gamma\subseteq\Cl(\Gamma_0;\Delta_0)$, no formula of which contains a variable of $\Bound(\Gamma_0)\cup\Bound(\Delta_0)$ free, and $\mathcal R^+$ is its single-rule expansion. To keep the notation of Section~\ref{sec:tableaux}, in this appendix $\Cl(\Gamma_0)$ denotes the tagged closure $\Cl(\Gamma_0;\Delta_0)$, $\Bound(\Gamma_0)$ the set of bound variables of $\Gamma_0$ and $\sim\Delta_0$ (disjoint by assumption), and $\Bound_\mu(\Gamma_0)$ those of fixpoint type $\mu$ on their side; $\Free$, $\Sub$, $\varphi_y$ and $\sqsubseteq$ are computed on the side of the formula concerned. Thus a token of a right formula is a $\nu$-variable of $\Delta_0$. The \emph{depth} of a node of $\mathcal R^+$ is its distance from the root; nodes on one branch have distinct depths.

\subsection{Kozen's context-recording rule}

Park's rule lets one \emph{record} a context when unfolding a least fixpoint, so that meeting the context again at a regeneration yields a propositional clash~\cite{Kozen83}.

\begin{lemma}[Kozen's context-recording rule {\cite[Proposition~5.7(vi)]{Kozen83}}]\label{lem:kozen}
Let $x$ be positive in $\alpha(x)$ and $x\notin\Free(\Gamma)$. The following rule is derivable in $\Koz$:
\[
\frac{\alpha\big(\mu x.(\sim\bigwedge\Gamma\land\alpha(x))\big),\Gamma\vdash}{\mu x.\alpha(x),\Gamma\vdash}\ (\mathrm{Record}).
\]
\end{lemma}

\begin{proof}
We first show, for $x\notin\Free(\vartheta)$: if $\alpha(\mu x.(\vartheta\land\alpha(x)))\vdash\vartheta$, then $\mu x.\alpha(x)\vdash\vartheta$. Put $M:=\mu x.(\vartheta\land\alpha(x))$. Working in the one-sided calculus via~\eqref{eq:two-one-sided}, the hypothesis is $\alpha(M),\sim\vartheta\vdash$.
\begin{enumerate}[(i)]
\item $\alpha(M),\sim(\vartheta\land\alpha(M))\vdash$: since $\sim(\vartheta\land\alpha(M))=\sim\vartheta\lor\sim\alpha(M)$, apply $(\lor)$; the left premise is the hypothesis, and the right premise $\alpha(M),\sim\alpha(M)\vdash$ is $(\mathrm{Tau})$.
\item $\vartheta\land\alpha(M),\sim M\vdash$ is an instance of $(\mathrm{Prefix})$ for the body $\vartheta\land\alpha(x)$.
\item $(\mathrm{Cut})$ on (i) and (ii) gives $\alpha(M),\sim M\vdash$, and $(\mathrm{Ind})$ gives $\mu x.\alpha(x),\sim M\vdash$.
\item $M,\sim\vartheta\vdash$: by $(\eta\text{-}\mathrm{unf})$ (Lemma~\ref{lem:standard}(2)) it suffices to show $\vartheta\land\alpha(M),\sim\vartheta\vdash$, which follows by $(\land)$, $(\mathrm{Tau})$ and $(\mathrm{Weak})$.
\item $(\mathrm{Cut})$ on (iii) and (iv) gives $\mu x.\alpha(x),\sim\vartheta\vdash$, i.e.\ $\mu x.\alpha(x)\vdash\vartheta$.
\end{enumerate}
Now take $\vartheta:=\sim\bigwedge\Gamma$, so that $M=\mu x.(\sim\bigwedge\Gamma\land\alpha(x))$ and $\sim\vartheta=\bigwedge\Gamma$. From the premise $\alpha(M),\Gamma\vdash$, repeated $(\land)$ (or $(\mathrm{Weak})$ if $\Gamma=\emptyset$) gives $\alpha(M),\bigwedge\Gamma\vdash$, i.e.\ the hypothesis, and the first part yields $\mu x.\alpha(x),\bigwedge\Gamma\vdash$. $(\mathrm{Cut})$ with $\Gamma,\sim\bigwedge\Gamma\vdash$, which follows from $(\mathrm{Bot})$ if $\Gamma=\emptyset$, and from $(\lor)$, $(\mathrm{Tau})$ and $(\mathrm{Weak})$ otherwise, gives $\mu x.\alpha(x),\Gamma\vdash$.
\end{proof}

\subsection{Dependency closure}

\begin{definition}[Dependency closure]\label{def:dep}
For $\psi\in\Cl(\Gamma_0)$, $\Dep(\psi)\subseteq\Bound(\Gamma_0)$ is the least set containing $\Free(\psi)\cap\Bound(\Gamma_0)$ and containing $\Free(\varphi_y)\cap\Bound(\Gamma_0)$ whenever $y\in\Dep(\psi)$. Put $\Dep_\mu(\psi):=\Dep(\psi)\cap\Bound_\mu(\Gamma_0)$.
\end{definition}

Thus $\Dep(\psi)$ is the set of variables whose defining formulas are needed to expand $\psi$ (Definition~\ref{def:expansion}), equivalently the variables that can be regenerated from $\psi$ along $E$-paths.

\begin{lemma}\label{lem:app-dep}
\begin{enumerate}
\item $\Dep(\psi)$ is a $\sqsubseteq$-chain, and every $x\in\Dep(\psi)$ is active in $\psi$.
\item If $\psi\to\psi'$ is an edge of $E$ other than an unfolding $\eta_xx.\varphi_x\to\varphi_x$, or a modal step $\nabla\Psi\to\theta$ ($\theta\in\Psi$) or $\nabla\Psi\to\bigvee\Psi$, then $\Dep(\psi')\subseteq\Dep(\psi)$; moreover $\Dep(\varphi_x)\subseteq\Dep(\eta_xx.\varphi_x)\cup\{x\}$.
\item If a $(\land)$-step of $\mathcal R^+$ keeps both conjuncts $\alpha,\beta$ of $\alpha\land\beta$, then $\Dep_\mu(\alpha)\cap\Dep_\mu(\beta)=\emptyset$.
\end{enumerate}
\end{lemma}

\begin{proof}
(1) Two variables free in a formula of $\Cl(\Gamma_0)$ have binders that both enclose an occurrence of that formula in its side, $\Gamma_0$ or $\sim\Delta_0$ (for a box-chain disjunction, of the corresponding $\nabla$-formula), hence are nested; and if $z\in\Free(\varphi_y)$ then $z\sqsubset y$ by cleanness. Hence $\Dep(\psi)$ is a chain. If $x\in\Dep(\psi)$, there are $y_1\in\Free(\psi)$ and $y_{i+1}\in\Free(\varphi_{y_i})$ with $x\in\Free(\varphi_{y_k})$ or $x=y_1$; then $x\preceq y_k\preceq\dots\preceq y_1$, so $x$ is active in $\psi$. (2) For subformula and modal steps, $\Free(\psi')\subseteq\Free(\psi)$; for a regeneration $x\to\varphi_x$, $\Dep(\varphi_x)\subseteq\Dep(x)$ by definition; for an unfolding, $\Free(\varphi_x)\subseteq\Free(\eta_xx.\varphi_x)\cup\{x\}$. (3) If $x\in\Dep_\mu(\alpha)\cap\Dep_\mu(\beta)$, then $x$ is active in both conjuncts by (1), and $\alpha\land\beta\in\Sub(\varphi_x)$ as in the proof of Lemma~\ref{lem:thin-reduction}(3), contradicting thinness of $\mathcal R^+$.
\end{proof}

\subsection{Tokens}

This subsection concerns only formulas and birth times; $\Koz$ does not occur in it.

\begin{definition}[Occurrences and tokens]\label{def:aged-occurrence}
An \emph{occurrence} is a pair $o=(\psi,b)$ with $\psi\in\Cl(\Gamma_0)$ and a \emph{birth function} $b:\Dep_\mu(\psi)\to\mathbb N$. The \emph{tokens} of $o$ are the pairs $(x,b(x))$, $x\in\Dep_\mu(\psi)$, and $o$ \emph{carries} them. If $\Dep_\mu(\psi)\neq\emptyset$, the \emph{age} of $o$ is $b(y_0)$ for the $\sqsubseteq$-least element $y_0$ of the chain $\Dep_\mu(\psi)$ (Lemma~\ref{lem:app-dep}(1)); of two occurrences with the same formula, the one with the smaller age is \emph{older}. For $\psi'$ with $\Dep(\psi')\subseteq\Dep(\psi)$ put $o{\restriction}\psi':=(\psi',b{\restriction}\Dep_\mu(\psi'))$.
\end{definition}

We assign to each node $n$ of $\mathcal R^+$ a set $\Phi(n)$ of occurrences with pairwise distinct formulas, whose set of formulas is $L(n)$, by induction on depth. The root receives $(\gamma,\emptyset)$ for $\gamma\in\Gamma$; by assumption $\Dep(\gamma)=\emptyset$. For a node $n$ with principal formula $\psi$, carried by $o=(\psi,b)\in\Phi(n)$, and $\Phi':=\Phi(n)\setminus\{o\}$, the rule at $n$ produces the \emph{candidate} occurrences for the children:
\begin{description}
\item[$(\lor)$] $\psi=\alpha\lor\beta$: $\Phi'\cup\{o{\restriction}\alpha\}$ for the child of $\alpha$, $\Phi'\cup\{o{\restriction}\beta\}$ for the child of $\beta$.
\item[$(\land)$] $\psi=\alpha\land\beta$: $\Phi'\cup\{o{\restriction}\chi\mid\chi\in\{\alpha,\beta\}\text{ kept}\}$.
\item[$(\eta)$] $\psi=\eta_xx.\varphi_x$: $\Phi'\cup\{(\varphi_x,b^+)\}$, where $b^+$ extends $b{\restriction}\Dep_\mu(\varphi_x)$ by $b^+(x):=\operatorname{depth}(n)$ if $\eta_x=\mu$; the token $(x,\operatorname{depth}(n))$ is \emph{born} at $n$.
\item[$(\mathrm{reg})$] $\psi=x$: $\Phi'\cup\{o{\restriction}\varphi_x\}$. If $\eta_x=\mu$, the token $(x,b(x))$ \emph{records} at $n$.
\item[$(\mathrm{Weak})$] $\Phi'$.
\item[$(\nabla)$] $L(n)$ modal, child the $(k,\theta)$-successor: the occurrence of $\nabla\Psi_k$ contributes its restriction to $\theta$, the occurrence of $\nabla\Psi_n$ ($n\neq k$) its restriction to $\bigvee\Psi_n$; literals are dropped.
\end{description}
All restrictions are legitimate by Lemma~\ref{lem:app-dep}(2). Whenever two candidates have the same formula $\psi'$, they are \emph{merged}: if $\Dep_\mu(\psi')=\emptyset$, both are $(\psi',\emptyset)$ and coincide; otherwise their ages differ (Lemma~\ref{lem:app-births}), the older one is kept and the other is dropped. A token is \emph{alive} at $n$ if some occurrence of $\Phi(n)$ carries it. Every token is born at an $(\eta)$-step, at most one at each node, and a token alive at $n$ was born at a proper ancestor of $n$; hence distinct tokens alive at the same node have distinct births. Every token of a candidate produced at $n$ is a token of its parent occurrence, except the token born at $n$. Hence, on a branch, the nodes at which a token $(y,t)$ is alive form an interval of depths, possibly empty, starting at $t+1$ (\emph{death is permanent}); in particular, a token that records infinitely often on a branch $P$ is alive at every node of $P$ of depth $>t$.

\begin{lemma}[Distinct carriers]\label{lem:app-births}
At every node $n$, no two distinct candidates produced at $n$ for the same child carry the same token, and no two distinct occurrences of $\Phi(n)$ do. Consequently, two such occurrences with the same formula $\psi$ and $\Dep_\mu(\psi)\neq\emptyset$ have different ages.
\end{lemma}

\begin{proof}
By induction on depth. At the root no occurrence carries a token. Each candidate produced at $n$ for a child is produced from one occurrence of $\Phi(n)$, and a token it carries is carried by that occurrence or is the token born at $n$, which only one candidate carries. The only rule producing two candidates from one occurrence is $(\land)$ with both conjuncts kept; by Lemma~\ref{lem:app-dep}(3) the two conjuncts have disjoint $\Dep_\mu$, so at most one of them carries a given token; and along a lineage, $\Dep_\mu$ can regain a variable $y$ only by an $(\eta)$-step for $y$, which creates a token with a new birth (Lemma~\ref{lem:app-dep}(2)). By the induction hypothesis for $\Phi(n)$, at most one candidate carries a given token, and the occurrences of the child form a subset of the candidates. Two occurrences with the same formula have the same $\Dep_\mu$; their tokens at its $\sqsubseteq$-least element $y_0$ are distinct, so their ages $b(y_0)$ differ.
\end{proof}

Fix an infinite branch $P=n_0n_1\cdots$ of $\mathcal R^+$. By Lemma~\ref{lem:thin-reduction}(2) it carries a $\mu$-trace $\tau=\psi_0\psi_1\cdots$, realised by single steps; let $x$ be the $\sqsubseteq$-least element of $\Reg^\infty(\tau)$, a $\mu$-variable, and let $o_j$ be the occurrence of $\psi_j$ in $\Phi(n_j)$.

\begin{lemma}[Trace preservation]\label{lem:app-trace}
There are $N$ and $\beta$ such that for all $j\ge N$: $x\in\Dep(\psi_j)$, the token $(x,\beta)$ is carried by $o_j$, and $o_{j+1}$ is produced from $o_j$ by the rule at $n_j$ (so $o_j$ is never dropped at a merge). In particular $(x,\beta)$ is alive at every $n_j$ with $j\ge N$, hence at every node of $P$ of depth $>\beta$, and it records infinitely often on $P$.
\end{lemma}

\begin{proof}
The proof of Lemma~\ref{lem:outermost} applies verbatim to the single-step realisation $\tau$: there is $N_0$ such that for $j\ge N_0$ every $\psi_j$ lies in $\Cl(\varphi_x)$ and every variable regenerated at or after $n_j$ is $\sqsupseteq x$. For $j\ge N_0$, $x\in\Dep(\psi_j)$: the trace later regenerates $x$, so an $E$-path leads from $\psi_j$ to the formula $x$, and Lemma~\ref{lem:app-dep}(2) propagates $x\in\Dep(x)$ backwards along it (the only step that could remove $x$ is an unfolding of $\eta_xx.\varphi_x$, which lies outside $\Cl(\varphi_x)$). Hence $D:=\{y\in\Dep_\mu(\psi_j)\mid y\sqsubseteq x\}$ is the same for all $j\ge N_0$ (it equals $\{y\in\Dep_\mu(\varphi_x)\mid y\sqsubset x\}\cup\{x\}$), and it is an initial segment of the chain $\Dep_\mu(\psi_j)$. Its least element $y_0$ is therefore the least element of $\Dep_\mu(\psi_j)$, and $b_{o_j}(y_0)$ is the age of $o_j$, writing $o_j=(\psi_j,b_{o_j})$. If $o_{j+1}$ is produced from $o_j$, then $b_{o_{j+1}}(y_0)=b_{o_j}(y_0)$: no $y\in D$ is rebound after $N_0$, since $\eta_yy.\varphi_y\notin\Cl(\varphi_x)$. Otherwise $o_{j+1}$ is the winner of a merge against the candidate produced from $o_j$; it has the same formula and a smaller age, so $b_{o_{j+1}}(y_0)<b_{o_j}(y_0)$. A sequence of natural numbers decreases only finitely often, so there is $N\ge N_0$ after which $o_{j+1}$ is produced from $o_j$, and $b_{o_j}(x)=:\beta$ is constant. The token $(x,\beta)$ records at every regeneration of $x$ along $\tau$ after $N$, which happens infinitely often.
\end{proof}

\subsection{Decorations and translation}

\begin{definition}[Decorations]\label{def:decoration}
A \emph{decorated occurrence} $o=(\psi,d,b)$ adds to an occurrence $(\psi,b)$ a \emph{decoration} $d$ assigning to each $x\in\Dep_\mu(\psi)$ a finite set $d(x)$ of closed formulas (its \emph{recorded contexts}); restrictions restrict $d$ as well. The \emph{translation} $\llangle o\rrangle$ is the closed formula obtained from the underlying formula of $\psi$ by substituting, for the variables of the chain $\Dep(\psi)$ from the innermost to the outermost, $y\mapsto\Fix_o(y)$, where
\[
\Fix_o(y):=\mu y.\Bigl(\bigwedge_{\theta\in d(y)}\sim\theta\;\land\;\varphi_y\Bigr)\ \text{ if } \eta_y=\mu,
\qquad
\Fix_o(y):=\nu y.\varphi_y\ \text{ if } \eta_y=\nu.
\]
For a set $\Phi$ of decorated occurrences, $\llangle\Phi\rrangle:=\{\llangle o\rrangle\mid o\in\Phi\}$, and $\Phi\vdash$ means $\llangle\Phi\rrangle\vdash$.
\end{definition}

If all recorded contexts are empty, $\llangle o\rrangle$ is provably equivalent to the expansion $\llangle\psi\rrangle$ of Definition~\ref{def:expansion}, because $\vdash\mu y.(\top\land\varphi_y)\leftrightarrow\mu y.\varphi_y$. The translation commutes with the connectives: $\llangle(\alpha\lor\beta,d,b)\rrangle=\llangle(\alpha,d,b){\restriction}\alpha\rrangle\lor\llangle(\beta,d,b){\restriction}\beta\rrangle$, and likewise for $\land$ and $\nabla$, since variables not in $\Dep$ of a subformula do not occur in it.

\begin{lemma}[Decorations]\label{lem:app-decoration}
Let $o=(\psi,d,b)$ and $o'=(\psi,d',b')$ with $d(x)\subseteq d'(x)$ for all $x$. Then $\vdash\llangle o'\rrangle\to\llangle o\rrangle$. Moreover, if $\psi=x\in\Bound_\mu(\Gamma_0)$ and $\theta\in d(x)$, then $\vdash\llangle o\rrangle\to\sim\theta$.
\end{lemma}

\begin{proof}
The first claim follows from Lemma~\ref{lem:standard}(3), applied at each substitution step to $\vdash\mu y.(\Theta'\land\varphi_y)\to\mu y.(\Theta\land\varphi_y)$ for conjunctions $\Theta'$ extending $\Theta$; the substituted variables occur positively. For the second, $\llangle o\rrangle=\mu x.(\bigwedge_{\theta'\in d(x)}\sim\theta'\land\varphi_x\sigma)$ for the substitution $\sigma$ of the outer variables (the $\sim\theta'$ are closed), and one unfolding (Lemma~\ref{lem:standard}(2)) gives $\sim\theta$.
\end{proof}

We now decorate the occurrences of $\Phi(n)$, by induction on depth; merges keep the decoration of the occurrence kept, which is determined by ages alone. At the root all decorations are empty, so $\llangle\Phi(\mathrm{root})\rrangle=\llangle\Gamma\rrangle$. At a node $n$ with principal occurrence $o=(\psi,d,b)$ and $\Phi':=\Phi(n)\setminus\{o\}$, every candidate keeps the decoration of the occurrence it is produced from (restricted), with two exceptions:
\begin{description}
\item[$(\eta)$] the new variable receives $d^+(x):=\emptyset$;
\item[$(\mathrm{reg})$, $\eta_x=\mu$] let $\Phi'_{\downarrow}$ be obtained from $\Phi'$ by \emph{resetting} every token younger than $(x,b(x))$, i.e.\ replacing $d'(w)$ by $\emptyset$ in every $(\psi',d',b')\in\Phi'$ with $b'(w)>b(x)$; let $\theta:=\bigwedge\llangle\Phi'_\downarrow\rrangle$; the child receives $\Phi'_\downarrow\cup\{(\varphi_x,d^\star,b{\restriction}\Dep_\mu(\varphi_x))\}$, where $d^\star(x):=d(x)\cup\{\theta\}$ and $d^\star$ agrees with $d$ elsewhere. The token $(x,b(x))$ \emph{records} $\theta$ at $n$.
\end{description}
While a token is alive, its decoration thus changes only when it records (it grows) or when it is reset at a record of a token with smaller birth.

\begin{lemma}[Local soundness]\label{lem:app-local}
If $n$ is a leaf, then $\Phi(n)\vdash$. If $n$ is not a leaf and $\Phi(m)\vdash$ for every child $m$ of $n$, then $\Phi(n)\vdash$.
\end{lemma}

\begin{proof}
A leaf is an inconsistent sequent; since a leaf is a modal sequent, its literals contain no variables from $\Bound(\Gamma_0)$ and are unchanged by $\llangle\cdot\rrangle$, so $\llangle\Phi(n)\rrangle$ contains $\bot$ or a pair $p,\neg p$, and is refuted by $(\mathrm{Bot})$ or $(\mathrm{Tau})$ followed by $(\mathrm{Weak})$. Dropping an occurrence at a merge or at $(\mathrm{Weak})$ is an instance of $(\mathrm{Weak})$. For $(\lor)$ and $(\land)$, the translation commutes with the connective, and the claims are instances of the rules $(\lor)$, $(\land)$ and $(\mathrm{Weak})$ of $\Koz$. For $(\eta)$, $\llangle(\eta_xx.\varphi_x,d,b)\rrangle$ is provably equivalent to the translation of the child occurrence by Lemma~\ref{lem:standard}(2),(3), the new decoration $d^+(x)=\emptyset$ contributing only $\top$. For $(\mathrm{reg})$ with $\eta_x=\nu$, this is an instance of $(\eta\text{-}\mathrm{unf})$. For $(\nabla)$, the translation commutes with $\nabla$ and $\bigvee$, and the derivable rule $(\nabla)$ of Lemma~\ref{lem:standard}(4) applies (in its variant if the label contains a twin of the principal formula with the other tag). For $(\mathrm{reg})$ with $\eta_x=\mu$, write $\Theta:=\bigwedge_{\theta'\in d(x)}\sim\theta'$, let $\sigma$ substitute the outer variables of $\varphi_x$ as in Lemma~\ref{lem:app-decoration}, and put $\alpha(x):=\Theta\land\varphi_x\sigma$, so that $\llangle o\rrangle=\mu x.\alpha(x)$. The translation of the new occurrence is $\varphi_x\sigma[M/x]$ with $M=\mu x.(\sim\theta\land\alpha(x))$. If the child is refutable, then $\{\alpha(M)\}\cup\llangle\Phi'_\downarrow\rrangle\vdash$ by $(\land)$ and $(\mathrm{Weak})$, discarding the conjunct $\Theta$, so rule $(\mathrm{Record})$ of Lemma~\ref{lem:kozen}, with $\Gamma:=\llangle\Phi'_\downarrow\rrangle$ (closed, so $x\notin\Free(\Gamma)$) and $\bigwedge\Gamma=\theta$, gives $\{\mu x.\alpha(x)\}\cup\llangle\Phi'_\downarrow\rrangle\vdash$. Finally $\vdash\llangle o'\rrangle\to\llangle o'_\downarrow\rrangle$ for every $o'\in\Phi'$ by Lemma~\ref{lem:app-decoration}, whence $\Phi(n)\vdash$ by $(\mathrm{Cut})$; resetting is thus justified by Lemma~\ref{lem:app-decoration}.
\end{proof}

\subsection{Closed nodes and the proof of Theorem~\ref{thm:thin-refutation}}

\begin{definition}[Closed nodes]\label{def:closed-node}
A node $n$ of $\mathcal R^+$ is \emph{closed} if the rule at $n$ is $(\mathrm{reg})$ with $\eta_x=\mu$, its principal formula $x$ is carried by $o=(x,d,b)\in\Phi(n)$, and the context $\theta=\bigwedge\llangle\Phi'_\downarrow\rrangle$ recorded at $n$, for $\Phi':=\Phi(n)\setminus\{o\}$, already belongs to $d(x)$.
\end{definition}

\begin{lemma}[Closed nodes are refutable]\label{lem:app-closed}
If $n$ is closed, then $\Phi(n)\vdash$.
\end{lemma}

\begin{proof}
By Lemma~\ref{lem:app-decoration}, $\vdash\llangle o\rrangle\to\sim\theta$ since $\theta\in d(x)$, and $\vdash\bigwedge\llangle\Phi'\rrangle\to\bigwedge\llangle\Phi'_\downarrow\rrangle=\theta$ since resetting only removes recorded contexts. As $\theta,\sim\theta\vdash$ by $(\mathrm{Tau})$, two applications of $(\mathrm{Cut})$ give $\Phi(n)\vdash$.
\end{proof}

\begin{lemma}[Closed nodes exist]\label{lem:app-closed-exists}
Every infinite branch of $\mathcal R^+$ contains a closed node.
\end{lemma}

\begin{proof}
Let $P=n_0n_1\cdots$ be an infinite branch, and let $S$ be the set of tokens that record infinitely often on $P$. By Lemma~\ref{lem:app-trace}, $S\neq\emptyset$. Let $\mathfrak t=(v,\beta^*)\in S$ have the least birth; it is unique, since two tokens of $S$ are alive at a common node of $P$ and therefore have distinct births. $\mathfrak t$ is alive at every node of $P$ of depth $>\beta^*$; for such $j$ let $o^{\mathfrak t}_j$ be the occurrence of $\Phi(n_j)$ carrying $\mathfrak t$ (unique by Lemma~\ref{lem:app-births}).

\emph{Older tokens are eventually frozen.} Every token alive on $P$ was born at a node of $P$, so at most $\beta^*$ tokens alive on $P$ have birth $<\beta^*$. Each of them records only finitely often on $P$, since otherwise it would belong to $S$ with a smaller birth. Let $N'>\beta^*$ be a depth after the last record on $P$ of any token with birth $<\beta^*$. As noted before Lemma~\ref{lem:app-local}, the decoration $d(w)$ of an alive token changes only when the token records (it grows) or when it is reset at a record of a token with smaller birth; restrictions keep $d(w)$ as long as $w$ stays in $\Dep_\mu$, and a token whose occurrence is dropped at a merge or at $(\mathrm{Weak})$ dies, since no other occurrence carries it (Lemma~\ref{lem:app-births}). Hence for $j\ge N'$: the decoration $d(v)$ of $\mathfrak t$ in $o^{\mathfrak t}_j$ only grows, and every token with birth $<\beta^*$ that is alive at $n_j$ has a decoration $D_{(w,s)}$ that does not depend on $j$.

\emph{Recorded contexts range over a finite set.} Let $n_j$ ($j\ge N'$) be a node at which $\mathfrak t$ records $\theta_j=\bigwedge\llangle\Phi'_\downarrow\rrangle$. In $\Phi'_\downarrow$ every token with birth $>\beta^*$ has been reset to $\emptyset$, and every token with birth $<\beta^*$ has its frozen decoration; no token of $\Phi'$ has birth $\beta^*$. Thus each occurrence of $\Phi'_\downarrow$ is determined by its formula $\psi'\in\Cl(\Gamma_0)$ and, for each $w\in\Dep_\mu(\psi')$, by whether $d(w)$ is $\emptyset$ or one of the finitely many frozen sets $D_{(w,s)}$. Hence the $\theta_j$ range over a finite set.

\emph{A recurrence.} Since $\mathfrak t$ records infinitely often after $N'$, there are $N'\le j<k$ such that $\mathfrak t$ records at $n_j$ and at $n_k$ with $\theta_j=\theta_k$. The context $\theta_j$ is added to $d(v)$ at $n_j$ and is never removed, so $\theta_k=\theta_j\in d(v)$ in $o^{\mathfrak t}_k=(v,d,b)$, and $n_k$ is closed.
\end{proof}

\begin{proof}[Proof of Theorem~\ref{thm:thin-refutation}]
Let $\mathcal D$ be the tree obtained from $\mathcal R^+$ by removing all proper descendants of closed nodes. It is finitely branching, and by Lemma~\ref{lem:app-closed-exists} it has no infinite branch; by K\"onig's lemma it is finite. Every leaf $n$ of $\mathcal D$ is a closed node or a leaf of $\mathcal R^+$, so $\Phi(n)\vdash$ by Lemma~\ref{lem:app-closed} or Lemma~\ref{lem:app-local}, and every inner node of $\mathcal D$ keeps all its children in $\mathcal R^+$. Lemma~\ref{lem:app-local}, applied from the leaves to the root, gives $\Phi(\mathrm{root})\vdash$, i.e.\ $\Gamma\vdash$. Reading each node $n$ as the sequent $\llangle\Phi(n)\rrangle\vdash$, $\mathcal D$ is the skeleton of a $\Koz$ derivation.
\end{proof}

\begin{remark}[Correspondence with Walukiewicz's proof]\label{rem:app-correspondence}
The following table lists the correspondences between the devices of~\cite[Section~5]{Walukiewicz00} and those used above. Walukiewicz argues by contradiction from an infinite branch of unprovable sequents; here closed nodes give a direct proof.
\begin{center}\small
\begin{tabular}{@{}p{0.47\textwidth}p{0.47\textwidth}@{}}
\toprule
Walukiewicz~\cite[Section~5]{Walukiewicz00} & Appendix~\ref{app:thin}\\
\midrule
token labelled $(\alpha,X)$ & token $(X,t)$ carried by the occurrence of $\alpha$\\
position in the token list (left = older) & birth time $t$ (smaller = older)\\
new token appended at the $(\mu)$-rule & token born at the $(\eta)$-step\\
replaceability: smallest variable first, then list position & older-wins: birth of the outermost token\\
step 7: delete tokens whose variable is not active in $\alpha$ & restriction of $d,b$ to $\Dep_\mu(\psi)$\\
counter of a token & number of contexts recorded since the last reset\\
counters to the right reset to $0$ at a regeneration & decorations of younger tokens reset at $(\mathrm{reg})_\mu$\\
$\mathcal D_{n,\beta}(X)=\mu X.\neg\gamma_1\land\dots\land\neg\gamma_k\land\alpha(X)$ & $\Fix_o(X)$ with $d(X)=\{\gamma_1,\dots,\gamma_k\}$\\
Observation~31.1 (support of the $\mu$-trace) & Lemma~\ref{lem:app-trace}\\
Observation~31.2 (one unprovable son) & Lemma~\ref{lem:app-local}\\
leftmost token with unbounded counter & token of $S$ with least birth\\
\bottomrule
\end{tabular}
\end{center}
The aged copies of Section~\ref{sec:pcc} are the counterpart of this bookkeeping for copies of a consequence relation $W : \mathcal{T}_B \Rightarrow \mathcal{T}_C$ entering $\mathcal{T}_C$ at its root: since $\Bound(C)$ is disjoint from $\Bound(A(X))$, a single birth time suffices, and ownership of a target formula is governed by the older-wins rule.
\end{remark}

\section*{Acknowledgements}
Microsoft Copilot, Anthropic Claude, and Google Gemini were used during manuscript preparation for proof checking, substantial editing and reorganisation, \LaTeX{} and code generation, and preparation of explanatory figures. The author directed their use, reviewed every output, independently checked the mathematical arguments and references, and accepts full responsibility for the accuracy and originality of the manuscript.

% Commented out for arXiv submission (uncomment if required for journal submission)
% \section*{Declarations}
% \noindent\textbf{Funding.} No funding was received to assist with the preparation of this manuscript.\\[2pt]
% \noindent\textbf{Competing interests.} The author has no competing interests to declare that are relevant to the content of this article.\\[2pt]
% \noindent\textbf{Data availability.} Data sharing is not applicable to this article, as no datasets were generated or analysed during the current study.

\end{document}